\documentclass{elsarticle}
\journal{Information and Computation}
\pdfoutput=1

\usepackage[FIGBOTCAP,TABBOTCAP,center,nooneline]{subfigure}
\usepackage[dvips]{rotating}
\usepackage{hyperref}
\usepackage[normalem]{ulem}
\usepackage{algorithm}
\usepackage[noend]{algpseudocode}
\usepackage{amscd}
\usepackage{amsfonts}
\usepackage{amsmath}
\usepackage{amssymb}
\usepackage{booktabs}
\usepackage{color}
\usepackage{comment}
\usepackage{datetime}
\usepackage{epsfig}
\usepackage{graphicx}
\usepackage{latexsym}
\usepackage{marginnote}
\usepackage{multirow}
\usepackage{paralist}
\usepackage{placeins}
\usepackage{rotate} 
\usepackage{subfigure}
\usepackage{tikz}
\usepackage{times} 
\usepackage{todonotes}
\usepackage{varwidth}
\usepackage{verbatim} 
\usepackage{wrapfig}
\usepackage{xcolor,colortbl}
\usepackage{xspace}

\usepackage[utf8]{inputenc}
\usepackage[english]{babel}
\usepackage{csquotes}
\usepackage{tikz}
\usepackage{float}
\usepackage{placeins}
\usepackage{adjustbox}
\usepackage{enumerate}

\usepackage{amsmath}
\usepackage{amssymb}
\usepackage{amsfonts}
\usepackage{amscd}
\usepackage{xspace}
\usepackage{siunitx}
\newcommand{\comm}[1]{}
\usetikzlibrary{graphs,shapes,decorations,positioning,backgrounds,fit}

\newcommand{\ts}{}

\definecolor {snow}                {rgb}{1.00,0.98,0.98}
\definecolor {ghostwhite}          {rgb}{0.97,0.97,1.00}
\definecolor {whitesmoke}          {rgb}{0.96,0.96,0.96}
\definecolor {gainsboro}           {rgb}{0.86,0.86,0.86}
\definecolor {floralwhite}         {rgb}{1.00,0.98,0.94}
\definecolor {oldlace}             {rgb}{0.99,0.96,0.90}
\definecolor {linen}               {rgb}{0.98,0.94,0.90}
\definecolor {antiquewhite}        {rgb}{0.98,0.92,0.84}
\definecolor {papayawhip}          {rgb}{1.00,0.94,0.84}
\definecolor {blanchedalmond}      {rgb}{1.00,0.92,0.80}
\definecolor {bisque}              {rgb}{1.00,0.89,0.77}
\definecolor {peachpuff}           {rgb}{1.00,0.85,0.73}
\definecolor {navajowhite}         {rgb}{1.00,0.87,0.68}
\definecolor {moccasin}            {rgb}{1.00,0.89,0.71}
\definecolor {cornsilk}            {rgb}{1.00,0.97,0.86}
\definecolor {ivory}               {rgb}{1.00,1.00,0.94}
\definecolor {lemonchiffon}        {rgb}{1.00,0.98,0.80}
\definecolor {seashell}            {rgb}{1.00,0.96,0.93}
\definecolor {honeydew}            {rgb}{0.94,1.00,0.94}
\definecolor {mintcream}           {rgb}{0.96,1.00,0.98}
\definecolor {azure}               {rgb}{0.94,1.00,1.00}
\definecolor {aliceblue}           {rgb}{0.94,0.97,1.00}
\definecolor {lavender}            {rgb}{0.90,0.90,0.98}
\definecolor {lavenderblush}       {rgb}{1.00,0.94,0.96}
\definecolor {mistyrose}           {rgb}{1.00,0.89,0.88}
\definecolor {white}               {rgb}{1.00,1.00,1.00}
\definecolor {black}               {rgb}{0.00,0.00,0.00}
\definecolor {darkslategray}       {rgb}{0.18,0.31,0.31}
\definecolor {dimgray}             {rgb}{0.41,0.41,0.41}
\definecolor {slategray}           {rgb}{0.44,0.50,0.56}
\definecolor {lightslategray}      {rgb}{0.47,0.53,0.60}
\definecolor {gray}                {rgb}{0.75,0.75,0.75}
\definecolor {lightgrey}           {rgb}{0.83,0.83,0.83}
\definecolor {midnightblue}        {rgb}{0.10,0.10,0.44}
\definecolor {navy}                {rgb}{0.00,0.00,0.50}
\definecolor {cornflowerblue}      {rgb}{0.39,0.58,0.93}
\definecolor {darkslateblue}       {rgb}{0.28,0.24,0.55}
\definecolor {slateblue}           {rgb}{0.42,0.35,0.80}
\definecolor {mediumslateblue}     {rgb}{0.48,0.41,0.93}
\definecolor {lightslateblue}      {rgb}{0.52,0.44,1.00}
\definecolor {mediumblue}          {rgb}{0.00,0.00,0.80}
\definecolor {royalblue}           {rgb}{0.25,0.41,0.88}
\definecolor {blue}                {rgb}{0.00,0.00,1.00}
\definecolor {dodgerblue}          {rgb}{0.12,0.56,1.00}
\definecolor {deepskyblue}         {rgb}{0.00,0.75,1.00}
\definecolor {skyblue}             {rgb}{0.53,0.81,0.92}
\definecolor {lightskyblue}        {rgb}{0.53,0.81,0.98}
\definecolor {steelblue}           {rgb}{0.27,0.51,0.71}
\definecolor {lightsteelblue}      {rgb}{0.69,0.77,0.87}
\definecolor {lightblue}           {rgb}{0.68,0.85,0.90}
\definecolor {powderblue}          {rgb}{0.69,0.88,0.90}
\definecolor {paleturquoise}       {rgb}{0.69,0.93,0.93}
\definecolor {darkturquoise}       {rgb}{0.00,0.81,0.82}
\definecolor {mediumturquoise}     {rgb}{0.28,0.82,0.80}
\definecolor {turquoise}           {rgb}{0.25,0.88,0.82}
\definecolor {cyan}                {rgb}{0.00,1.00,1.00}
\definecolor {lightcyan}           {rgb}{0.88,1.00,1.00}
\definecolor {cadetblue}           {rgb}{0.37,0.62,0.63}
\definecolor {mediumaquamarine}    {rgb}{0.40,0.80,0.67}
\definecolor {aquamarine}          {rgb}{0.50,1.00,0.83}
\definecolor {darkgreen}           {rgb}{0.00,0.39,0.00}
\definecolor {darkolivegreen}      {rgb}{0.33,0.42,0.18}
\definecolor {darkseagreen}        {rgb}{0.56,0.74,0.56}
\definecolor {seagreen}            {rgb}{0.18,0.55,0.34}
\definecolor {mediumseagreen}      {rgb}{0.24,0.70,0.44}
\definecolor {lightseagreen}       {rgb}{0.13,0.70,0.67}
\definecolor {palegreen}           {rgb}{0.60,0.98,0.60}
\definecolor {springgreen}         {rgb}{0.00,1.00,0.50}
\definecolor {lawngreen}           {rgb}{0.49,0.99,0.00}
\definecolor {green}               {rgb}{0.00,1.00,0.00}
\definecolor {chartreuse}          {rgb}{0.50,1.00,0.00}
\definecolor {mediumspringgreen}   {rgb}{0.00,0.98,0.60}
\definecolor {greenyellow}         {rgb}{0.68,1.00,0.18}
\definecolor {limegreen}           {rgb}{0.20,0.80,0.20}
\definecolor {yellowgreen}         {rgb}{0.60,0.80,0.20}
\definecolor {forestgreen}         {rgb}{0.13,0.55,0.13}
\definecolor {olivedrab}           {rgb}{0.42,0.56,0.14}
\definecolor {darkkhaki}           {rgb}{0.74,0.72,0.42}
\definecolor {khaki}               {rgb}{0.94,0.90,0.55}
\definecolor {palegoldenrod}       {rgb}{0.93,0.91,0.67}
\definecolor {lightgoldenrodyellow} {rgb}{0.98,0.98,0.82}
\definecolor {lightyellow}         {rgb}{1.00,1.00,0.88}
\definecolor {yellow}              {rgb}{1.00,1.00,0.00}
\definecolor {gold}                {rgb}{1.00,0.84,0.00}
\definecolor {lightgoldenrod}      {rgb}{0.93,0.87,0.51}
\definecolor {goldenrod}           {rgb}{0.85,0.65,0.13}
\definecolor {darkgoldenrod}       {rgb}{0.72,0.53,0.04}
\definecolor {rosybrown}           {rgb}{0.74,0.56,0.56}
\definecolor {indianred}           {rgb}{0.80,0.36,0.36}
\definecolor {saddlebrown}         {rgb}{0.55,0.27,0.07}
\definecolor {sienna}              {rgb}{0.63,0.32,0.18}
\definecolor {peru}                {rgb}{0.80,0.52,0.25}
\definecolor {burlywood}           {rgb}{0.87,0.72,0.53}
\definecolor {beige}               {rgb}{0.96,0.96,0.86}
\definecolor {wheat}               {rgb}{0.96,0.87,0.70}
\definecolor {sandybrown}          {rgb}{0.96,0.64,0.38}
\definecolor {tan}                 {rgb}{0.82,0.71,0.55}
\definecolor {chocolate}           {rgb}{0.82,0.41,0.12}
\definecolor {firebrick}           {rgb}{0.70,0.13,0.13}
\definecolor {brown}               {rgb}{0.65,0.16,0.16}
\definecolor {darksalmon}          {rgb}{0.91,0.59,0.48}
\definecolor {salmon}              {rgb}{0.98,0.50,0.45}
\definecolor {lightsalmon}         {rgb}{1.00,0.63,0.48}
\definecolor {orange}              {rgb}{1.00,0.65,0.00}
\definecolor {darkorange}          {rgb}{1.00,0.55,0.00}
\definecolor {coral}               {rgb}{1.00,0.50,0.31}
\definecolor {lightcoral}          {rgb}{0.94,0.50,0.50}
\definecolor {tomato}              {rgb}{1.00,0.39,0.28}
\definecolor {orangered}           {rgb}{1.00,0.27,0.00}
\definecolor {red}                 {rgb}{1.00,0.00,0.00}
\definecolor {hotpink}             {rgb}{1.00,0.41,0.71}
\definecolor {deeppink}            {rgb}{1.00,0.08,0.58}
\definecolor {pink}                {rgb}{1.00,0.75,0.80}
\definecolor {lightpink}           {rgb}{1.00,0.71,0.76}
\definecolor {palevioletred}       {rgb}{0.86,0.44,0.58}
\definecolor {maroon}              {rgb}{0.69,0.19,0.38}
\definecolor {mediumvioletred}     {rgb}{0.78,0.08,0.52}
\definecolor {violetred}           {rgb}{0.82,0.13,0.56}
\definecolor {magenta}             {rgb}{1.00,0.00,1.00}
\definecolor {violet}              {rgb}{0.93,0.51,0.93}
\definecolor {plum}                {rgb}{0.87,0.63,0.87}
\definecolor {orchid}              {rgb}{0.85,0.44,0.84}
\definecolor {mediumorchid}        {rgb}{0.73,0.33,0.83}
\definecolor {darkorchid}          {rgb}{0.60,0.20,0.80}
\definecolor {darkviolet}          {rgb}{0.58,0.00,0.83}
\definecolor {blueviolet}          {rgb}{0.54,0.17,0.89}
\definecolor {purple}              {rgb}{0.63,0.13,0.94}
\definecolor {mediumpurple}        {rgb}{0.58,0.44,0.86}
\definecolor {thistle}             {rgb}{0.85,0.75,0.85}
\definecolor {snow2}               {rgb}{0.93,0.91,0.91}
\definecolor {snow3}               {rgb}{0.80,0.79,0.79}
\definecolor {snow4}               {rgb}{0.55,0.54,0.54}
\definecolor {seashell2}           {rgb}{0.93,0.90,0.87}
\definecolor {seashell3}           {rgb}{0.80,0.77,0.75}
\definecolor {seashell4}           {rgb}{0.55,0.53,0.51}
\definecolor {antiquewhite1}       {rgb}{1.00,0.94,0.86}
\definecolor {antiquewhite2}       {rgb}{0.93,0.87,0.80}
\definecolor {antiquewhite3}       {rgb}{0.80,0.75,0.69}
\definecolor {antiquewhite4}       {rgb}{0.55,0.51,0.47}
\definecolor {bisque2}             {rgb}{0.93,0.84,0.72}
\definecolor {bisque3}             {rgb}{0.80,0.72,0.62}
\definecolor {bisque4}             {rgb}{0.55,0.49,0.42}
\definecolor {peachpuff2}          {rgb}{0.93,0.80,0.68}
\definecolor {peachpuff3}          {rgb}{0.80,0.69,0.58}
\definecolor {peachpuff4}          {rgb}{0.55,0.47,0.40}
\definecolor {navajowhite2}        {rgb}{0.93,0.81,0.63}
\definecolor {navajowhite3}        {rgb}{0.80,0.70,0.55}
\definecolor {navajowhite4}        {rgb}{0.55,0.47,0.37}
\definecolor {lemonchiffon2}       {rgb}{0.93,0.91,0.75}
\definecolor {lemonchiffon3}       {rgb}{0.80,0.79,0.65}
\definecolor {lemonchiffon4}       {rgb}{0.55,0.54,0.44}
\definecolor {cornsilk2}           {rgb}{0.93,0.91,0.80}
\definecolor {cornsilk3}           {rgb}{0.80,0.78,0.69}
\definecolor {cornsilk4}           {rgb}{0.55,0.53,0.47}
\definecolor {ivory2}              {rgb}{0.93,0.93,0.88}
\definecolor {ivory3}              {rgb}{0.80,0.80,0.76}
\definecolor {ivory4}              {rgb}{0.55,0.55,0.51}
\definecolor {honeydew2}           {rgb}{0.88,0.93,0.88}
\definecolor {honeydew3}           {rgb}{0.76,0.80,0.76}
\definecolor {honeydew4}           {rgb}{0.51,0.55,0.51}
\definecolor {lavenderblush2}      {rgb}{0.93,0.88,0.90}
\definecolor {lavenderblush3}      {rgb}{0.80,0.76,0.77}
\definecolor {lavenderblush4}      {rgb}{0.55,0.51,0.53}
\definecolor {mistyrose2}          {rgb}{0.93,0.84,0.82}
\definecolor {mistyrose3}          {rgb}{0.80,0.72,0.71}
\definecolor {mistyrose4}          {rgb}{0.55,0.49,0.48}
\definecolor {azure2}              {rgb}{0.88,0.93,0.93}
\definecolor {azure3}              {rgb}{0.76,0.80,0.80}
\definecolor {azure4}              {rgb}{0.51,0.55,0.55}
\definecolor {slateblue1}          {rgb}{0.51,0.44,1.00}
\definecolor {slateblue2}          {rgb}{0.48,0.40,0.93}
\definecolor {slateblue3}          {rgb}{0.41,0.35,0.80}
\definecolor {slateblue4}          {rgb}{0.28,0.24,0.55}
\definecolor {royalblue1}          {rgb}{0.28,0.46,1.00}
\definecolor {royalblue2}          {rgb}{0.26,0.43,0.93}
\definecolor {royalblue3}          {rgb}{0.23,0.37,0.80}
\definecolor {royalblue4}          {rgb}{0.15,0.25,0.55}
\definecolor {blue2}               {rgb}{0.00,0.00,0.93}
\definecolor {blue4}               {rgb}{0.00,0.00,0.55}
\definecolor {dodgerblue2}         {rgb}{0.11,0.53,0.93}
\definecolor {dodgerblue3}         {rgb}{0.09,0.45,0.80}
\definecolor {dodgerblue4}         {rgb}{0.06,0.31,0.55}
\definecolor {steelblue1}          {rgb}{0.39,0.72,1.00}
\definecolor {steelblue2}          {rgb}{0.36,0.67,0.93}
\definecolor {steelblue3}          {rgb}{0.31,0.58,0.80}
\definecolor {steelblue4}          {rgb}{0.21,0.39,0.55}
\definecolor {deepskyblue2}        {rgb}{0.00,0.70,0.93}
\definecolor {deepskyblue3}        {rgb}{0.00,0.60,0.80}
\definecolor {deepskyblue4}        {rgb}{0.00,0.41,0.55}
\definecolor {skyblue1}            {rgb}{0.53,0.81,1.00}
\definecolor {skyblue2}            {rgb}{0.49,0.75,0.93}
\definecolor {skyblue3}            {rgb}{0.42,0.65,0.80}
\definecolor {skyblue4}            {rgb}{0.29,0.44,0.55}
\definecolor {lightskyblue1}       {rgb}{0.69,0.89,1.00}
\definecolor {lightskyblue2}       {rgb}{0.64,0.83,0.93}
\definecolor {lightskyblue3}       {rgb}{0.55,0.71,0.80}
\definecolor {lightskyblue4}       {rgb}{0.38,0.48,0.55}
\definecolor {slategray1}          {rgb}{0.78,0.89,1.00}
\definecolor {slategray2}          {rgb}{0.73,0.83,0.93}
\definecolor {slategray3}          {rgb}{0.62,0.71,0.80}
\definecolor {slategray4}          {rgb}{0.42,0.48,0.55}
\definecolor {lightsteelblue1}     {rgb}{0.79,0.88,1.00}
\definecolor {lightsteelblue2}     {rgb}{0.74,0.82,0.93}
\definecolor {lightsteelblue3}     {rgb}{0.64,0.71,0.80}
\definecolor {lightsteelblue4}     {rgb}{0.43,0.48,0.55}
\definecolor {lightblue1}          {rgb}{0.75,0.94,1.00}
\definecolor {lightblue2}          {rgb}{0.70,0.87,0.93}
\definecolor {lightblue3}          {rgb}{0.60,0.75,0.80}
\definecolor {lightblue4}          {rgb}{0.41,0.51,0.55}
\definecolor {lightcyan2}          {rgb}{0.82,0.93,0.93}
\definecolor {lightcyan3}          {rgb}{0.71,0.80,0.80}
\definecolor {lightcyan4}          {rgb}{0.48,0.55,0.55}
\definecolor {paleturquoise1}      {rgb}{0.73,1.00,1.00}
\definecolor {paleturquoise2}      {rgb}{0.68,0.93,0.93}
\definecolor {paleturquoise3}      {rgb}{0.59,0.80,0.80}
\definecolor {paleturquoise4}      {rgb}{0.40,0.55,0.55}
\definecolor {cadetblue1}          {rgb}{0.60,0.96,1.00}
\definecolor {cadetblue2}          {rgb}{0.56,0.90,0.93}
\definecolor {cadetblue3}          {rgb}{0.48,0.77,0.80}
\definecolor {cadetblue4}          {rgb}{0.33,0.53,0.55}
\definecolor {turquoise1}          {rgb}{0.00,0.96,1.00}
\definecolor {turquoise2}          {rgb}{0.00,0.90,0.93}
\definecolor {turquoise3}          {rgb}{0.00,0.77,0.80}
\definecolor {turquoise4}          {rgb}{0.00,0.53,0.55}
\definecolor {cyan2}               {rgb}{0.00,0.93,0.93}
\definecolor {cyan3}               {rgb}{0.00,0.80,0.80}
\definecolor {cyan4}               {rgb}{0.00,0.55,0.55}
\definecolor {darkslategray1}      {rgb}{0.59,1.00,1.00}
\definecolor {darkslategray2}      {rgb}{0.55,0.93,0.93}
\definecolor {darkslategray3}      {rgb}{0.47,0.80,0.80}
\definecolor {darkslategray4}      {rgb}{0.32,0.55,0.55}
\definecolor {aquamarine2}         {rgb}{0.46,0.93,0.78}
\definecolor {aquamarine4}         {rgb}{0.27,0.55,0.45}
\definecolor {darkseagreen1}       {rgb}{0.76,1.00,0.76}
\definecolor {darkseagreen2}       {rgb}{0.71,0.93,0.71}
\definecolor {darkseagreen3}       {rgb}{0.61,0.80,0.61}
\definecolor {darkseagreen4}       {rgb}{0.41,0.55,0.41}
\definecolor {seagreen1}           {rgb}{0.33,1.00,0.62}
\definecolor {seagreen2}           {rgb}{0.31,0.93,0.58}
\definecolor {seagreen3}           {rgb}{0.26,0.80,0.50}
\definecolor {palegreen1}          {rgb}{0.60,1.00,0.60}
\definecolor {palegreen2}          {rgb}{0.56,0.93,0.56}
\definecolor {palegreen3}          {rgb}{0.49,0.80,0.49}
\definecolor {palegreen4}          {rgb}{0.33,0.55,0.33}
\definecolor {springgreen2}        {rgb}{0.00,0.93,0.46}
\definecolor {springgreen3}        {rgb}{0.00,0.80,0.40}
\definecolor {springgreen4}        {rgb}{0.00,0.55,0.27}
\definecolor {green2}              {rgb}{0.00,0.93,0.00}
\definecolor {green3}              {rgb}{0.00,0.80,0.00}
\definecolor {green4}              {rgb}{0.00,0.55,0.00}
\definecolor {chartreuse2}         {rgb}{0.46,0.93,0.00}
\definecolor {chartreuse3}         {rgb}{0.40,0.80,0.00}
\definecolor {chartreuse4}         {rgb}{0.27,0.55,0.00}
\definecolor {olivedrab1}          {rgb}{0.75,1.00,0.24}
\definecolor {olivedrab2}          {rgb}{0.70,0.93,0.23}
\definecolor {olivedrab4}          {rgb}{0.41,0.55,0.13}
\definecolor {darkolivegreen1}     {rgb}{0.79,1.00,0.44}
\definecolor {darkolivegreen2}     {rgb}{0.74,0.93,0.41}
\definecolor {darkolivegreen3}     {rgb}{0.64,0.80,0.35}
\definecolor {darkolivegreen4}     {rgb}{0.43,0.55,0.24}
\definecolor {khaki1}              {rgb}{1.00,0.96,0.56}
\definecolor {khaki2}              {rgb}{0.93,0.90,0.52}
\definecolor {khaki3}              {rgb}{0.80,0.78,0.45}
\definecolor {khaki4}              {rgb}{0.55,0.53,0.31}
\definecolor {lightgoldenrod1}     {rgb}{1.00,0.93,0.55}
\definecolor {lightgoldenrod2}     {rgb}{0.93,0.86,0.51}
\definecolor {lightgoldenrod3}     {rgb}{0.80,0.75,0.44}
\definecolor {lightgoldenrod4}     {rgb}{0.55,0.51,0.30}
\definecolor {lightyellow2}        {rgb}{0.93,0.93,0.82}
\definecolor {lightyellow3}        {rgb}{0.80,0.80,0.71}
\definecolor {lightyellow4}        {rgb}{0.55,0.55,0.48}
\definecolor {yellow2}             {rgb}{0.93,0.93,0.00}
\definecolor {yellow3}             {rgb}{0.80,0.80,0.00}
\definecolor {yellow4}             {rgb}{0.55,0.55,0.00}
\definecolor {gold2}               {rgb}{0.93,0.79,0.00}
\definecolor {gold3}               {rgb}{0.80,0.68,0.00}
\definecolor {gold4}               {rgb}{0.55,0.46,0.00}
\definecolor {goldenrod1}          {rgb}{1.00,0.76,0.15}
\definecolor {goldenrod2}          {rgb}{0.93,0.71,0.13}
\definecolor {goldenrod3}          {rgb}{0.80,0.61,0.11}
\definecolor {goldenrod4}          {rgb}{0.55,0.41,0.08}
\definecolor {darkgoldenrod1}      {rgb}{1.00,0.73,0.06}
\definecolor {darkgoldenrod2}      {rgb}{0.93,0.68,0.05}
\definecolor {darkgoldenrod3}      {rgb}{0.80,0.58,0.05}
\definecolor {darkgoldenrod4}      {rgb}{0.55,0.40,0.03}
\definecolor {rosybrown1}          {rgb}{1.00,0.76,0.76}
\definecolor {rosybrown2}          {rgb}{0.93,0.71,0.71}
\definecolor {rosybrown3}          {rgb}{0.80,0.61,0.61}
\definecolor {rosybrown4}          {rgb}{0.55,0.41,0.41}
\definecolor {indianred1}          {rgb}{1.00,0.42,0.42}
\definecolor {indianred2}          {rgb}{0.93,0.39,0.39}
\definecolor {indianred3}          {rgb}{0.80,0.33,0.33}
\definecolor {indianred4}          {rgb}{0.55,0.23,0.23}
\definecolor {sienna1}             {rgb}{1.00,0.51,0.28}
\definecolor {sienna2}             {rgb}{0.93,0.47,0.26}
\definecolor {sienna3}             {rgb}{0.80,0.41,0.22}
\definecolor {sienna4}             {rgb}{0.55,0.28,0.15}
\definecolor {burlywood1}          {rgb}{1.00,0.83,0.61}
\definecolor {burlywood2}          {rgb}{0.93,0.77,0.57}
\definecolor {burlywood3}          {rgb}{0.80,0.67,0.49}
\definecolor {burlywood4}          {rgb}{0.55,0.45,0.33}
\definecolor {wheat1}              {rgb}{1.00,0.91,0.73}
\definecolor {wheat2}              {rgb}{0.93,0.85,0.68}
\definecolor {wheat3}              {rgb}{0.80,0.73,0.59}
\definecolor {wheat4}              {rgb}{0.55,0.49,0.40}
\definecolor {tan1}                {rgb}{1.00,0.65,0.31}
\definecolor {tan2}                {rgb}{0.93,0.60,0.29}
\definecolor {tan4}                {rgb}{0.55,0.35,0.17}
\definecolor {chocolate1}          {rgb}{1.00,0.50,0.14}
\definecolor {chocolate2}          {rgb}{0.93,0.46,0.13}
\definecolor {chocolate3}          {rgb}{0.80,0.40,0.11}
\definecolor {firebrick1}          {rgb}{1.00,0.19,0.19}
\definecolor {firebrick2}          {rgb}{0.93,0.17,0.17}
\definecolor {firebrick3}          {rgb}{0.80,0.15,0.15}
\definecolor {firebrick4}          {rgb}{0.55,0.10,0.10}
\definecolor {brown1}              {rgb}{1.00,0.25,0.25}
\definecolor {brown2}              {rgb}{0.93,0.23,0.23}
\definecolor {brown3}              {rgb}{0.80,0.20,0.20}
\definecolor {brown4}              {rgb}{0.55,0.14,0.14}
\definecolor {salmon1}             {rgb}{1.00,0.55,0.41}
\definecolor {salmon2}             {rgb}{0.93,0.51,0.38}
\definecolor {salmon3}             {rgb}{0.80,0.44,0.33}
\definecolor {salmon4}             {rgb}{0.55,0.30,0.22}
\definecolor {lightsalmon2}        {rgb}{0.93,0.58,0.45}
\definecolor {lightsalmon3}        {rgb}{0.80,0.51,0.38}
\definecolor {lightsalmon4}        {rgb}{0.55,0.34,0.26}
\definecolor {orange2}             {rgb}{0.93,0.60,0.00}
\definecolor {orange3}             {rgb}{0.80,0.52,0.00}
\definecolor {orange4}             {rgb}{0.55,0.35,0.00}
\definecolor {darkorange1}         {rgb}{1.00,0.50,0.00}
\definecolor {darkorange2}         {rgb}{0.93,0.46,0.00}
\definecolor {darkorange3}         {rgb}{0.80,0.40,0.00}
\definecolor {darkorange4}         {rgb}{0.55,0.27,0.00}
\definecolor {coral1}              {rgb}{1.00,0.45,0.34}
\definecolor {coral2}              {rgb}{0.93,0.42,0.31}
\definecolor {coral3}              {rgb}{0.80,0.36,0.27}
\definecolor {coral4}              {rgb}{0.55,0.24,0.18}
\definecolor {tomato2}             {rgb}{0.93,0.36,0.26}
\definecolor {tomato3}             {rgb}{0.80,0.31,0.22}
\definecolor {tomato4}             {rgb}{0.55,0.21,0.15}
\definecolor {orangered2}          {rgb}{0.93,0.25,0.00}
\definecolor {orangered3}          {rgb}{0.80,0.22,0.00}
\definecolor {orangered4}          {rgb}{0.55,0.15,0.00}
\definecolor {red2}                {rgb}{0.93,0.00,0.00}
\definecolor {red3}                {rgb}{0.80,0.00,0.00}
\definecolor {red4}                {rgb}{0.55,0.00,0.00}
\definecolor {deeppink2}           {rgb}{0.93,0.07,0.54}
\definecolor {deeppink3}           {rgb}{0.80,0.06,0.46}
\definecolor {deeppink4}           {rgb}{0.55,0.04,0.31}
\definecolor {hotpink1}            {rgb}{1.00,0.43,0.71}
\definecolor {hotpink2}            {rgb}{0.93,0.42,0.65}
\definecolor {hotpink3}            {rgb}{0.80,0.38,0.56}
\definecolor {hotpink4}            {rgb}{0.55,0.23,0.38}
\definecolor {pink1}               {rgb}{1.00,0.71,0.77}
\definecolor {pink2}               {rgb}{0.93,0.66,0.72}
\definecolor {pink3}               {rgb}{0.80,0.57,0.62}
\definecolor {pink4}               {rgb}{0.55,0.39,0.42}
\definecolor {lightpink1}          {rgb}{1.00,0.68,0.73}
\definecolor {lightpink2}          {rgb}{0.93,0.64,0.68}
\definecolor {lightpink3}          {rgb}{0.80,0.55,0.58}
\definecolor {lightpink4}          {rgb}{0.55,0.37,0.40}
\definecolor {palevioletred1}      {rgb}{1.00,0.51,0.67}
\definecolor {palevioletred2}      {rgb}{0.93,0.47,0.62}
\definecolor {palevioletred3}      {rgb}{0.80,0.41,0.54}
\definecolor {palevioletred4}      {rgb}{0.55,0.28,0.36}
\definecolor {maroon1}             {rgb}{1.00,0.20,0.70}
\definecolor {maroon2}             {rgb}{0.93,0.19,0.65}
\definecolor {maroon3}             {rgb}{0.80,0.16,0.56}
\definecolor {maroon4}             {rgb}{0.55,0.11,0.38}
\definecolor {violetred1}          {rgb}{1.00,0.24,0.59}
\definecolor {violetred2}          {rgb}{0.93,0.23,0.55}
\definecolor {violetred3}          {rgb}{0.80,0.20,0.47}
\definecolor {violetred4}          {rgb}{0.55,0.13,0.32}
\definecolor {magenta2}            {rgb}{0.93,0.00,0.93}
\definecolor {magenta3}            {rgb}{0.80,0.00,0.80}
\definecolor {magenta4}            {rgb}{0.55,0.00,0.55}
\definecolor {orchid1}             {rgb}{1.00,0.51,0.98}
\definecolor {orchid2}             {rgb}{0.93,0.48,0.91}
\definecolor {orchid3}             {rgb}{0.80,0.41,0.79}
\definecolor {orchid4}             {rgb}{0.55,0.28,0.54}
\definecolor {plum1}               {rgb}{1.00,0.73,1.00}
\definecolor {plum2}               {rgb}{0.93,0.68,0.93}
\definecolor {plum3}               {rgb}{0.80,0.59,0.80}
\definecolor {plum4}               {rgb}{0.55,0.40,0.55}
\definecolor {mediumorchid1}       {rgb}{0.88,0.40,1.00}
\definecolor {mediumorchid2}       {rgb}{0.82,0.37,0.93}
\definecolor {mediumorchid3}       {rgb}{0.71,0.32,0.80}
\definecolor {mediumorchid4}       {rgb}{0.48,0.22,0.55}
\definecolor {darkorchid1}         {rgb}{0.75,0.24,1.00}
\definecolor {darkorchid2}         {rgb}{0.70,0.23,0.93}
\definecolor {darkorchid3}         {rgb}{0.60,0.20,0.80}
\definecolor {darkorchid4}         {rgb}{0.41,0.13,0.55}
\definecolor {purple1}             {rgb}{0.61,0.19,1.00}
\definecolor {purple2}             {rgb}{0.57,0.17,0.93}
\definecolor {purple3}             {rgb}{0.49,0.15,0.80}
\definecolor {purple4}             {rgb}{0.33,0.10,0.55}
\definecolor {mediumpurple1}       {rgb}{0.67,0.51,1.00}
\definecolor {mediumpurple2}       {rgb}{0.62,0.47,0.93}
\definecolor {mediumpurple3}       {rgb}{0.54,0.41,0.80}
\definecolor {mediumpurple4}       {rgb}{0.36,0.28,0.55}
\definecolor {thistle1}            {rgb}{1.00,0.88,1.00}
\definecolor {thistle2}            {rgb}{0.93,0.82,0.93}
\definecolor {thistle3}            {rgb}{0.80,0.71,0.80}
\definecolor {thistle4}            {rgb}{0.55,0.48,0.55}
\definecolor {gray1}               {rgb}{0.01,0.01,0.01}
\definecolor {gray2}               {rgb}{0.02,0.02,0.02}
\definecolor {gray3}               {rgb}{0.03,0.03,0.03}
\definecolor {gray4}               {rgb}{0.04,0.04,0.04}
\definecolor {gray5}               {rgb}{0.05,0.05,0.05}
\definecolor {gray6}               {rgb}{0.06,0.06,0.06}
\definecolor {gray7}               {rgb}{0.07,0.07,0.07}
\definecolor {gray8}               {rgb}{0.08,0.08,0.08}
\definecolor {gray9}               {rgb}{0.09,0.09,0.09}
\definecolor {gray10}              {rgb}{0.10,0.10,0.10}
\definecolor {gray11}              {rgb}{0.11,0.11,0.11}
\definecolor {gray12}              {rgb}{0.12,0.12,0.12}
\definecolor {gray13}              {rgb}{0.13,0.13,0.13}
\definecolor {gray14}              {rgb}{0.14,0.14,0.14}
\definecolor {gray15}              {rgb}{0.15,0.15,0.15}
\definecolor {gray16}              {rgb}{0.16,0.16,0.16}
\definecolor {gray17}              {rgb}{0.17,0.17,0.17}
\definecolor {gray18}              {rgb}{0.18,0.18,0.18}
\definecolor {gray19}              {rgb}{0.19,0.19,0.19}
\definecolor {gray20}              {rgb}{0.20,0.20,0.20}
\definecolor {gray21}              {rgb}{0.21,0.21,0.21}
\definecolor {gray22}              {rgb}{0.22,0.22,0.22}
\definecolor {gray23}              {rgb}{0.23,0.23,0.23}
\definecolor {gray24}              {rgb}{0.24,0.24,0.24}
\definecolor {gray25}              {rgb}{0.25,0.25,0.25}
\definecolor {gray26}              {rgb}{0.26,0.26,0.26}
\definecolor {gray27}              {rgb}{0.27,0.27,0.27}
\definecolor {gray28}              {rgb}{0.28,0.28,0.28}
\definecolor {gray29}              {rgb}{0.29,0.29,0.29}
\definecolor {gray30}              {rgb}{0.30,0.30,0.30}
\definecolor {gray31}              {rgb}{0.31,0.31,0.31}
\definecolor {gray32}              {rgb}{0.32,0.32,0.32}
\definecolor {gray33}              {rgb}{0.33,0.33,0.33}
\definecolor {gray34}              {rgb}{0.34,0.34,0.34}
\definecolor {gray35}              {rgb}{0.35,0.35,0.35}
\definecolor {gray36}              {rgb}{0.36,0.36,0.36}
\definecolor {gray37}              {rgb}{0.37,0.37,0.37}
\definecolor {gray38}              {rgb}{0.38,0.38,0.38}
\definecolor {gray39}              {rgb}{0.39,0.39,0.39}
\definecolor {gray40}              {rgb}{0.40,0.40,0.40}
\definecolor {gray42}              {rgb}{0.42,0.42,0.42}
\definecolor {gray43}              {rgb}{0.43,0.43,0.43}
\definecolor {gray44}              {rgb}{0.44,0.44,0.44}
\definecolor {gray45}              {rgb}{0.45,0.45,0.45}
\definecolor {gray46}              {rgb}{0.46,0.46,0.46}
\definecolor {gray47}              {rgb}{0.47,0.47,0.47}
\definecolor {gray48}              {rgb}{0.48,0.48,0.48}
\definecolor {gray49}              {rgb}{0.49,0.49,0.49}
\definecolor {gray50}              {rgb}{0.50,0.50,0.50}
\definecolor {gray51}              {rgb}{0.51,0.51,0.51}
\definecolor {gray52}              {rgb}{0.52,0.52,0.52}
\definecolor {gray53}              {rgb}{0.53,0.53,0.53}
\definecolor {gray54}              {rgb}{0.54,0.54,0.54}
\definecolor {gray55}              {rgb}{0.55,0.55,0.55}
\definecolor {gray56}              {rgb}{0.56,0.56,0.56}
\definecolor {gray57}              {rgb}{0.57,0.57,0.57}
\definecolor {gray58}              {rgb}{0.58,0.58,0.58}
\definecolor {gray59}              {rgb}{0.59,0.59,0.59}
\definecolor {gray60}              {rgb}{0.60,0.60,0.60}
\definecolor {gray61}              {rgb}{0.61,0.61,0.61}
\definecolor {gray62}              {rgb}{0.62,0.62,0.62}
\definecolor {gray63}              {rgb}{0.63,0.63,0.63}
\definecolor {gray64}              {rgb}{0.64,0.64,0.64}
\definecolor {gray65}              {rgb}{0.65,0.65,0.65}
\definecolor {gray66}              {rgb}{0.66,0.66,0.66}
\definecolor {gray67}              {rgb}{0.67,0.67,0.67}
\definecolor {gray68}              {rgb}{0.68,0.68,0.68}
\definecolor {gray69}              {rgb}{0.69,0.69,0.69}
\definecolor {gray70}              {rgb}{0.70,0.70,0.70}
\definecolor {gray71}              {rgb}{0.71,0.71,0.71}
\definecolor {gray72}              {rgb}{0.72,0.72,0.72}
\definecolor {gray73}              {rgb}{0.73,0.73,0.73}
\definecolor {gray74}              {rgb}{0.74,0.74,0.74}
\definecolor {gray75}              {rgb}{0.75,0.75,0.75}
\definecolor {gray76}              {rgb}{0.76,0.76,0.76}
\definecolor {gray77}              {rgb}{0.77,0.77,0.77}
\definecolor {gray78}              {rgb}{0.78,0.78,0.78}
\definecolor {gray79}              {rgb}{0.79,0.79,0.79}
\definecolor {gray80}              {rgb}{0.80,0.80,0.80}
\definecolor {gray81}              {rgb}{0.81,0.81,0.81}
\definecolor {gray82}              {rgb}{0.82,0.82,0.82}
\definecolor {gray83}              {rgb}{0.83,0.83,0.83}
\definecolor {gray84}              {rgb}{0.84,0.84,0.84}
\definecolor {gray85}              {rgb}{0.85,0.85,0.85}
\definecolor {gray86}              {rgb}{0.86,0.86,0.86}
\definecolor {gray87}              {rgb}{0.87,0.87,0.87}
\definecolor {gray88}              {rgb}{0.88,0.88,0.88}
\definecolor {gray89}              {rgb}{0.89,0.89,0.89}
\definecolor {gray90}              {rgb}{0.90,0.90,0.90}
\definecolor {gray91}              {rgb}{0.91,0.91,0.91}
\definecolor {gray92}              {rgb}{0.92,0.92,0.92}
\definecolor {gray93}              {rgb}{0.93,0.93,0.93}
\definecolor {gray94}              {rgb}{0.94,0.94,0.94}
\definecolor {gray95}              {rgb}{0.95,0.95,0.95}
\definecolor {gray97}              {rgb}{0.97,0.97,0.97}
\definecolor {gray98}              {rgb}{0.98,0.98,0.98}
\definecolor {gray99}              {rgb}{0.99,0.99,0.99}
\definecolor {darkgrey}            {rgb}{0.66,0.66,0.66}

\newcommand{\resp}[1]{[resp. #1]}

\newcommand{\TODO}[1]{{}}

\newcommand{\ignore}[1]{}

\newcommand{\RSTODO}[1]{{\bf \textcolor{darkgreen}{{\fbox{RS TODO:} #1}}}}
\renewcommand{\RSTODO}[1]{\todo{RS: #1}}

\newenvironment{rschange}{\color{darkgreen}}{\normalcolor}

\newcommand{\RSNOTE}[1]{\marginpar{\textcolor{darkgreen}{\textbf{RS: }
      {\footnotesize #1}}}}
\renewcommand{\RSNOTE}[1]{\textcolor{darkgreen}{{\bf RS: #1}}}

 \newcommand{\ignoreinshort}[1]{}
 \newcommand{\ignoreinlong}[1]{{#1}}

\def\makenewenumerate#1#2{%
\newcounter{cnt#1}
\newenvironment{#1}%
{\begin{list}{\makebox[0pt][r]{#2}}%
{\setlength{\itemsep}{0pt}%
 \setlength{\parsep}{.2em}%
 \setlength{\leftmargin}{1.5em}%
 \setlength{\labelwidth}{.4em}%
 \usecounter{cnt#1}}}
{\end{list}}}

\makenewenumerate{myenumerate}{\arabic{cntmyenumerate}.}

\makenewenumerate{renumerate}{\rm(\roman{cntrenumerate})}
\makenewenumerate{renumerateprime}{\rm(\roman{cntrenumerateprime}$'$)}
\makenewenumerate{renumeratesecond}{\rm(\roman{cntrenumeratesecond}$''$)}

\makenewenumerate{aenumerate}{({\it\alph{cntaenumerate}})}
\makenewenumerate{aenumerateprime}{({\it\alph{cntaenumerateprime}$'$})}
\makenewenumerate{aenumeratesecond}{({\it\alph{cntaenumeratesecond}$''$})}

\def\newplaintheorem#1#2{%
\newtheorem{#1plain}{#2}[section]%
\newenvironment{#1}{\begin{#1plain}\rm }{\end{#1plain}}}

\newtheorem{definition}{Definition}

\newtheorem{proposition}{Proposition}
\newplaintheorem{notation}{Notation}
\newplaintheorem{cexample}{Counter-Example}

\newcommand{\sref}[1]{\S{}\ref{#1}}
\newcommand{\noi}{\noindent}

\newcommand{\tuple}[1]{\ensuremath{\langle{#1}\rangle}\xspace}
\newcommand{\set}[1]{\ensuremath{\{{#1}\}}\xspace}
\newcommand{\imp}{\ensuremath{\rightarrow}\xspace}

\renewcommand{\iff}{\ensuremath{\leftrightarrow}\xspace}
\newcommand{\defas}{\ensuremath{\stackrel{\text{\tiny def}}{=}}\xspace}
\newcommand{\thus}{\ensuremath{\Longrightarrow}\xspace}

\newcommand{\pos}{\phantom{\neg}}

\newcommand\calb{\ensuremath{\mathcal{B}}\xspace}

\newcommand\cali{\ensuremath{\mathcal{I}}\xspace}

\newcommand{\true}{\top}

\newcommand{\F}{{\bf F\/}}

\newcommand{\omt}{\ensuremath{\text{OMT}}\xspace}

\newcommand{\omlarat}{\ensuremath{\text{OMT}(\larat)}\xspace}

\newcommand\mysout{\bgroup \markoverwith{{-}}\ULon}
\newcommand\nosout{\bgroup \markoverwith{{ }}\ULon}
\definecolor{mygray}{rgb}{0.90,0.90,0.90}
\definecolor{mywhite}{rgb}{1.00,1.00,1.00}

\newcommand{\optimathsat}{\textsc{OptiMathSAT}\xspace}

\newcommand{\T}{\ensuremath{\mathcal{T}}\xspace}

\newcommand{\smt}{SMT\xspace}

\newcommand{\smttt}[1]{\ensuremath{\text{SMT}(#1)}\xspace}

\newcommand{\la}{\ensuremath{\mathcal{LA}}\xspace}
\newcommand{\larat}{\ensuremath{\mathcal{LA}(\mathbb{Q})}\xspace}

\renewcommand{\la}{\ensuremath{\mathcal{LA}}\xspace}
\renewcommand{\larat}{\ensuremath{\mathcal{LRA}}\xspace}

\newcommand{\smtlarat}{\smttt{\larat}}

\newcommand{\mathsatfive}{\textsc{MathSAT5}\xspace}

\newcommand{\PTTODO}[1]{}

\renewcommand{\RSTODO}[1]{\noindent{\bf \textcolor{blue}{{\fbox{RS TODO:} #1}}}}
\renewcommand{\RSTODO}[1]{\todo[]{RS TODO: #1}}

\renewcommand{\RSNOTE}[1]{\todo[inline]{RS: #1}}

\renewcommand{\TODO}[1]{\noindent{\bf \textcolor{blue}{{\fbox{TODO:} #1}}}}

\newcommand{\bias}{\ensuremath{{\sf b}}}
\newcommand{\coup}{\ensuremath{{\sf c}}}
\newcommand{\offset}{\ensuremath{{\sf o}}}
\renewcommand{\offset}{\ensuremath{\theta_{0}}}

\newcommand{\DWbiasgen}[1]{\ensuremath{{\sf b}_{#1}}}
\renewcommand{\DWbiasgen}[1]{\ensuremath{h_{#1}}}
\newcommand{\DWbiasi}{\DWbiasgen{i}}
\newcommand{\DWcoupgen}[1]{\ensuremath{{\sf c}_{#1}}}
\renewcommand{\DWcoupgen}[1]{\ensuremath{J_{#1}}}
\newcommand{\DWcoupij}{\DWcoupgen{ij}}

\newcommand{\localbounds}[1]{}
\newcommand{\globalbounds}[1]{}

\newcommand{\uncutifnecessary}[1]{}

\renewcommand{\la}{\ensuremath{\mathcal{LRIA}}\xspace}

\newcommand{\sgen}{\textsc{sgen}\xspace}

\newcommand{\xs}{\ensuremath{\underline{\mathbf{x}}}\xspace}
\newcommand{\ys}{\ensuremath{\underline{\mathbf{y}}}\xspace}

\newcommand{\zs}{\ensuremath{\underline{\mathbf{z}}}\xspace}
\newcommand{\qs}{\ensuremath{\underline{\mathbf{q}}}\xspace}
\newcommand{\as}{\ensuremath{\underline{\mathbf{a}}}\xspace}
\newcommand{\vs}{\ensuremath{\underline{\mathbf{v}}}\xspace}

\renewcommand{\ts}{\ensuremath{\underline{\boldsymbol{\theta}}}\xspace}
\newcommand{\Fx}{\ensuremath{F(\xs)}\xspace}

\newcommand{\Fxy}{\ensuremath{F^*(\xs,\ys)}\xspace}
\newcommand{\Fixy}{\ensuremath{F_i(\xs,\yis)}\xspace}
\renewcommand{\Fixy}{\ensuremath{F_i(\xs^i,\ys^i)}\xspace}
\newcommand{\Fmxy}{\ensuremath{F_m(\xs^m,\ys^m)}\xspace}
\newcommand{\Px}{\ensuremath{P_F(\xs|\ts)}\xspace}
\newcommand{\Pxa}{\ensuremath{P_F(\xs,\as|\ts)}\xspace}
\newcommand{\Pxauf}{\ensuremath{P_F(\xs,\as|\theta_b,\theta_c,\offset{},\qs)}\xspace}

\newcommand{\lowb}{\ensuremath{B_{l}}\xspace}
\newcommand{\upb}{\ensuremath{B_{u}}\xspace}
\newcommand{\lowc}{\ensuremath{C_{l}}\xspace}
\newcommand{\upc}{\ensuremath{C_{u}}\xspace}
\renewcommand{\lowb}{\ensuremath{-2}\xspace}
\renewcommand{\upb}{\ensuremath{2}\xspace}
\renewcommand{\lowc}{\ensuremath{-1}\xspace}
\renewcommand{\upc}{\ensuremath{1}\xspace}

\newcommand{\uf}{\ensuremath{\mathcal{UF}}\xspace}

\newcommand{\laeuf}{\ensuremath{\la\cup\uf}\xspace}

\newcommand{\includedin}[3]{\ensuremath{[ #1 \in [#2,#3]]}}
\renewcommand{\includedin}[3]{\ensuremath{(#2\le #1)\wedge (#1 \le #3) }}

\newcommand{\rangebfull}{\ensuremath{\bigwedge_{z\in \xs,\as} \includedin{\theta_z}{\lowb}{\upb}}\xspace}
\newcommand{\rangecfull}{\ensuremath{\bigwedge_{\substack{z,z'\in\xs,\as\\z<z'}} \includedin{\theta_{z,z'}}{\lowc}{\upc}}\xspace}

\renewcommand{\rangebfull}{\ensuremath{\bigwedge_{z_i\in \xs,\as} \includedin{\theta_i}{\lowb}{\upb}}\xspace}
\renewcommand{\rangecfull}{\ensuremath{\bigwedge_{\substack{z_i,z_j\in\xs,\as\\i<j}} \includedin{\theta_{ij}}{\lowc}{\upc}}\xspace}

\renewcommand{\Pxauf}{\ensuremath{P_F(\xs,\as|\offset{},\bias,\coup,\qs)}\xspace}

\newcommand{\qof}[1]{\ensuremath{q_{#1}}}
\newcommand{\qz}[1]{\ensuremath{q_{z_{#1}}}}
\renewcommand{\qof}[1]{\ensuremath{v_{#1}}}
\renewcommand{\qz}[1]{\ensuremath{v_{#1}}}
\renewcommand{\qs}{\ensuremath{\underline{\mathbf{v}}}\xspace}

\newcommand{\Pxaufexpanded}{\ensuremath{\offset{} + 
\sum_{z\in\xs,\as} \bias(q_{z})  {z} + 
\sum_{\substack{z,z'\in\xs,\as\\z<z'}} \coup(t_{z},t_{z'})  {z} {z'}}}
\renewcommand{\Pxaufexpanded}{\ensuremath{\offset{} + 
\sum_{1\le j\le n+h} \hspace{-.4cm}
\bias(\qz{j}) \cdot {z_j} + 
\sum_{1\le i<j\le n+h} \hspace{-.5cm}
\coup(\qz{i},\qz{j})\cdot  {z_i} \cdot{z_j}}}

\newcommand{\sattoqubo}{{\sf SATtoQUBO}}
\newcommand{\maxsattoqubo}{{\sf MaxSATtoQUBO}}

\renewcommand{\sattoqubo}{{SATtoIsing}}
\renewcommand{\maxsattoqubo}{{MaxSATtoIsing}}

\input{sv_macros}
\DeclareMathOperator*{\argmin}{argmin}

% ...
\newcommand{\longversion}{true}

\excludecomment{IGNORE}
\ifthenelse{\equal{\longversion}{true}}
{%
  \renewcommand{\ignoreinshort}[1]{{\textcolor{midnightblue}{#1}}}
  \renewcommand{\ignoreinlong}[1]{}
  \specialcomment{IGNOREINSHORT}{\begingroup\color{midnightblue}}{\endgroup}
  \excludecomment{IGNOREINLONG}
}%
{%
 \renewcommand{\ignoreinshort}[1]{}
 \renewcommand{\ignoreinlong}[1]{#1}
 \excludecomment{IGNOREINSHORT}
  \specialcomment{IGNOREINLONG}{\begingroup}{\endgroup}
}
\excludecomment{IGNORE}
\specialcomment{TOBEREWRITTEN}{\begingroup\color{darkviolet}}{\endgroup}

\specialcomment{ADDED}{\begingroup}{\endgroup}

\newtheorem{property}{Property}
\newtheorem{example}{Example}
\newdefinition{remark}{Remark}

\begin{document}

\ignore{
\pagenumbering{Roman}
\setcounter{tocdepth}{3}
\setcounter{page}{1}
\tableofcontents
\begin{center}
\noi
{\em Latest update: \today, \currenttime}
\end{center}
\newpage
\subsection*{TODO}
\begin{enumerate}
\item {Motivations \& Goals:}
  \begin{itemize}
  \item DONE add part ``the quest for quantum computing''
  \item DONE add paragraph on quantum annealing
  \item DONE move most technical part to sec 2
  \end{itemize}
\item {Background}
  \begin{itemize}
  \item DONE expand section on QA, import from previous introduction
  \item expand SAT, maxSAT, SMT and OMT
  \end{itemize}
\item {Foundations}
  \begin{itemize}
  \item DONE make it more formal (definitions, properties, etc)
  \item DONE add subsection about exact penalties \& MaxSAT
  \end{itemize}
\item {Encoding}
  \begin{itemize}
  \item DONE rewrite structure: 
  \item DONE 2 sections, raise level: subsubsections \thus subsections
  \end{itemize}

\item {Experimental evaluation}
  \begin{itemize}
  \item rewrite significantly
  \end{itemize}
\end{enumerate}

\newpage  
}
\pagenumbering{arabic}

\begin{frontmatter}
\title{%
Solving SAT
and MaxSAT 
with a Quantum Annealer:\\
Foundations, Encodings, and Preliminary Results
}

%

\author[dwave]{Zhengbing Bian}
\author[dwave]{Fabian Chudak}
\author[dwave]{William Macready}
\author[dwave]{Aidan Roy}
\address[dwave]{D-Wave Systems Inc., Burnaby, Canada }

\author[unitn]{\mbox{Roberto Sebastiani}}
\author[unitn]{Stefano Varotti}
\address[unitn]{DISI, University of Trento, Italy}

\begin{abstract}
Quantum annealers (QAs) are specialized quantum computers that minimize
objective functions over discrete variables by physically
exploiting quantum effects. Current QA platforms allow for the optimization of quadratic objectives
defined over binary variables (qubits), also known as Ising problems. In the last
decade, QA systems as implemented by D-Wave have scaled with Moore-like growth. Current
architectures provide 2048 sparsely-connected qubits,
and continued exponential growth is anticipated, {together
  with increased connectivity.}

We explore the feasibility of such architectures for solving SAT
and MaxSAT problems as QA systems scale. We develop techniques
for effectively encoding SAT {--and, with some limitations, MaxSAT--}
 into Ising problems
compatible with sparse QA architectures. We provide the
theoretical foundations for this mapping, and present encoding
techniques that combine offline Satisfiability and Optimization
Modulo Theories with on-the-fly placement and routing.
Preliminary empirical tests on a current generation 2048-qubit
D-Wave system support the feasibility of the approach for certain SAT and MaxSAT problems. 
%

\end{abstract}

  
\end{frontmatter}

\section{Motivations and Goals}
\label{sec:intro}
%


{\em Quantum Computing (QC)} promises significant computational
speedups by exploiting the quantum-mechanical phenomena of
\emph{superposition}, \emph{entanglement} and \emph{tunneling}.
%
QC relies on \emph{quantum bits (qubits)}. As opposed to bits, qubits
can be in a superposition state of 0 and 1.\footnote{Superposition is
  perhaps the best-known and most surprising aspect of quantum physics
  (e.g. the famous Scr\"odinger's cat which is both dead and alive
  prior to observation).}  Theoretically, quantum algorithms can
outperform their classical counterparts. Examples of this are Shor's
algorithm \cite{Shor1997} for prime-number factoring and Grover's
algorithm \cite{Grover1996} for unstructured search.  Once the
technology is fully developed, it is expected that quantum computing
will replace classical computing for some complex computational tasks.

\begin{figure}[t]
  \centering
\includegraphics[width=.8\textwidth]{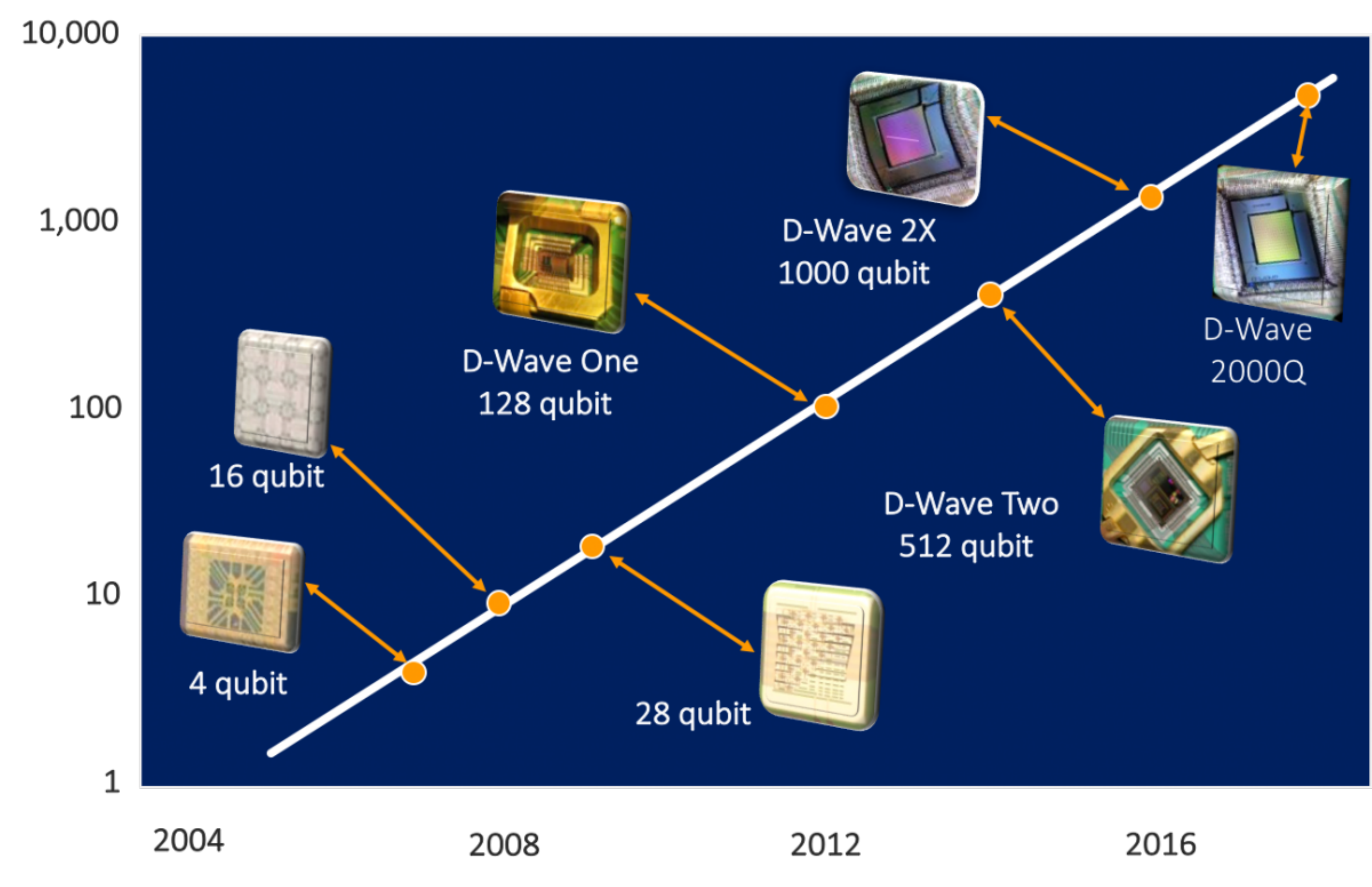}
\\
\includegraphics[width=.7\textwidth]{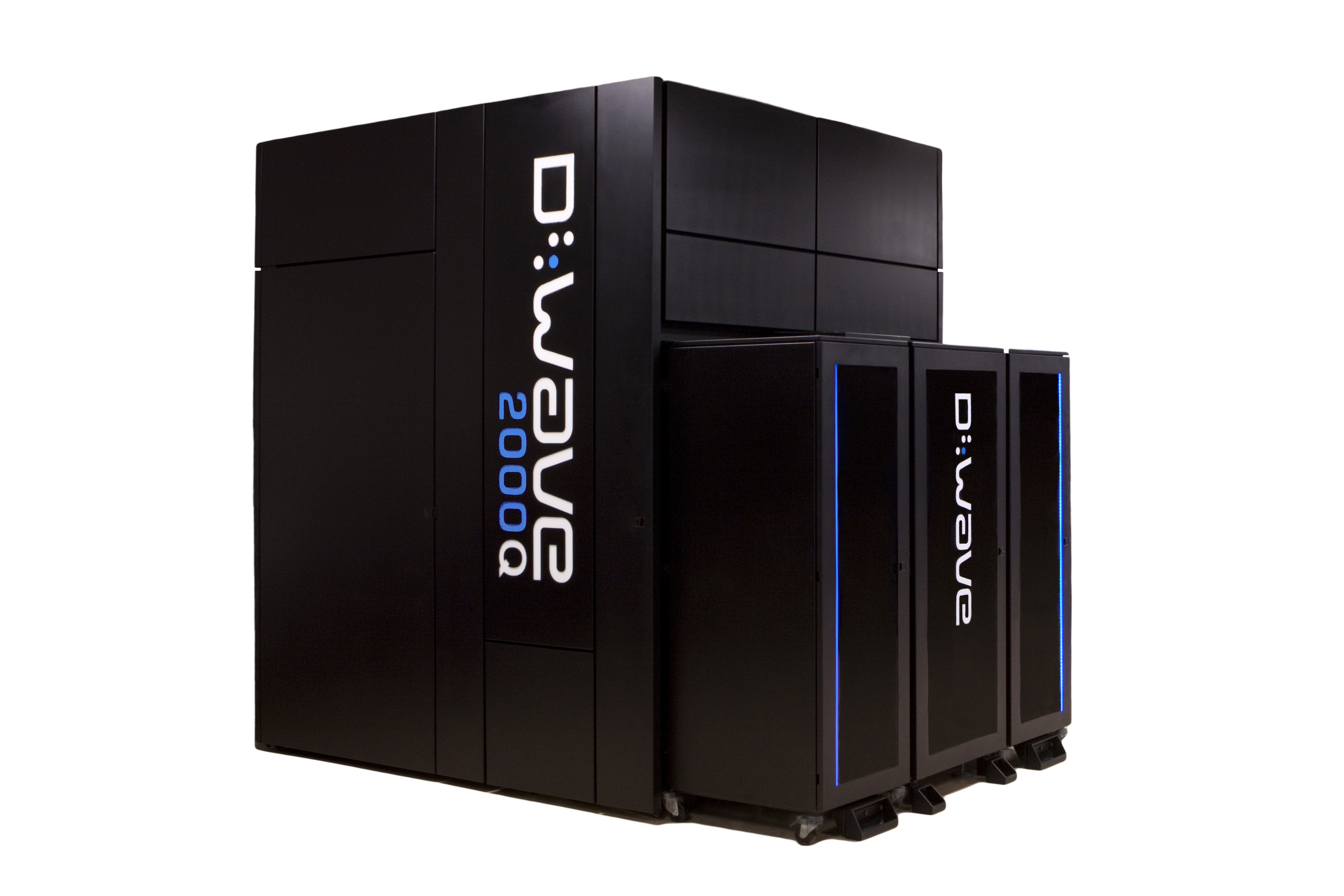}
  \caption{Top: Moore-like progress diagram of the development of D-Wave's Quantum
    annealers. \newline
X axis: year of release. Y axis: \# of qubits.
Notice the logarithmic scale of the Y axis.
\newline
Bottom: The state-of-the-art D-Wave 2000Q quantum annealer. (Courtesy
D-Wave Systems Inc.)
  \label{fig:moorelaw}}
\end{figure}

However, despite large investment, the development of practical
gate-model quantum computers is still in its infancy and current
prototypes are limited to less than $20$ qubits.  An alternative
approach to standard gate-model QC is \emph{Quantum Annealing}, a form
of computation that efficiently samples the low-energy configurations
of a quantum system\cite{Finilla94, Kadowaki98, farhi2000quantum}. In
particular, D-Wave Systems
Inc.\footnote{\url{http://www.dwavesys.com}} has developed
special-purpose {Quantum Annealers (QAs)} which draw optima or
near-optima from certain quadratic cost functions on binary variables.
Since 2007, this approach has allowed D-Wave to improve QAs at a
Moore-like rate, doubling the number of qubits roughly every 1.2
years, and reaching 2048 qubits in the state-of-the-art D-Wave 2000Q
annealer in January 2017 (Figure~\ref{fig:moorelaw}).  These
sophisticated devices are nearly-completely shielded from magnetic
fields ($\le 10^{-9}$ Tesla) and are cooled to cryogenic temperatures
($\le 20$ mK).

\begin{figure}[t]
\includegraphics[width=\textwidth]{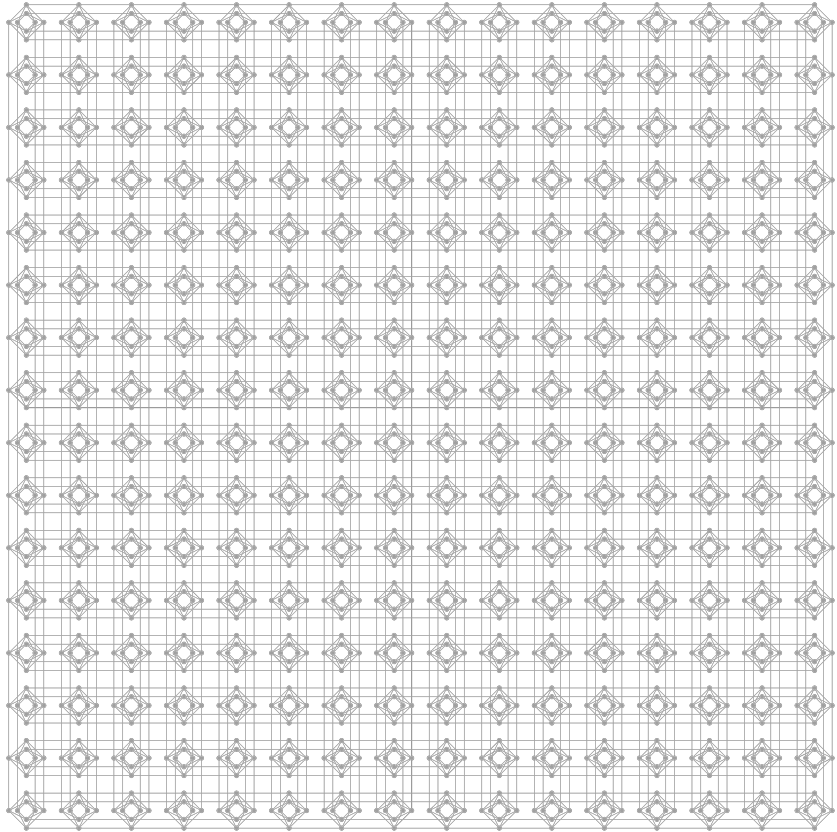}
\caption{\label{fig:full16chimera} The 2048-qubit connection graph
  of the D-Wave 2000Q quantum annealer architecture.}
\end{figure}

\begin{figure}[t]
  \centering
  \includegraphics[width=0.7\columnwidth]{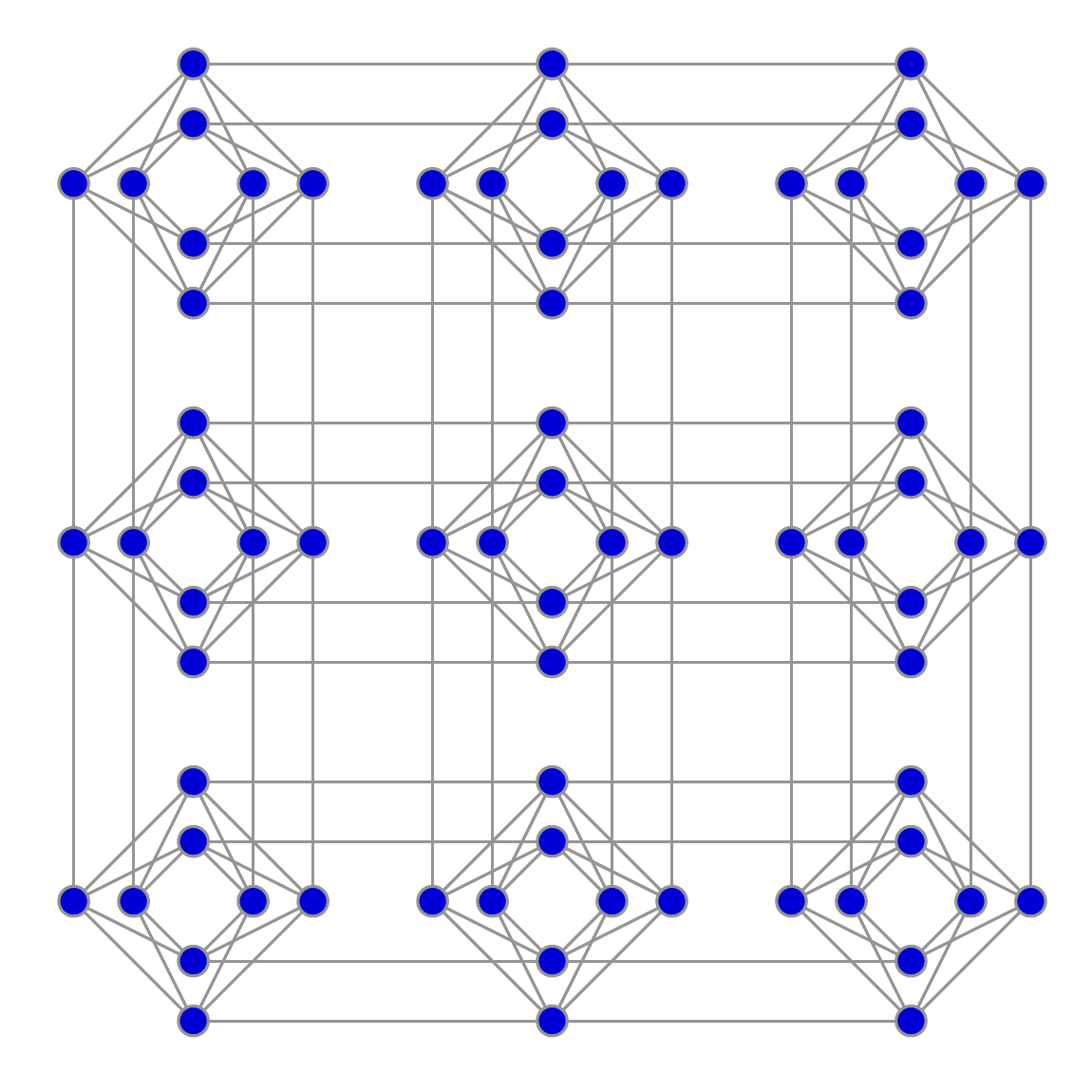}
  \caption{\label{fig:chimera}
    Example of the Chimera topology: the hardware graph for system
    of 72 qubits in a 3-by-3 grid of 8-qubit tiles. (D-Wave 2000Q systems
    have 2048 qubits in a 16-by-16 grid.)
\ignore{
    This topology consists of a lattice of strongly-connected
    components of 8 qubits, called \emph{tiles}. Each tile consists of
    a complete bipartite graph between two sets of four qubits. One
    set, the ``vertical" set, is connected to the tiles above and
    below; the other set, the ``horizontal" set, is connected to the
    tiles to the left and to the right. Notice that each qubit is
    connected with at most six other qubits. In other words, each
    variable $z_i$ in the Ising model \eqref{eq:isingmodel} has at
    most $6$ non-zero $\DWcoupij{}$ interactions with other
    variables.}}
\end{figure}

D-Wave's QAs can be used as specialized hardware for
solving the \emph{Ising {problem}}:
\begin{eqnarray}
\ignore{
\label{eq:isingmodel_minimization}
 \argmin_{\zs \in \set{-1,1}^{|V|}} H(\zs),
 \label{eq:isingmodel}
 &\ \ \ &
 H(\zs) \defas \sum_{i \in V} \DWbiasi z_i +
                  \sum_{\substack{(i,j)\in E}} \DWcoupij{} z_i z_j
}
          \label{eq:isingmodel_minimization}
 && \argmin_{\zs \in \set{-1,1}^{|V|}} H(\zs),
  \\
\label{eq:isingmodel}
H(\zs) &\defas & \sum_{i \in V} \DWbiasi z_i +
                 \sum_{\substack{(i,j)\in E}} \DWcoupij{} z_i z_j,
\end{eqnarray}
where each variable $z_i\in\set{-1,1}$ is associated with a qubit;
$G=(V,E)$ is an undirected graph, {\em the hardware graph}, whose
edges correspond to the physically allowed qubit interactions; and
$h_i$, $J_{ij}$ are programmable real-valued parameters.
%
$H(\zs)$ is known as the {\em Ising Hamiltonian} or {{\em Ising model}}. Ising
{problems} 
are equivalent to \emph{Quadratic Unconstrained Binary Optimization}
(QUBO) problems, which use $\{0,1\}$-valued variables rather than
$\{-1,1\}$-valued ones.\footnote{Ising variables $z_i$ are related
  to QUBO variables $x_i$ through $z_i = 2x_i-1$.}  In current 2000Q
systems, $\DWbiasi$ and $\DWcoupij{}$ must be within the ranges
$[-2,2]$ and $[-1,1]$ respectively, and $G$ is a lattice of
$16\times16$ 8-qubit bipartite modules ({\em tiles}) known as the {\em
  Chimera} topology, shown in Figures~\ref{fig:full16chimera} and~\ref{fig:chimera}. The quadratic term in
\eqref{eq:isingmodel} is restricted to the edges of $G$, which is very
sparse (vertices have degree at most $6$).  Despite this restriction, the Chimera Ising problem
\eqref{eq:isingmodel_minimization} is NP-hard
\cite{bunic-architecture14}.

Theory suggests that quantum annealing may solve certain optimization
problems faster than state-of-the-art algorithms on classical
computers \cite{farhi2000quantum}.  Quantum effects such as tunneling
and superposition provide QAs with novel mechanisms for escaping local
minima, thereby potentially avoiding sub-optimal solutions commonly
found by classical algorithms based on bit-flip operations (including
WalkSAT, simulated annealing and others
\cite{selman-wsat-96,Spears93simulatedannealing,tompkins_ubcsat:_2004}). Although
practical QA systems do not return optimal solutions with probability
1, the D-Wave processor has been shown to outperform a range of
classical algorithms on certain problems designed to match its
hardware structure \cite{denchev16google,king17fcl}.  This suggests
the possible use of QAs to address hard combinatorial
decision/optimization problems, in particular NP-hard problems like
SAT and MaxSAT \cite{HandbookOfSAT2009,LM09HBSAT}, by encoding them
into the Ising problem \eqref{eq:isingmodel_minimization}.

Our goal is to exploit quantum annealing as an engine for solving SAT,
MaxSAT, and other NP-hard problems. Since current QAs have a limited
number of qubits and connections, we target problem instances which
are relatively small but computationally hard enough to be out of the
reach of state-of-the-art classical solvers.  Since QAs are not
guaranteed to find an optimum and hence cannot certify the
unsatisfiability of an encoded formula (\sref{sec:background_qa}), we
target SAT problems such as cryptanalysis
\cite{Massacci2000,zhang_sat06,lafitte_JSAT14} or radio bandwidth
repacking \cite{Frechette2016SolvingTS} which are surely or
most-likely satisfiable, but whose solution is hard to find.

\medskip



\medskip {In this paper, we investigate the problem of encoding the
  satisfiability of an input Boolean formula $\Fx$ into an Ising
  problem \eqref{eq:isingmodel_minimization} from both theoretical and
  practical perspectives.  In principle, converting SAT to Ising with
  an unbounded number of fully-connected qubits is straightforward.
  In practice, these encodings must be done both effectively (i.e., in
  a way that uses only the limited number of qubits and connections
  available within the QA architecture, while optimizing performance
  of the QA algorithm), and efficiently (i.e., using a limited
  computational budget for computing the encoding).
}
%
We provide the necessary theoretical foundations, in which we analyze
and formalize the problem and its properties. Based on this analysis,
we then provide and implement practical encoding procedures. Finally,
we empirically evaluate the effectiveness of these encodings on a
D-Wave 2000Q quantum annealer.

We start from the observation that \sattoqubo{} can be formulated as a
problem in Satisfiability or Optimization Modulo Theories (SMT/OMT)
\cite{barrettsst09,sebastiani15_optimathsat} on the theory of linear
rational arithmetic, possibly enriched with uninterpreted function
symbols. \sattoqubo{} is an intrinsically over-constrained problem, so
a direct ``monolithic'' solution, encoding the whole input Boolean
formula $\Fx$ in one step, would typically require the introduction of
many additional ancillary Boolean variables.  These extra variables,
in addition to wasting many qubits, would result in very large SMT/OMT
formulas: solving the \sattoqubo{} via SMT would become
computationally very hard, possibly even harder than the original SAT
problem.

To cope with these issues, we adopt a scalable ``divide-and-conquer''
approach {to \sattoqubo{}}. First, we decompose the input Boolean
formula into a conjunction of smaller subformulas. Then, we encode
each subformula into an Ising model and place each subformula model
into a disjoint subgraph of the hardware graph. Finally, we connect
the qubits representing common variables from different subformulas
using chains of qubits that are constrained to be logically identical.

To exploit the intrinsic modularity of the architecture graph
(Figures~\ref{fig:full16chimera}, and \ref{fig:chimera}), we partition
the input formula $\Fx$ into subformulas which can be naturally
encoded and placed into one or two adjacent 8-qubit tiles of the
architecture, so that the encoding of each subformula is small enough
to be handled efficiently by an SMT/OMT solver, and the encoded
(sub)problems can be placed and interconnected within the modular
structure of the graph.  More concretely, we generate a library of
encodings of commonly-used and relatively-small Boolean
subfunctions. This library is only built once and consequently can use
a large amount of computational resources.  When presented with a SAT
formula $\Fx$, we decompose it, use the library to obtain encoded
(sub)functions and use place-and-routing techniques to place and
connect the encoded (sub)functions within the QA hardware graph.

We have implemented and made publicly available prototype encoders
built on top of the SMT/OMT tool \optimathsat{} \cite{st_cav15}.  We
present an empirical evaluation, in which we have run
\sattoqubo{}-encoded problems and \maxsattoqubo{}-encoded problems on
a D-Wave~2000Q system.  We have chosen input problems that are small
enough to fit into the current architecture but are very hard with
respect to their limited size, requiring some computational effort
using a state-of-the-art solver.

We stress the fact that this evaluation is not meant to present a
comparison with state-of-the-art of classic computing; rather, it is
intended as a preliminary assessment of the challenges and potential
of QAs to impact SAT and MaxSAT solving.  This assessment is
``preliminary'' due to the limitations in number of qubits and
qubit-connections of current QAs; however novel QAs currently under
development at D-Wave have a more interconnected tile structure and
higher per-qubit connectivity (degree $15$ instead of $6$, see also
\sref{sec:future}).\footnote{See
  \url{https://www.dwavesys.com/sites/default/files/mwj_dwave_qubits2018.pdf}.}

Empirical evaluation shows that most
encoded SAT problems
are
solved by the quantum annealer within negligible annealing time
($\approx10\mu s$).
%
Although preliminary, the results confirm the feasibility of the
approach. They also suggest that quantum annealers run on
\sattoqubo{}-encoded problems {(and to a lower extent,
\maxsattoqubo{}-encoded ones)} might outperform standard algorithms
on classical computers for certain difficult classes of relevant
problems as soon as QA systems contain enough qubits and connections.

\paragraph{Content of the paper}
The rest of the paper is organized as follows:
\sref{sec:background} presents necessary background on quantum
annealing, SAT, MaxSAT, SMT and OMT;
\sref{sec:foundations} presents the theoretical foundations of this
work;
\sref{sec:small_boolean} describes SMT/OMT-based encoding encoding
techniques for small Boolean formulas;
\sref{sec:larger-boolean} describes the process of encoding larger Boolean
formulas by formula decomposition, encoding, placement and routing;
\sref{sec:related} summarizes the related work;
\sref{sec:expaval} presents preliminary empirical evaluation;
\sref{sec:future} suggests future developments.

\paragraph{Disclaimer}
A preliminary and much shorter version of this paper was presented at
the 11th International Symposium on Frontiers of Combining Systems,
FroCoS'17 \cite{bian_frocos17}.

\section{Background}
\label{sec:background}
We provide the necessary background on quantum annealing
(\sref{sec:background_qa}) SAT, MaxSAT, SMT and OMT
(\sref{sec:background_smt}).

\subsection{Quantum Annealing}
\label{sec:background_qa}

As mentioned in \sref{sec:intro}, quantum annealers as currently
implemented by D-Wave Systems are specialized chips that use quantum
effects to sample or minimize energy configurations over binary
variables (qubits) in the form of an Ising model
\eqref{eq:isingmodel_minimization} \cite{bunic-architecture14,
  Harris_2010, Johnson2011}.  The qubits are interconnected in a grid
of tightly connected groups of 8 qubits, called \emph{tiles}, as
displayed in Figures~\ref{fig:full16chimera} and
\ref{fig:chimera}. Each tile consists of a complete bipartite graph
between two sets of four qubits: the ``vertical" set, which is
connected to the tiles above and below, and the ``horizontal" set,
which is connected to the tiles to the left and to the right. Each
qubit is connected to at most six other qubits, so that each variable
$z_i$ occurs in \eqref{eq:isingmodel} in at most 6 non-zero quadratic
terms $\DWcoupij{}z_iz_j$ (or $\DWcoupgen{ji}z_jz_i$).  The graphs in
Figures~\ref{fig:full16chimera} and \ref{fig:chimera} are known as
{\em Chimera} graphs.

\begin{figure}[t]
  \centering
  \begin{tabular}{l}
\includegraphics[width=.7\textwidth]{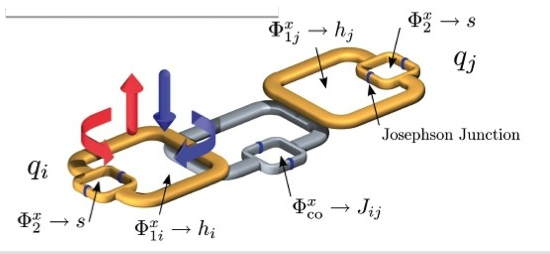}
\\ \ \\
\includegraphics[width=.7\textwidth]{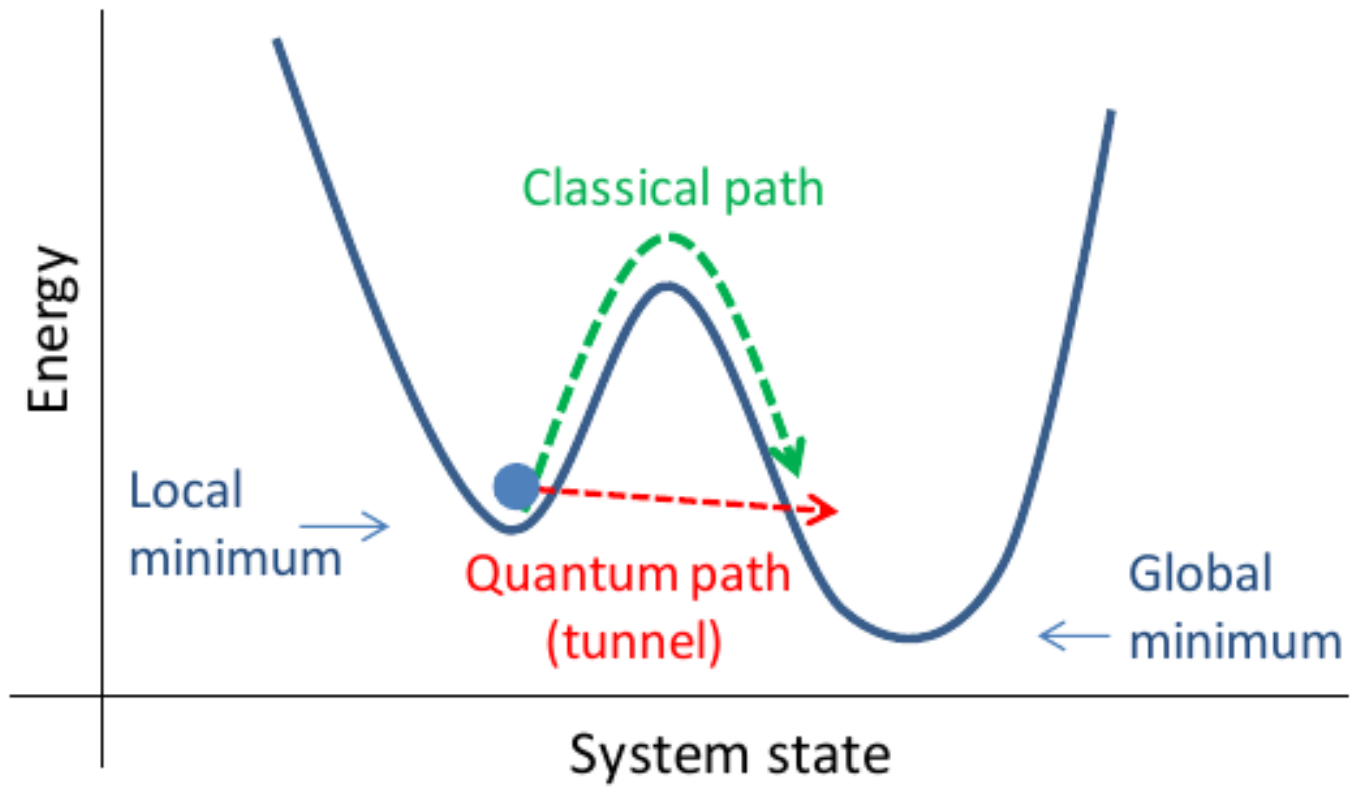}
  \end{tabular}
  \caption{\label{fig:implementation_tunneling}
Top: implementation of two coupled qubits. (Courtesy of
D-Wave Systems Inc.)\newline
 Bottom: graphical representation of the  tunneling effect within an energy landscape.}
\end{figure}

Single qubits $z_i$ are implemented as inter-connected superconducting
rings (Figure~\ref{fig:implementation_tunneling}, top), and a qubit's
$\pm1$-value represents the direction of current in its ring.  The
user-programmable values $h_i \in [-2,2]$ ({\em biases}) and
$J_{ij} \in [-1,1]$ ({\em couplings}) in
\eqref{eq:isingmodel_minimization} are real values within the
specified interval, and are set by applying magnetic flux to the
rings.\footnote{We consider normalized bounds without units of measure
  and scale because the only relevant information for us is that both
  ranges are symmetric wrt. zero and that the bounds for the
  \DWbiasi{}s are twice as big as these for the \DWcoupij{}s in
  \eqref{eq:isingmodel}.}  Overall, $H(\zs)$ in \eqref{eq:isingmodel}
defines the energy landscape for a system of qubits whose global
minimum correspond to the solutions of problem
\eqref{eq:isingmodel_minimization}.

During quantum annealing, the state of a qubit will be in a
superposition of $+1$ and $-1$ simultaneously. The system of $|V|$
qubits is evolved from an initial Hamiltonian, whose lowest energy
state is an equal superposition of all $2^{|V|}$ classical states,
to a final, user-defined Hamiltonian as in \eqref{eq:isingmodel}.  At
the end of the annealing, the system is measured, and a single,
classical state $\zs\in\set{-1,1}^{|V|}$ is observed. In theory, if
the evolution is sufficiently slow,\footnote{Notice that here and
  elsewhere ``slow'' is intended in a quantum-physics sense, which is
  definitely not ``slow'' from a computer-science perspective: e.g., a
  complete annealing process on a D-Wave 2000Q annealer may typically
  take \mbox{$\approx10\mu s$.}}  then the lowest energy state (the
\emph{ground state}) is maintained throughout. As a result, the final
state $\zs$ is a solution to the Ising problem
\eqref{eq:isingmodel_minimization} (with some probability, see below).
Unlike classical minimization techniques such as simulated annealing
\cite{Spears93simulatedannealing}, the QA energy-minimization process can use {\em quantum tunneling} \cite{lanting10tunneling} to pass
through tall, thin energy barriers, thereby avoiding trapping in
certain classical local minima
(Figure~\ref{fig:implementation_tunneling}, bottom).

\medskip QA theory shows that in the limit of arbitrarily low
temperature, arbitrarily small noise, and arbitrarily slow annealing,
the probability of obtaining a minimum energy solution converges to
1. In practice, these conditions cannot be achieved, and minimum
energy solutions are not guaranteed.
%
%
Indeed, practical QA systems are physical, analog devices, subject to
engineering limitations, and the optimal annealing rate is often
determined empirically. Moreover, hardware performance is dramatically
affected by the choice of Ising model.  Among the most relevant
factors are:

\begin{itemize}
\item[\emph{Thermal and electromagnetic noise}.]  Despite cooling and
  shielding, thermal and electromagnetic noise still have noticeable
  effects. One (approximate) model of these effects is based on
  Boltzmann sampling, in which the probability of seeing a state $\zs$
  with energy $H(\zs)$ in $\eqref{eq:isingmodel}$ is proportional to
  $e^{-\beta H(\zs)}$, with $\beta \in [3,5]$ being observed for
  certain problem classes
  \cite{amin15quasistatic,kulchytskyy2016qbm,raymond2016warming}.


\item[\emph{Intrinsic parameter errors}.] Due to engineering
  limitations and sources of environmental noise, the Ising model
  realized in QA hardware is not exactly the one programmed by the
  user. A simplified model of error is that each specified
  $h_i \in [-2,2]$ and $J_{ij} \in [-1,1]$ value is subject to
  additive Gaussian noise with standard deviation $0.03$ and $0.02$
  respectively.  \ignore{
    As a result, all other things being equal, Ising models with fewer
    qubits are preferable because they are subject to less noise.  }

\item[\emph{Freeze-out}.]
Because of the limited connectivity, we often use
chains of several interconnected qubits to represent a single Boolean
variable (\sref{sec:problems_embedding}).
However, the {quantum tunneling} effect on which quantum annealing is based is
diminished for chains \cite{lanting10tunneling}, thereby reducing the
hardware's ability to find global minima.  This effect can be
mitigated by constructing Ising models with chains that are as small
as possible.

\item[\emph{Energy gaps}.]  From the Boltzmann model, we see that a
  larger energy gap $g_{min}$ between ground and excited states leads
  to a higher probability of an optimal solution, as a ground state is
  $e^{\beta g_{min}}$ times more likely than a first excited state.
This suggests producing Ising models with large $g_{min}$ in order to
maximize the probability of obtaining an optimal solution.

\ignore{
\item[\emph{Energy gaps}.]
By Boltzmann sampling we can provide an (inaccurate
  \cite{amin15quasistatic,kulchytskyy2016qbm,raymond2016warming})
  approximation of the hardware's behaviour:
state $\zs$ is returned with probability
\begin{equation}
p(\zs) \sim e^{-\beta H(\zs)}
\label{eqn:boltzmann}
\end{equation}
for some fixed inverse temperature $\beta > 0$. From
\eqref{eqn:boltzmann}, we see that a larger energy gap $g_{min}$
between ground and excited states leads to a higher probability of an
optimal solution, as a ground state is $e^{\beta g_{min}}$ times more
likely than a first excited state.
This suggests to produce Ising models maximizing $g_{min}$ in order to
maximize the probability of obtaining an optimum solution.
}

\end{itemize}

%
The fact that QAs are not guaranteed
to return a minimum-energy solution is partially addressed by
taking a sequence of $N$ samples from the same Ising model and selecting the result with smallest energy.
Distinct samples are statistically independent, so the probability $P_{{\sf min}}[N]$ of
obtaining at least one minimum solution over
 $N$ samples converges exponentially to 1 with $N$:
 \begin{eqnarray}
P_{{\sf min}}[N] = 1 - (1 - P_{{\sf min}}[1])^N.
 \end{eqnarray}
%
Typical annealing times and readout times are very short
($\approx10\mu s$ and $\approx120\mu s$ respectively), and many samples can be drawn from the same Ising model within a single programming cycle, so is possible to obtain a large number of samples in reasonable time.



\begin{IGNORE} 
\newpage
\RSNOTE{vedi anche qualifying}
\RSTODO{Sposta e fondi in intro?}
Quantum Annealing (QA) is a specialized form of computation that uses
quantum mechanical effects to efficiently sample low-energy
configurations of particular quadratic cost functions on binary
variables.
Currently, the largest QA system  D-Wave 2000Q
heuristically minimizes an
Ising cost function given by
\begin{eqnarray}
  H(\zs) &\defas & \sum_{i \in V} \DWbiasi z_i +
                   \sum_{\substack{(i,j)\in E}} \DWcoupij{} z_i z_j
  \\
  && \argmin_{\zs \in \set{-1,1}^{|V|}} H(\zs).
\end{eqnarray}
where $G=(V,E)$ is an undirected graph of allowed variable interactions.  Ising
models are equivalent to \emph{Quadratic Unconstrained Binary
  Optimization} (QUBO) problems, which use $\{0,1\}$-valued
variables rather than $\pm 1$-valued variables.  \footnote{The bijective
  transformation between $z_i\in\set{-1,1}$ and $x_i\in\{0,1\}$ is
  $z_i = 2x_i-1$.}  The decision version of
the Ising problem on most graphs $G$ is NP-complete \cite{TODO}.

The connectivity graph $G$ of current D-Wave
processors is shown in Figure~\ref{chimera}, and is called the
\emph{Chimera} graph.
Further, because the Ising model is solved on a physical, analog
device, it is subject to engineering limitations. The D-Wave 2000Q
system currently requires $\DWbiasi \in [-2,2]$ and
$\DWcoupij{} \in [-1,1]$ and there are limits on the precision to
which these parameters may be specified.  Parameter imprecisions
act as small additive noise sources on parameter values, and arise
from operating quantum mechanical systems in real-world
environments. These real-world practicalities necessitate a carefully
defined SAT-to-Ising encoding.


\RSNOTE{IGNORARE DA QUI IN POI. SPOSTARE PARTE IN INTRO?}

The D-Wave 2000X system attempts to minimize an energy function
$H(\zs)$ known
as an \emph{Ising model} \eqref{eq:isingmodel}. In this formula,
$\DWcoupij{}$
and $\DWbiasi$ are rational values specified by a user, and $z_i$ values are
binary variables in $\set{-1,1}$ to be optimized,
and $G\defas\tuple{V,E}$ is the hardware graph (see below).
With some probability, the D-Wave system produces a state $\zs$ of
lowest energy (the \emph{ground state}), and therefore provides a
solution of the problem \eqref{eq:isingmodel_minimization}:
\begin{eqnarray}
  H(\zs) &\defas & \sum_{i \in V} \DWbiasi{} z_i +
                   \sum_{\substack{(i,j)\in E, \ i<j}} \DWcoupij{} z_i z_j
  \\
  && argmin_{\zs \in \set{-1,1}^{|V|}} H(\zs).
\end{eqnarray}

Finding the ground state of an Ising model is equivalent to
a \emph{Quadratic Unconstrained Binary Optimization} (QUBO)
problem.~%
\footnote{In the literature
QUBO problems are described as polynomials
  over variables $x_i\in\set{0,1}$, but the transformation from/to
  $z_i\in\set{-1,1}$ is
  trivial:
  $x_i= (z_i   +1) / 2 $ and $z_i = 2x_i-1$.}
The decision version of the QUBO problem is NP-complete, and using the D-Wave processor offers the possibility of solving faster than any classical algorithm.

Physically, the D-Wave system uses a process called \emph{quantum
annealing (QA)}. During QA, a quantum mechanical system is
transitioned from an initial, easy to prepare state to a final system
whose lowest energy state is the ground state of the Ising model
above.
The system (represented by the values $\DWbiasi$'s and $\DWcoupij{}$'s) is programmed by applying magnetic fields to a circuit
and the $z_i$ values (the \emph{qubits}) represent the direction of the current in the circuit at the end of the annealing process.

\begin{figure}[th]
  \centering
  \includegraphics[width=0.7\columnwidth]{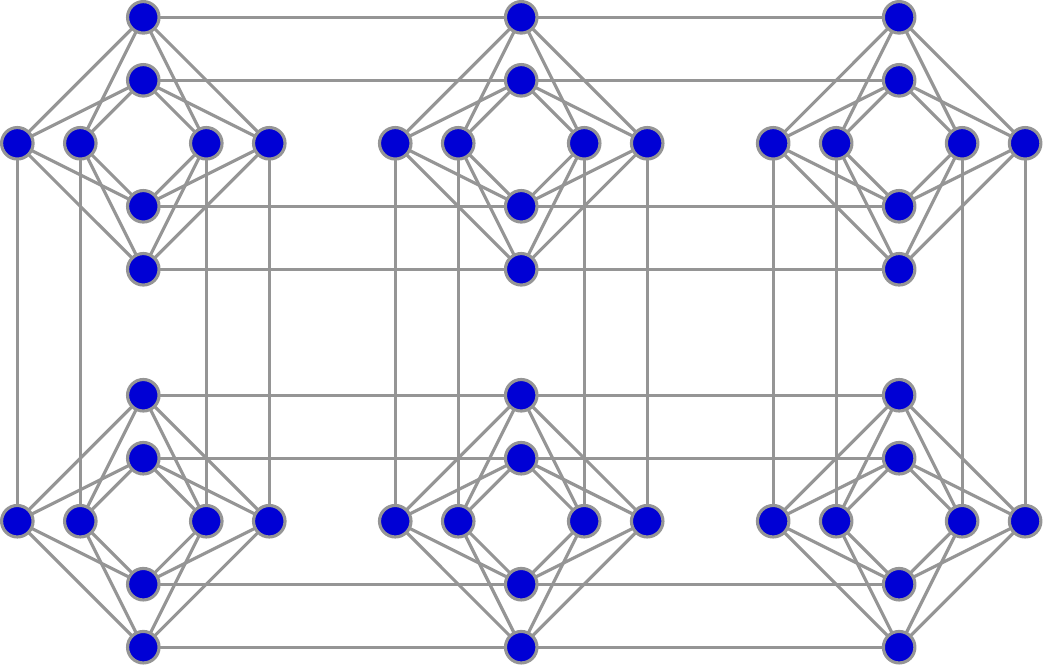}
  \caption{Example of Chimera topology: the hardware graph for system of
    48 qubits in a 2-by-3 grid of tiles (The D-Wave 2000X system has
2048 qubits in a 16-by-16 grid
\cite{dwavewebsite2000q}). 
This topology consists of a grid of strongly-connected components of 8 qubits, called \emph{tiles}. Each tile consists of a complete bipartite graph between two sets of four qubits. One set, the 'vertical' set, is connected to the tiles above and below; the other set, the 'horizontal', set is connected to the tiles to the left and to the right. Notice that each qubit is connected with at most six other qubits. In other words, each variable $z_i$ in the Ising model \eqref{eq:isingmodel} has at most $6$ non-zero $\DWcoupij{}$ interactions with other variables.}
\end{figure}

Because the Ising model is solved on a physical, analogue device, it is subject to several engineering limitations. Firstly, the Ising model must be specified within a certain range of rational numbers. The D-Wave 2000X system currently requires $\DWbiasi \in [-2,2]$ and $\DWcoupij{} \in [-1,1]$. Secondly, there are certain errors that occur as a result of programming analogue circuits and operating quantum mechanical systems in a real-world environment; as a result of these noise sources, the system is not guaranteed to return an optimal solution. Thirdly, the system has only a finite number of variables and limited number of non-zero interactions between those variables. In the Ising model \eqref{eq:isingmodel}, the entries $\DWcoupij{}$ that are allowed to be non-zero are determined by the edges $(i,j)$ in a graph $\tuple{V,E}$ referred to as the \emph{hardware graph}. In current D-Wave machines, qubits are connected in the pattern shown in Figure \ref{fig:chimera}, in what is called a \emph{Chimera topology}.
\ignore{
Hereafter, if not specified otherwise,
we implicitly assume that the graph $G\defas\tuple{V,E}$ is
the 2048-qubit hardware graph of the D-Wave 2000X machine, that is,
the $16\times 16$-tile extension of the graph in Figure~\ref{fig:chimera}.
}

\ignore{
Adapting generic problems to the architecture constraints is not trivial. Whereas the number of available qubits is ever increasing, it is still very limited and optimizing usage is difficult.
}
\end{IGNORE}

\subsection{SAT, MaxSAT, SMT and OMT} 
\label{sec:background_smt}
We assume the reader is familiar with the basic syntax, semantics and properties
 of Boolean and first-order logic and theories.
In the following we recall the main concepts of interest for our
 purposes,
referring the reader to
 \cite{HandbookOfSAT2009,MSLM09HBSAT,LM09HBSAT,barrettsst09,sebastiani15_optimathsat}
 for more details.

\paragraph{SAT \& MaxSAT}
\ignore{%
In particular, given some finite set of Boolean variables \xs{} (aka
 Boolean atoms)
a Boolean formula  is defined inductively as follows:
a Boolean atom is a Boolean formula; if $F_1,F_2$ are Boolean
 formulas, then
$\neg F_1$ (not $F_1$) and
$F_1\ op\ F_2$ are Boolean formulas, $op \in
 \set{\wedge,\vee,\imp,\iff,\oplus}$
(and, or, imply, iff, xor) with their standard meaning.
}
Given some finite set of Boolean variables \xs{} (aka Boolean atoms)
the language of Boolean
logic (\calb) is the set of formulas containing the atoms in \xs{}
 and closed under the standard propositional connectives
\set{\neg,\wedge,\vee,\imp,\iff,\oplus} (not, and, or, imply, iff,
xor)  with their usual meaning.
A {\em literal} is an atom (positive literal) or its
negation (negative literal).
We implicitly remove double negations: e.g., if $l$ is the negative literal
$\neg x_i$, then by $\neg l$ we mean $x_i$ rather than $\neg\neg x_i$.
A {\em clause} is a disjunction of literals. A formula is in
 {\em conjunctive normal form (CNF)} iff it is written as a
 conjunctions of clauses.

A truth value assignment \xs{} {\em satisfies} \Fx{} iff it makes it
evaluate to true. If so, \xs{} is called a {\em model} for \Fx{}.
A formula \Fx{} is {\em satisfiable} iff at least one truth
assignment satisfies it, {\em unsatisfiable} otherwise.
\Fx{} is {\em valid} iff all truth assignments satisfy it.
$F_1(\xs), F_2(\xs)$ are {\em equivalent} iff they are satisfied by exactly
the same truth assignments.

A formula \Fx which is not a conjunction can
always be decomposed 
into a conjunction of
smaller formulas \Fxy{} { by means of Tseitin's transformation
  \cite{tseitin68}}:
\begin{equation}
\label{eq:decomposition}
\Fxy \defas \bigwedge_{i=1}^{m-1} (y_i \iff \Fixy) \wedge  \Fmxy,
\end{equation}
where the $F_i$s are formulas which decompose the original
formula \Fx, and the
$y_i$s are fresh Boolean variables each labeling the corresponding
$F_i$.
%
(If the input formula is itself a conjunction, then Tseitin's
transformation can be applied recursively to each conjunct.)
Tseitin's
  transformation~\eqref{eq:decomposition} guarantees that $\Fx$ is
  satisfiable if and only if \Fxy{} is satisfiable, and that if $\xs,\ys$ is a  model for \Fxy, then \xs is a model for \Fx.
To this extent, it is pervasively used also as a main recursive step
for efficient CNF conversion of formulas \cite{tseitin68}.

A {\em quantified Boolean formula (QBF)} is defined inductively as follows:
a Boolean formula is a QBF; if \Fx{} is a QBF, then $\forall x_i \Fx$
and $\exists x_i \Fx$ are QBFs.
$\forall x_i \Fx$
is equivalent to $(\Fx_{x_i=\top} \wedge \Fx_{x_i=\bot})$
and {$\exists x_i \Fx$} is equivalent to {$(\Fx_{x_i=\top} \vee
\Fx_{x_i=\bot})$} (aka {\em Shannon's expansion}).

{\em Propositional Satisfiability (SAT)} is the problem of
establishing whether an input Boolean formula is satisfiable or not.  SAT
is NP-complete \cite{cook1}.  Efficient SAT solvers are publicly
available, most notably those based on {\em Conflict-driven
clause-learning (CDCL)} \cite{MSLM09HBSAT} and on {\em stochastic
local search} \cite{Maj09HBSAT}.  Most solvers require the input
formula to be in CNF, implementing a CNF pre-conversion based on
Tseitin's transformation~\eqref{eq:decomposition} when this is not the
case.  See \cite{HandbookOfSAT2009} for a survey of SAT-related
problems and techniques.

Weighted \emph{MaxSAT} $\set{\tuple{F_k,c_k}}_k$ is an optimization
extension of SAT, in which the input formula is a (typically
unsatisfiable) conjunction of subformulas $F\defas \bigwedge_k F_k$
such that each conjunct $F_k$ is given a positive penalty $c_k$ if
$F_k$ is not satisfied, and an assignment minimizing the sum of the
penalties is sought. (Often $F$ is in CNF and the $F_k$s are single
clauses or conjunctions of clauses.)  {\em Partial Weighted MaxSAT} is
an extension of Weighted MaxSAT in which some conjuncts, called {\em
  hard constraints}, have penalty $+\infty$.  Efficient MaxSAT tools
are publicly available (see, e.g.,
\cite{LM09HBSAT,tompkins_ubcsat:_2004}).


\paragraph{SMT and OMT} {Satisfiability Modulo Theories (SMT)} is the
problem of checking the satisfiability of first order formulas in a
background theory \T (or combinations of theories thereof). We focus
on the theories of interest for our purposes.  Given \xs{} as above
and some finite set of rational-valued variables \vs{}, the language
of the theory of {\em Linear Rational Arithmetic} (\larat) extends
that of Boolean logics with
{\larat-atoms} in
the form $(\sum_i c_i v_i \bowtie c)$, $c_i$ being rational
values, $v_i\in \vs$  and $\bowtie\ \in\set{=,\neq,<,>,\le,\ge}$, with
their usual meaning.
In the theory of {\em linear rational-integer arithmetic with
uninterpreted functions symbols}  (\laeuf)
 the \larat language is extended by adding integer-valued variables
to \vs (\la) and
 {uninterpreted
function symbols}.~\footnote{A n-ary function symbol $f()$ is
said to be {\em uninterpreted} if its interpretations have no constraint,
except that of being a function (congruence): if
$t_1=s_1$, ..., $t_n=s_n$ then $f(t_1,...,t_n)=f(s_1,...,s_n)$.}
(E.g., $(x_i \imp (3v_1 + f(2v_2) \le f(v_3)))$ is a \laeuf
formula.)
Notice that \calb is a sub-theory of  \larat and \larat is a
sub-theory  of \laeuf. The notions of literal, assignment,
clause and CNF, satisfiability, equivalence and validity,
Tseitin's transformation and
quantified formulas extend straightforwardly to \larat and \laeuf.

\emph{Satisfiability Modulo \laeuf{} (\smt{(\laeuf)})} \cite{barrettsst09}
is the problem of deciding the
satisfiability of arbitrary formulas on {\laeuf} and its sub-theories.
Efficient \smt{(\laeuf)} solvers are
 available, including \mathsatfive \cite{mathsat5_tacas13}.

\emph{Optimization Modulo \laeuf{}
(\omt(\laeuf))} \cite{sebastiani15_optimathsat} extends \smt{(\laeuf)}
searching solutions which optimize some \la{} objective(s).
Efficient \omlarat solvers like \optimathsat \cite{st_cav15} are available.

\section{Foundations}
\label{sec:foundations}
Let \Fx be a Boolean function on a set of $n$ {\em 
Boolean
variables} $\xs\defas\set{x_1,...,x_n}$. 
We represent Boolean value $\bot$ with $-1$ and $\top$ with $+1$, 
so that we can assume that each $x_i\in \set{-1,1}$.
Suppose first that we have a QA system with $n$ qubits defined on a hardware 
graph $G=(V,E)$, for instance, any $n$-vertex subgraph of the Chimera graph
of Figures~\ref{fig:full16chimera} and \ref{fig:chimera}. Furthermore,
we assume that the state of each qubit 
$z_i$ corresponds to the value of variable $x_i$, $i=1,\ldots,n=|V|$. 
One way to determine whether \Fx 
is satisfiable using the QA system is to find 
an energy function as
in \eqref{eq:isingmodel} whose ground states $\zs$ correspond to the
satisfying assignments $\xs$ of \Fx.

\begin{example}
Suppose $F(\xs) \defas x_1 \oplus x_2$. Since $F(\xs)=\top$ if and only if $x_1+x_2=0$, 
the Ising model
 $H(z_1,z_2) = z_1\cdot z_2$ 
in a graph containing 2 qubits $z_1$, $z_2$ joined by an edge
$(1,2)\in E$ s.t. $J_{12}=1$  has two ground states $(+1,-1)$ and $(-1,+1)$, 
which correspond to  the satisfying assignments of $F$, and two
excited states $(+1,+1)$ and $(-1,-1)$, corresponding to the
non-satisfying ones.
\end{example}

Because the energy $H(\zs)$ in \eqref{eq:isingmodel} is restricted to quadratic terms and graph $G$ is 
typically sparse, the number of functions \Fx that can be solved with
this approach is very limited.
To deal in part with this difficulty, we can use a larger QA system with 
a number of additional qubits, say $h$,  representing {\em ancillary Boolean variables} (or \emph{ancillas} for short)
$\as\defas\set{a_1,...,a_h}$, so that $|V|=n+h$. 
%
%
A {\em variable placement} is a mapping of the $n+h$ input and ancillary
variables into the qubits of $V$. 
Since $G$ is not a complete graph, different variable placements will produce energy functions with different properties.
We use {\em Ising encoding} to refer to the \DWbiasi{} and
  \DWcoupij{} parameters in \eqref{eq:isingmodel} that are provided to
  the QA hardware together 
with a variable placement. 
The {\em gap} 
of an Ising encoding is the minimum energy difference between ground states 
(i.e., satisfying assignments) and the other states (i.e.,
non-satisfying assignments). 
In general, larger gaps lead to higher success rates in the QA process \cite{bian2014discrete}. Thus, we define
the {\em encoding problem} for $\Fx$ as the problem of finding an Ising encoding
with maximum gap.

%

Note that the encoding
problem is typically over-constrained. The Ising model
\eqref{eq:isingmodel} has to discriminate between $m$ satisfying
 assignments and $k$ non-satisfying assignments, with $m+k=2^n$, 
whereas the number of degrees of freedom is given by the number of the
\DWbiasi{} and \DWcoupij{} parameters, which grows as $O(n+h)$ in the Chimera
architecture. Thus, in order to have a solution, the number of
ancilla variables needed ($h$) may grow exponentially with the number of $\xs$ variables ($n$). 

In the rest of this section, we assume that a Boolean function \Fx is
given and that $h$ qubits are used for ancillary variables $\as$.

\subsection{Penalty Functions}

%

%
Here we assume that a variable placement is given,
placing $\xs\cup\as$ into the subgraph $G$.
%
Thus, we can identify each variable $z_j$
representing the binary value of the qubit
associated with the $j$th vertex in $V$
with either an original variable $x_k\in\xs$ or as an ancilla variable
$a_\ell\in\as$, writing $\zs = \xs\cup\as$.

\begin{definition}
\label{def:penalty}
A {\bf penalty function} $\Pxa$ is an Ising model
\begin{align}
  \label{eq:penfunction}
\Pxa \defas \offset{} + \sum_{\substack{i\in V}} \theta_{i} z_i +
\sum_{\substack{(i,j) \in E}}
\theta_{ij} z_i z_j
\end{align}
with the property that for some $g_{min}>0$,
\begin{align}
  \text{ }\quad  \forall \xs\ \ min_{\set{\as}} \Pxa
  \begin{cases}
= 0 &\text{ if }   F(\xs)=\top \\
   \geq g_{min} &\text{ if } F(\xs)=\bot
  \end{cases}\label{eq:pencriteria}
\end{align}
where $\offset{}\in (-\infty,+\infty)$ (``{\em offset}''),
$\theta_i \in [-2,2]$ (``{\em biases}'') and
$\theta_{ij} \in [-1,1]$ (``{\em couplers}'')
such that $z_i,z_j\in\zs$,
and $g_{min}$ are rational-valued parameters.~%
The largest $g_{min}$ such that $\Pxa$ satisfies \eqref{eq:pencriteria} is called the {\bf gap} of $\Pxa$.
\end{definition}

\noindent
Notice that a penalty function 
separates satisfying assignments from non-satisfying ones by a gap of
at least 
$g_{min}$. 
The offset value \offset{} is added to set the value of
\Pxa to zero when $F(\xs)=\top$, so that $-\offset$ corresponds to the
energy of the ground states of \eqref{eq:isingmodel}.

To simplify the notation we assume that $\theta_{ij}=0$ when
$(i,j) \not\in E$, and use $\Px$ when $\as=\emptyset$.

\begin{example}
\label{es:exampleiff}
  The equivalence between two variables,
  $\Fx\defas (x_1 \leftrightarrow
  x_2)$, can be encoded without ancillas by means of a single
  coupling between two connected vertices, with zero biases:
$\Px \defas 1 - x_1x_2$, so that
  $g_{min}=2$. In fact, $\Px=0$ if $x_1,x_2$ have the same value;
  $\Px=2$ otherwise.
\end{example}
\noi
Penalty \Px in Example~\ref{es:exampleiff} is also called a (equivalence) \emph{chain} connecting $x_1,x_2$,
because it forces $x_1,x_2$ to have the same value.

The following examples show that ancillary variables are needed, even
for small Boolean functions \Fx and even when $G$ is a complete graph.

\begin{figure}[t]
  \subfigure[\label{fig:mappings_and}$x_3 \iff (x_1 \wedge x_2)$ with
  one ancilla.]%
  {\includegraphics[width=0.21\columnwidth]{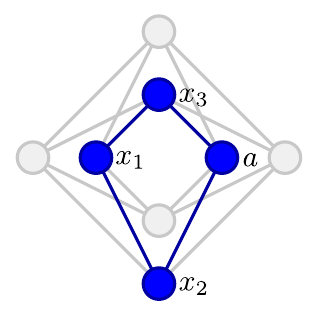} }
  \subfigure[\label{fig:mappings_xor}$x_3 \iff (x_1 \oplus x_2)$ with
  three ancillas.]%
  {\includegraphics[width=0.21\columnwidth]{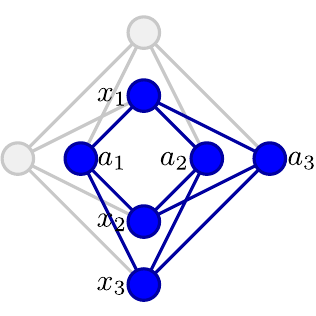}\ \ \ \ \ }
  \subfigure[\label{fig:mappings_two}$x_4 \iff (x_3 \wedge (x_1 \oplus
  x_2))$ \hspace{\textwidth} obtained by combining
  \ref{fig:mappings_xor} and \ref{fig:mappings_and}.]%
  {\includegraphics[width=0.5\columnwidth]{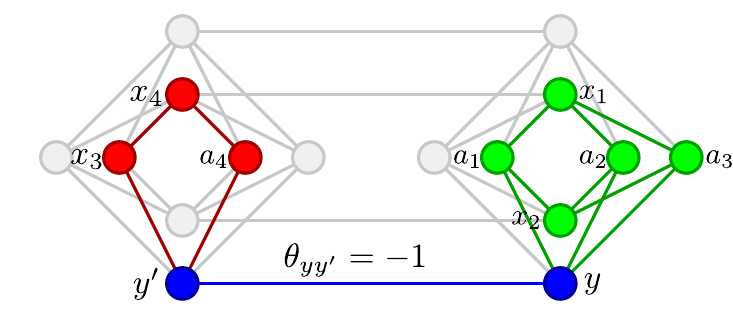}\ \ \ \ \ }
\caption{\label{fig:mappings} Mappings within the Chimera graph, penalty functions use only colored edges. \ref{fig:mappings_two} combines \ref{fig:mappings_and} and
  \ref{fig:mappings_xor} using chained proxy variables $y,y'$.
The resulting penalty function is obtained by rewriting
$x_4 \iff (x_3 \wedge (x_1\oplus x_2))$
into its equi-satisfiable formula
$
(x_4 \iff (x_3\wedge y'))
\wedge
(y \iff (x_1 \oplus x_2))
\wedge
(y' \iff y)
$.
}
\end{figure}

\begin{example}\label{es:exampleand}
  Consider the AND function $\Fx\defas x_3 \leftrightarrow
  (x_1 \wedge x_2)$.
  If $x_1,\ x_2,\ x_3$ could be all connected in a 3-clique,
  then \Fx could be encoded without ancillas by setting
$\Px = \frac{3}{2} -\frac{1}{2} x_1 - \frac{1}{2}x_2 + x_3 +\frac{1}{2}x_1 x_2 - x_1
x_3 - x_2 x_3 $, so that
  $g_{min}=2$.
{
  In fact, 
  $\Px=0$ if $x_1,x_2,x_3$
  verify \Fx, $\Px=6$ if $x_1=x_2=-1$ and $x_3=1$,
  $\Px=2$ otherwise.
}
Since the Chimera graph has no cliques, 
the above AND function needs (at least) one ancilla $a$ to be encoded  as:
$\Pxa =  \frac{5}{2} - \frac{1}{2} x_1 - \frac{1}{2} x_2 + x_3 + \frac{1}{2}
x_1 x_2 - x_1 x_3 - x_2a - x_3a $, which still has gap $g_{min} =2$
and can be embedded, e.g., as in Figure~\ref{fig:mappings_and}.
%
%
%
\end{example}
\begin{example}
\label{es:examplexor}
Consider the XOR function
$\Fx\defas x_3 \leftrightarrow (x_1 \oplus x_2)$. Even within a
3-clique, \Fx has no ancilla-free encoding.
Within the Chimera graph, \Fx
can be encoded with three ancillas $a_1, a_2, a_3$ as:
  $
  \Pxa =  5 + x_3 + a_2 - a_3 + x_1 a_1  - x_1 a_2 - x_1 a_3 - x_2 a_1 - x_2 a_2
  - x_2 a_3 + x_3 a_2 - x_3 a_3
  $, which has gap $g_{min} = 2$ and is embedded, e.g., as in Figure
  ~\ref{fig:mappings_xor}. 
\end{example}

The following fact is a straightforward consequence of
Definition~\ref{def:penalty}.
\begin{proposition} Let \Pxa be a penalty function of \Fx as in
\label{prop:penalty}
  Definition~\ref{def:penalty}. Then:
  \begin{itemize}
  \item
  If \xs{},\as{} is such that $\Pxa=0$, then \Fx is satisfiable and
  \xs satisfies it.
  \item
  If \xs{},\as{} minimizes \Pxa and $\Pxa\ge g_{min}$, then \Fx is unsatisfiable.
  \end{itemize}

\end{proposition}

Proposition~\ref{prop:penalty} shows that the QA hardware can used as
a satisfiability checker for \Fx by minimizing the Ising model defined
by penalty function $\Pxa$.  A returned value of $\Pxa = 0$ implies
that $\Fx$ is satisfiable.  {If the QA hardware guaranteed minimality, then a
  returned value of $\Pxa \ge g_{min}$ would imply that $\Fx$ is
  unsatisfiable.}  However, since QAs do not guarantee minimality
(\sref{sec:background_qa}), if $\Pxa\ge g_{min}$ then there is still a
chance that $\Fx$ is satisfiable. Nevertheless, the larger $g_{min}$
is, the less likely this false negative case occurs
\cite{bian2014discrete}.

A penalty function $\Pxa$ is \emph{normal} if
$|\theta_i|=2$ for at least one $\theta_i$ or
$|\theta_{ij}|=1$ for at least one $\theta_{ij}$.
In order to maximize $g_{min}$, it is important to use normal penalty
functions so that to exploit the full range of the \ts parameters.
Any penalty function \Pxa can be normalized
by multiplying all its coefficients by a \emph{normalization factor}:
\begin{eqnarray}
\label{eq:normalization-factor}
  c \defas
  \min\left\{\min_i\left(\frac{2}{|\theta_{i}|}\right),\min_{\tuple{ij}}\left(\frac{1}{|\theta_{ij}|}\right)\right\}.
\end{eqnarray}
Note that if \Pxa is non-normal, then $c>1$, so that the resulting gap
$c\cdot g_{min}>g_{min}$. Normalization also works in the opposite
direction to scale down some \Pxa whose $\theta$'s do not fit into the
allowable ranges (in which case $c<1)$.

Hereafter we assume w.l.o.g. that all penalty functions are normal.

\label{sec:penalty}

\subsection{Properties of Penalty Functions and Problem Decomposition}
\label{sec:problems_decomposition}

As it will be made clear in \sref{sec:computing_offline}, after a
variable placement is set, finding the values for the
$\theta$'s implicitly requires solving a set of equations whose size grows with the number of models of \Fx plus 
a number of inequalities whose size grows with the number of
counter-models of \Fx.
Thus, the $\theta$'s must satisfy 
a number of linear constraints that grows exponentially in $n$.
Since the $\theta$'s grow approximately as $4(n+h)$, the number of ancillary variables 
needed to satisfy \eqref{eq:pencriteria} can also grow very rapidly.
This seriously limits the scalability of a solution method based on
\eqref{eq:penfunction}-\eqref{eq:pencriteria}.
%
%
%
%
%
%
We address this issue by showing how to construct penalty functions by
combining smaller penalty functions, 
{albeit at the expense of introducing extra variables.}

The following properties are straightforward consequence of Definition~\ref{def:penalty}.

\begin{property}
\label{prop:equivalence}
  Let \Pxa be a penalty function for \Fx and let $F^*(\xs)$ be logically
equivalent to \Fx. Then \Pxa is a penalty function also for
$F^*(\xs)$ with the same gap $g_{min}$.
\end{property}

Property \ref{prop:equivalence} states that a penalty function \Pxa
does not depend on the syntactic structure of \Fx but only on its
semantics. 

\begin{property}
\label{prop:npn}
Let $F^*(\xs)\defas
F(x_1,...,x_{r-1},\neg x_r,x_{r+1},...,x_n)$ for some index $r$.
Assume a variable placement of \xs into $V$ s.t.  $\Pxa$ is a penalty
function for \Fx of gap $g_{min}$.
Then $P_{F^*}(\xs,\as|\ts) = P_{F}(\xs,\as|\ts^*)$,
  where $\ts^*$ is defined as follows for every $z_i,z_j\in\xs,\as$:
\begin{eqnarray*}
  \theta^*_{i} =
\left \{
  \begin{array}{ll}
    -\theta_{i} & \text{ if}\ z_i=x_r\\
     \theta_{i} & \text{ otherwise;}\\    
  \end{array}
\right .
&\ \ \ \ \ & 
  \theta^*_{ij} =
\left \{
  \begin{array}{ll}
    -\theta_{ij} & \text{ if}\ z_i=x_r\ \text{ or}\  z_j=x_r\\
     \theta_{ij} & \text{ otherwise.}\\    
  \end{array}
\right .
\end{eqnarray*}
Notice that since the previously defined bounds over $\ts$ (namely $\theta_i \in [-2,2]$ and
$\theta_{ij} \in [-1,1]$)  are symmetric, if $\ts$ is in range then $\ts^*$ is
as well. 

\end{property}

Two Boolean functions that become equivalent by permuting or negating some
of their variables are called
\emph{NPN-equivalent} \cite{correia2001classifying}. Thus,
given
the penalty function for a Boolean formula, any other NPN equivalent formula can
be encoded trivially by repeatedly applying Property~\ref{prop:npn}. 
Notice  that checking NPN equivalence is a
hard problem in theory, but it is fast in practice for small $n$ (i.e., $n \leq 16$) \cite{huang2013fast}. The process of negating a single variable in an Ising model as in Property \ref{prop:npn} is known as a \emph{spin-reversal transform}.

\begin{example}\label{es:exampleor}
  Consider the OR function $\Fx\defas x_3 \leftrightarrow
  (x_1 \vee x_2)$. We notice that this can be rewritten as
  $\Fx= \neg x_3 \leftrightarrow
  (\neg x_1 \wedge \neg x_2)$, that is, it is NPN-equivalent to that
  of Example~\ref{es:exampleand}. Thus, by Property~\ref{prop:npn} a penalty function for
  \Fx can be placed as in Figure~\ref{fig:mappings_and} and defined by
  taking that in Example~\ref{es:exampleand} and toggling the signs of the coefficients of the $x_i$'s:
  $\Pxa=\frac{5}{2} + \frac{1}{2} x_1 + \frac{1}{2} x_2 - x_3 + \frac{1}{2}
x_1 x_2 - x_1 x_3 + x_2a + x_3a $, which still has gap $g_{min}
=2$. 
  
\end{example}
\ignore{
  \begin{example}\label{es:exampleand}
  Consider the AND function $\Fx\defas x_3 \leftrightarrow
  (x_1 \wedge x_2)$.
  If $x_1,\ x_2,\ x_3$ could be all connected in a 3-clique,
  then \Fx could be encoded without ancillas by setting
\end{example}
}

\newcommand{\ask}{\ensuremath{\underline{\mathbf{a}^k}}\xspace}
\newcommand{\xsk}{\ensuremath{\underline{\mathbf{x}^k}}\xspace}
\newcommand{\xskp}{\ensuremath{\underline{\mathbf{x}^{k'}}}\xspace}


\begin{property}
\label{prop:decomposition}
  Let $\Fx = \bigwedge_{k=1}^K F_k(\xsk)$ be a Boolean formula such that $\xs=\cup_k\xsk$, the \xsk{}s may be non-disjoint, and each sub-formula
  $F_k$ has a penalty function $P_{F_k}(\xsk,\ask|\ts^k)$ with minimum
  gap $g_{min}^k$ where 
the \ask{}s are all disjoint. Given a list $w_k$ of positive rational values such that, for every
  $z_i,z_j\in \xs\cup\bigcup_{k=1}^K \ask$:
\begin{eqnarray}
\label{eq:parameterranges}
\theta_i \defas \sum_{k=1}^K w_k \theta_i^k \in [-2,2], \quad \theta_{ij} \defas \sum_{k=1}^K w_k
  \theta_{ij}^k \in [-1,1],
\end{eqnarray}
then a penalty function for \Fx is:
\begin{eqnarray}
P_{F}(\xs,\as^1...\as^K|\ts) = \sum_{k=1}^K w_k
P_{F_k}(\xsk,\ask|\ts^k).
\end{eqnarray}
%
  The gap for $P_{F}$ is 
  $g_{min}\ge \min_{k=1}^K w_k g_{min}^k$.
%
\label{prop:anddecompose} 
\end{property}

\noindent
The choice of the set of weights $w_k$ in Property~\ref{prop:anddecompose} is not unique in
general. 
Also note that $g_{min}$ may be greater than $\min_{k=1}^K w_k g_{min}^k$, because, for example, it might be the
case that $g_{min}=w_k g_{min}^k$ for some unique $k$ and no truth
assignment violating $F_k$ with cost $w_k g_{min}^k$ satisfies all
other $F_i$'s. 

  Property~\ref{prop:anddecompose} states that a penalty function for the
  conjunction of sub-formulas can be obtained as
a (weighted) sum of the penalty
  functions of the sub-formulas. 
The weights $w_k$ are needed because
 penalty functions of formulas that share variables sum up biases or
couplings, possibly resulting into
  out-of-range values \eqref{eq:parameterranges}. 
If the $w_k$'s are smaller than 1, then
the gap $g_{min}$ of the final penalty function may become smaller.
Also, Property~\ref{prop:anddecompose} requires placing variables
into qubits that
are shared among conjunct subformulas. This may restrict the chances
of finding suitable placements for the variables in the graph.

An alternative way of coping with this problem is to map shared variables into 
distinct qubits which are connected by  chains of equivalences.
Consider $\Fx = \bigwedge_{k=1}^K F_k(\xsk)$
as in Property~\ref{prop:decomposition}. 
For every variable $x_i$ and for every  $F_k$ where
$x_i$ occurs, we can replace the occurrences 
of $x_i$ in $F_k$ with a fresh variable ${{x_i}^k}^*$, obtaining a formula
$\bigwedge_{k=1}^K F_k(\xsk^*)$ such that the sets $\xsk^*$ are
all disjoint.
Let 
\begin{eqnarray}
\label{eq:decomposition2}
F^*(\xs^*)\defas\bigwedge_{k=1}^K F_k(\xsk^*)\wedge
\bigwedge_{\tuple{{{x_i}^k}^*,{{x_i}^{k'}}^*}\in Eq(x_i)}
( {{x_i}^k}^* \iff  {{x_i}^{k'}}^*)
\end{eqnarray}
where  $\xs^*=\cup_k\xsk^*$, 
and $ Eq(x_i)$ is any set of pairs 
\tuple{{{x_i}^k}^*,{{x_i}^{k'}}^*} of the variables replacing $x_i$ 
such that the conjunction of equivalences  in 
\eqref{eq:decomposition2} 
states that of all of them are equivalent.
%
By construction, 
$\Fx$ is
  satisfiable if and only if $F^*(\xs^*)$ is satisfiable, and 
 from every model
  $\xs^*$ for $F^*(\xs^*)$ we have a model \xs   for \Fx by simply assigning
  to each $x_i$ the value of the corresponding
  ${{x_i}^k}^*$s.

Now assume we have a penalty function $P_{F_k}(\xs^{k^*},\ask|\ts^k)$
for each $k$ with disjoint $\ask$.
We recall from  Example~\ref{es:exampleiff} that 
$( 1 - {{x_i}^k}^* {{x_i}^{k'}}^*)$
are 
penalty functions of gap 2 for the
$( {{x_i}^k}^* \iff {{x_i}^{k'}}^*)$ subformulas in
\eqref{eq:decomposition2}.
Thus we can apply Property~\ref{prop:decomposition} with all weights
$w_k=1$ and write a penalty function for $F^*(\xs^*)$ in the
following way:
\begin{eqnarray}
  P_{F^*}(\xs^*,\as|\ts) = \sum_{k=1}^K P_{F_k}({\xs^k}^*,\ask|\ts^k)  + 
\sum_{\tuple{{{x_i}^k}^*,{{x_i}^{k'}}^*}\in Eq(x_i)}
( 1 - {{x_i}^k}^*
 {{x_i}^{k'}}^*).
\label{eqn:distinct_qubits}
\end{eqnarray}
Note that the $\theta$'s stay within valid range because the 
$\xsk^*$s and $\ask$s are
all disjoint and the biases of the $( 1 - {{x_i}^k}^*
 {{x_i}^{k'}}^*)$ terms are zero, so distinct sub-penalty
 functions in \eqref{eqn:distinct_qubits} involve disjoint groups of
 biases and couplings. 
Thus we have the following.

\begin{property}
\label{prop:decompositionwithrenaming}
$P_{F^*}(\xs^*,\as|\ts)$ in \eqref{eqn:distinct_qubits} is a penalty
function for $F^*(\xs^*)$ in \eqref{eq:decomposition2}.
%
The gap of $P_{F^*}(\xs^*,\as|\ts)$ is $g_{min}\ge\min(\min_{k=1}^K g_{min}^k,2)$.
\end{property}
\noindent
%
Thus, we can represent a single variable $x_i$
with a series of qubits connected by strong couplings
$(1-x_i x_i')$. 
(For $x_i \iff \neg x_i'$, we use $(1+x_i x_i')$.) 
%
Notice that it is not necessary that every copy of variable $x_i$ be
connected to every other one; rather, to enforce the condition that
all copies of $x_i$ are logically equivalent, it suffices that the
copies of $x_i$ induce a connected graph. 
Moreover, additional copies of $x_i$ may be introduced on unused
vertices of the hardware graph $G$ to facilitate connectedness. A set
of qubits all representing the same variable in this way is called a
\emph{chain} and is the subject of
\sref{sec:problems_embedding}.
Thus, it is possible to implement $P_{F^*}(\xs^*,\as|\ts)$ in
\eqref{eqn:distinct_qubits} by placing 
the distinct penalty functions $P_{F_k}({\xs^k}^*,\ask|\ts^k)$
into sub-graphs and connect them with chains. 
 

%
Recall from \sref{sec:background_smt} that a formula \Fx which is not a conjunction can
always be decomposed 
into a conjunction of
smaller formulas \Fxy{} { by means of Tseitin's transformation
 \eqref{eq:decomposition}.
\ignore{
\begin{equation}
\label{eq:decomposition}
\Fxy \defas \bigwedge_{i=1}^{m-1} (y_i \iff \Fixy) \wedge  \Fmxy,
\end{equation}
where the $F_i$s are Boolean functions which decompose the original
formula \Fx, and the 
$y_i$s are fresh Boolean variables each labeling the corresponding
$F_i$.
%
(If the input formula is itself a conjunction, then Tseitin's
transformation can be applied recursively to each conjunct.)
Tseitin's
  transformation~\eqref{eq:decomposition} guarantees that $\Fx$ is
  satisfiable if and only if \Fxy{} is satisfiable, and that if $\xs,\ys$ is a model for \Fxy, then \xs is a model for \Fx.
}
By Properties~\ref{prop:decomposition} and \ref{prop:decompositionwithrenaming}, this allows us to AND-decompose \Fx into
multiple and smaller conjuncts that can be encoded separately and recombined. 
The problem thus reduces to choosing Boolean functions $(y_i\iff\Fixy)$ and
$\Fmxy$ whose penalty
functions are easy to compute, have large gap, and whose
combination keeps the gap of the penalty function for the original
function as large as possible.

\begin{example}
  Let   $\Fx\defas x_4 \iff (x_3 \wedge (x_1 \oplus x_2))$.
  Applying \eqref{eq:decomposition} and \eqref{eq:decomposition2}
  this can be rewritten as
  $\F^*(\xs,y,y') = 
(x_4 \iff (x_3\wedge y'))
\wedge 
(y \iff (x_1 \oplus x_2))
\wedge 
(y' \iff y)
$.  The penalty
functions of the three conjuncts can be produced as in
Examples \ref{es:exampleand}, \ref{es:examplexor} and
\ref{es:exampleiff} respectively, and summed as in
Property~\ref{prop:decompositionwithrenaming}:
\begin{eqnarray}
  \label{eq:example_combination}
\nonumber 
&&P_{F^*}(\xs,y,y',\as|\ts)\\
\nonumber
&=&
\frac{5}{2} - \frac{1}{2} x_3 - \frac{1}{2} y' + x_4  
+ \frac{1}{2} x_3 y' - x_3 x_4 - y'a_4 - x_4a_4 \\
\nonumber 
&+&
5 + y + a_2 - a_3 
+ x_1 a_1  - x_1 a_2 - x_1 a_3 - x_2 a_1 - x_2 a_2   - x_2 a_3 + y a_2 - y a_3
\\
\nonumber 
&+&
  1 - yy'  
\\
&=&
\nonumber 
\frac{17}{2}
- \frac{1}{2} x_3  + x_4 + y - \frac{1}{2} y' + a_2 - a_3 
+ x_1 a_1  - x_1 a_2 - x_1 a_3 - x_2 a_1 - x_2 a_2 
\\
\nonumber &&
  - x_2 a_3 - x_3
             x_4 + \frac{1}{2} x_3 y' - x_4a_4 + y a_2 - y a_3 
- yy'  - y'a_4 
\end{eqnarray}
Notice that there is no interaction between the biases and couplings
of the three components, only the offsets are summed up.
The resulting gap is $min\set{2,2,2}=2$.
Then they can be placed, e.g., as in
Figure~\ref{fig:mappings_two}. 
\end{example}

Overall, these facts suggest a  ``divide-and-conquer'' approach for
addressing the \sattoqubo{} problem: 
\begin{renumerate}
\item AND-decompose the input 
formula, by rewriting every conjunct \Fx  which 
is not small enough
into an equivalently-satisfiable one \Fxy as in
\eqref{eq:decomposition} such that penalty functions for all its
conjuncts can be easily computed; 
\item
rename shared variables and compute the global
penalty functions as in 
Property~\ref{prop:decompositionwithrenaming};
\item
place the sub-penalty
functions into subgraphs and connect  by chains equivalent
qubits representing shared variables. 
\end{renumerate}

\subsection{Exact Penalty Functions and MaxSAT}
\label{sec:maxsat}
In order to encode MaxSAT, we require
 a stronger version of the penalty function in Definition \ref{def:penalty}.

\begin{definition}
  A penalty function $\Pxa$ is {\bf exact} if for all $\xs$ such that $F(\xs)=\bot$,
  \[
  \min_{\set{\as}} \Pxa = g_{min}.
  \]
  \label{def:exact}
\end{definition}


\noindent
That is, an exact penalty function separates satisfying assignments
from all
non-satisfying ones by exactly the same gap 
$g_{min}$.

\begin{example}
  The penalty function of $\Fx\defas (x_1 \leftrightarrow
  x_2)$ in Example~\ref{es:exampleiff} is exact,
  whereas those of $\Fx\defas x_3 \leftrightarrow
  (x_1 \wedge x_2)$ and $\Fx\defas x_3 \leftrightarrow (x_1 \oplus
  x_2)$ in Examples~\ref{es:exampleand} and
  \ref{es:examplexor} are not exact. 
\end{example}

Exact penalty functions allow for the encoding of weighted MaxSAT problems, {with some restrictions}. The
following fact is a straightforward consequence of Property
\ref{prop:decomposition} and Definition \ref{def:exact}.

\begin{proposition}
  \label{prop:maxsat}
  Let $\Fx = \bigwedge_{k=1}^K F_k(\xsk)$ be a Boolean formula
  s.t. $\xs=\cup_k\xsk$, and 
$
    \Pxa \defas \sum_{k=1}^K P_{F_k}(\xsk,\ask|\ts^k),
$
where 
$\as\defas\cup_k\as^k$
s.t. the $\as^k$ are all disjoint, each $P_{F_k}(\xsk,\ask|\ts^k) $ is
an \underline{exact} penalty function for $F_k$ of gap $g_k$.
 Let \xs{},\as{} be a truth assignment which minimizes  \Pxa.
 Then \xs{} is a solution for the weighted MaxSAT problem
 $\set{\tuple{F_k,g_k}}_k$.
\end{proposition}

Proposition~\ref{prop:maxsat} allows for encoding a generic weighted
MaxSAT problem $\set{\tuple{F_k,c_k}}_k$ by setting
$\Pxa \defas \sum_{k=1}^K w_k P_{F_k}(\xsk,\ask|\ts^k)$ where
$w_k \defas \frac{c_k}{g_k}\cdot c$ and $c$ is a normalization
factor \eqref{eq:normalization-factor}.
Notice that in Proposition~\ref{prop:maxsat} the penalty functions
$P_{F_k}(\xsk,\ask|\ts^k)$ must be exact; otherwise, a solution \xs{},\as{} that is optimal for MaxSAT but violates
some $F_k$ might not minimize \Pxa if
$P_{F_k}(\xsk,\ask|\ts^k) > g_k$.  
%


In \sref{sec:problems_decomposition} we outlined a
``divide-and-conquer'' approach for \sattoqubo{} based on the idea of
mapping shared variables into distinct qubits which are then connected
by chains of equivalences.  Applying the same
approach to MaxSAT is not as
straightforward, because
Property~\ref{prop:decompositionwithrenaming} cannot always be
combined with Proposition~\ref{prop:maxsat} in a useful way.  Consider
the scenario in Property~\ref{prop:decompositionwithrenaming}, and
suppose we want to use \eqref{eqn:distinct_qubits} to solve the MaxSAT
problem $\set{\tuple{F_k,g_k}}_k$ as with
Proposition~\ref{prop:maxsat}.  As the following example shows, there may be
minimum-energy solutions of \eqref{eqn:distinct_qubits} which violate some
equivalence $({{x_i}^k}^* \iff {{x_i}^{k'}}^*)$ in
\eqref{eq:decomposition2} if this avoids violating one or more of the
$F_k$'s whose sum of gaps is greater than 2.  Such a solution is {\em
  not} a solution of the MaxSAT problem, because it corresponds to
assigning different truth values to distinct instances of the same
variable in the original problem.

\begin{example}
Consider the trivial MaxSAT problem $\set{\tuple{F_i(x),c}}_{i=1}^4$
for some penalty value $c>0$ where
$F_1(x)=F_2(x)\defas x$, and $F_3(x)=F_4(x)\defas \neg x$.
The two possible solutions $x=\top$ and $x=\bot$ are both optimum with
penalty $2c$ and 
falsify $F_3,F_4$ and $F_1,F_2$ 
respectively.
We have the following normal and exact penalty functions: $P_{F_1}(x)=P_{F_2}(x)=2-2x$ and 
$P_{F_3}(x)=P_{F_4}(x)=2+2x$, each of gap $g_i=4$.
Suppose we want to encode the problem in such a way to fit into a
linear chain of 4 qubits adopting the encoding in
Property~\ref{prop:decompositionwithrenaming}.
 We introduce 
 four copies of $x$, namely $x^1, x^2, x^3, x^4$, and obtain:
\begin{eqnarray}
  \label{eq:example-nomaxsat-formula}
  \nonumber
{F^*}(x^1,x^2,x^3,x^4) &=& 
x^1 \wedge x^2 \wedge \neg x^3 \wedge \neg x^4 \wedge 
(x^1\iff x^2) \wedge (x^2\iff x^3) \wedge (x^3\iff x^4) 
\\
  \label{eq:example-nomaxsat-penalty}
  \nonumber
P_{F^*}(x^1,x^2,x^3,x^4) &=& 
(2-2 x^1) + (2-2 x^2) +(2+2 x^3) + (2+2 x^4) + 
\\
\nonumber
&& (1-x^1x^2) + (1-x^2x^3) +  (1-x^3x^4) 
\\
\nonumber &=& 11 - 2 x^1 -2 x^2 +2 x^3 +2 x^4 -x^1x^2 -x^2x^3 -x^3x^4.
\end{eqnarray}
The minimum-energy solution to $P_{F^*}$ is $x^1=x^2=1$
and $x^3=x^4=-1$ with $P_{F^*}(...)=2$, which violates the equivalence $(x^2\iff
x^3)$. The correct MaxSAT solutions $x^1=x^2=x^3=x^4=1$ and
$x^1=x^2=x^3=x^4=-1$ both have $P_{F^*}(...)=8$.
\end{example}

In general, the problem arises when it is energetically cheaper to violate some  equivalence $({x_i^k}^*\iff
{x_i^{k'}}^*)$ in a chain in \eqref{eq:decomposition2} than to violate all the penalty functions $\{F_k(\xsk): x_i \in \xsk\}$ on one side of the equivalence.
One solution to this problem is to multiply the $P_{F_k}$'s by sufficiently small 
weights $w_k<1$, at the cost reducing their gaps $g_k$.
In the following we discuss the bounds that can be placed on $w_k$.

Let $\cali$ denote the indices of the functions $F_k(\xsk)$ that use the variable $x_i$; that is, $\mathcal{I} = \{k: x_i \in \xsk\}$. An equivalence $({x_i^k}^*\iff {x_i^{k'}}^*)$ in the chain of $x_i$ splits the chain into two subchains, and splits $\mathcal{I}$ into two subsets $\cali_k$ and $\cali_{k'}$ such that $({x_i^k}^*\iff {x_i^{k'}}^*)$ connects the functions of $\cali_k$ to the functions of $\cali_{k'}$. Assume we have a desired gap $g_{desired}>0$ separating solutions with broken chains from true solutions. Then a sufficiently large gap for the equivalence $({x_i^k}^*\iff {x_i^{k'}}^*)$ is
\[
g_{(k,k')} = \min\left(\sum_{j \in \cali_k}  g_j, \sum_{j \in \cali_{k'}} g_j\right) + g_{desired},
\]
as this gap ensures that it is $g_{desired}$ cheaper to violate all the constraints in $\cali_k$ or $\cali_{k'}$ then to violate $({x_i^k}^*\iff {x_i^{k'}}^*)$. Recall from \eqref{eq:decomposition2} that $Eq(x_i)$ is the set of variable pairs $({x_i^k}^*,{x_i^{k'}}^*)$ that form equivalences $({x_i^k}^*\iff {x_i^{k'}}^*)$ in the chain of $x_i$. To ensure that all equivalence constraints are not violated, a sufficient gap for the entire chain is
\begin{equation}
  \label{eq:maxsat-weight-bound}
g_{chain} = \max_{({x_i^k}^*,{x_i^{k'}}^*) \in Eq(x_i)} g_{(k,k')}.
\end{equation}

Finally, recalling that each equivalence has gap 2, we update the weight
definition in Proposition~\ref{prop:maxsat} for each $k \in \cali$:\footnote{Note that the normalization factor $c$ here is $1$
  as chains are normal.}
\begin{equation}
w_k  = \frac{2 \cdot c_k}{g_k \cdot g_{chain}}
\end{equation}



An alternative bound on $g_{chain}$ is given in \cite{Choi08minor-embeddingin}. In the paper, the author bounds the chain strength required to ensure that all minima of an
embedded QUBO problem can be mapped to a minimum of the original QUBO problem (see \sref{sec:problems_embedding} below).
Let $\theta_i^* = \sum_k w_k\theta_i$ be the bias value obtained by sharing the $x_i$ variable as in Property~\ref{prop:anddecompose}\footnote{For simplicity, we assume to share a single $x_i$, so each $\theta_{ij}^* = w_k\theta_{ij}^k$ for some unique $k$.}. If $x_i$ is substituted by a chain with $l_i$ endpoints, QUBO minima are preserved if the chain gap is the following:

\begin{equation}
  \label{eq:maxsat-weight-bound-choi}
  g_{chain} = 2  \frac{l_i-1}{l_i}\left( \sum_{(i,j) \in E} |\theta_{ij}^*|- |\theta_i^*| \right) + g_{desired}
\end{equation} 

This alternative bound is sometimes lower than \eqref{eq:maxsat-weight-bound}, especially when $|\theta_i^*|$ is high. 
Note that, as the original paper explains, if the bound value is negative then $P_{F^*}$ is monotonic on $x_i$. If that is the case, then $x_i=-sgn(\theta_i^*)$ always minimizes $P_{F^*}$, so we can fix the value of $x_i$ and there is no need for a chain.


In general, neither \eqref{eq:maxsat-weight-bound} nor \eqref{eq:maxsat-weight-bound-choi} are typically very tight bounds on required chain gap, and finding the smallest viable chain gap analytically appears to be a difficult problem. In practice $g_{chain}$ is often determined empirically; this is discussed further in \sref{sec:expaval}.

\smallskip
Overall, the \maxsattoqubo{} problem 
is subject to some intrinsic limitations.
Firstly, it requires the usage of {\em exact} penalty functions for its
sub-formulas, which are more difficult to obtain. 
Secondly, the need to re-weight penalty functions to ensure chain equivalences are not violated typically results in smaller gaps.
Thirdly, it is difficult to directly encode hard
constraints in a MaxSAT problem; this again requires re-weighting soft constraints by very small factors, reducing their gaps accordingly.



\subsection{Embedding}
\label{sec:problems_embedding}
The process of representing a single variable $x_i$ by a collection of qubits connected in chains of strong couplings is known as \emph{embedding}, in reference to the minor embedding problem of graph theory \cite{Choi08minor-embeddingin,choi2011minor}. More precisely, let $P_F(\xs|\ts)$ be a penalty function whose interactions define a graph $G_F$ (so $x_i$ and $x_j$ are adjacent iff $\theta_{ij} \neq 0$) and let $G_H$ be a QA hardware graph. A \emph{minor embedding} of $G_F$ in $G_H$ is a function $\Phi : V_{G_F} \rightarrow 2^{V_{G_H}}$ such that:
\begin{itemize}
\item for each $G_F$-vertex $x_i$, the subgraph induced by $\Phi(x_i)$ is connected;
\item for all distinct $G_F$-vertices $x_i$ and $x_j$, $\Phi(x_i)$ and $\Phi(x_j)$ are
  disjoint;
\item for each edge $(x_i,x_j)$ in $G_F$, there is at least one edge between $\Phi(x_i)$
  and $\Phi(x_j)$.
\end{itemize}
The image $\Phi(x_i)$ of a $G_F$-vertex is a chain, and the set of qubits in a chain are constrained to be equivalent using $(1-{{x_i}^k}^*{{x_i}^{k'}}^*)$ couplings as in Equation \eqref{eqn:distinct_qubits}.

Embedding generic graphs is a computationally  difficult problem \cite{adler11embed}, although certain structured problem graphs may be easily embedded in the Chimera graph \cite{boothby2016fast,zaribafiyan2016sys} and heuristic algorithms may also be used \cite{cai2014practical}.  A reasonable goal in embedding is to minimize the sizes of the chains, as quantum annealing
becomes less effective as more qubits are included in chains \cite{lanting10tunneling}. 


A different approach to finding models for $\Fx$, \emph{global embedding}, is based on first finding a penalty function 
on a complete graph $G_F$ on $n+h$ variables, and secondly, embedding $G_F$ into a hardware graph $G_H$ using chains (e.g., using \cite{boothby2016fast}).
Following \cite{bian2014discrete}, global embeddings usually need fewer qubits than the methods presented in this paper;
however, the final gap of the penalty function obtained in this way is generally smaller and difficult to compute exactly.


\section{Encoding Small Boolean Sub-Formulas}
\label{sec:small_boolean}
In this section we present general SMT/OMT-based techniques to address
the encoding problem for small Boolean formulas \Fx.  

\subsection{Computing Penalty Functions via SMT/OMT(\larat).}
\label{sec:computing_offline}
Given $\xs\defas\set{x_1,...,x_n}$, 
$\as\defas\set{a_1,...,a_h}$,
$\Fx$ as in Section \ref{sec:foundations},
a variable placement in a Chimera subgraph 
s.t. $\zs=\xs\cup\as$, and some gap $g_{min}>0$,
the problem of finding a penalty 
function \Pxa as in \eqref{eq:penfunction} 
corresponds to solving the following problem:\footnote{As in \eqref{eq:penfunction}, we implicitly assume 
$\theta_{ij}=0$ when $(i,j)\not\in E$.}
\begin{eqnarray}
\nonumber
&&   
    \text{For every } i\ j,\ \text{ find }\ \theta_i\in[-2,2],\
                \theta_{ij}\in[-1,1]\ \text{ such that }\\
\label{eq:encoding1}
&&
\ \forall \xs. 
    \left [ 
    \begin{array}{ll}
      {(\pos\Fx \imp \exists \as. (\Pxa=0 ))\  \wedge}\\
      (\pos\Fx \imp \forall \as. (\Pxa\ge 0 ))\  \wedge\\
      (\neg\Fx \imp \forall \as. (\Pxa\ge g_{min} ))\  
    \end{array}
    \right ] .
  \end{eqnarray}
\noindent
%
By applying Shannon's expansion~(\sref{sec:background_smt})%
\ignore{\footnote{We recall that Shannon's expansion eliminates Boolean
  quantifiers by recursively applying the following rewriting rules:
$\forall x. \Psi \thus (\Psi_{x=\top} \wedge \Psi_{x=\bot})$ and
$\exists x. \Psi \thus (\Psi_{x=\top} \vee \Psi_{x=\bot})$.
}}
to the quantifiers in
  \eqref{eq:encoding1}, the problem reduces
  straightforwardly to solving 
the following \smtlarat problem:
\begin{eqnarray}
\label{eq:encoding2-unrolled}
\label{eq:encoding2-range}
\Phi(\ts) &\defas& \rangebfull \wedge \rangecfull
%
\\
\label{eq:encoding2-unrolled_in}
&\wedge&\bigwedge_{\set{\xs\in\set{-1,1}^n|F(\xs)=\top}} \ \ 
\bigvee_{\as\in\set{-1,1}^h} (\Pxa=0 )\ 
\\
\label{eq:encoding2-unrolled_inout}
&\wedge&\bigwedge_{\set{\xs\in\set{-1,1}^n|F(\xs)=\top}} \ \ 
\bigwedge_{\as\in\set{-1,1}^h} (\Pxa \ge 0 ) 
\\
\label{eq:encoding2-unrolled_out}
&\wedge&\bigwedge_{\set{\xs\in\set{-1,1}^n|F(\xs)=\bot}} \ \
\bigwedge_{\as\in\set{-1,1}^h} (\Pxa\ge g_{min} ). \label{eq:encoding2-unrolled-end}
\end{eqnarray}
Consequently, the problem of finding the penalty function \Pxa that
maximizes the 
gap $g_{min}$ reduces to solving the \omlarat{} maximization problem 
\tuple{\Phi(\ts),g_{min}}. Notice that, since $g_{min}$ is maximum, \Pxa is also normal.

Intuitively,
\eqref{eq:encoding2-range} states the ranges of the \ts;
\eqref{eq:encoding2-unrolled_in} and
\eqref{eq:encoding2-unrolled_inout} 
state that, for every 
 \xs satisfying \Fx, 
\Pxa must be zero for at least one ``minimum'' \as and 
nonnegative for all the others;
\eqref{eq:encoding2-unrolled_out} states that 
for every \xs not satisfying \Fx, 
\Pxa must be greater than or equal to the gap.
Consequently, if the values of the \ts in \Pxa satisfy $\Phi(\ts)$, 
then \Pxa complies with \eqref{eq:pencriteria};
{if $\Phi(\ts)$ is unsatisfiable, then there is no \Pxa complying 
with \eqref{eq:pencriteria} for the given placement.}

Note that, if $\as=\emptyset$, then the \omlarat{} maximization problem 
\tuple{\Phi(\ts),g_{min}} reduces to a linear program because the disjunctions in 
\eqref{eq:encoding2-unrolled_in} disappear.

To force \Pxa to be an \emph{exact} penalty function,
we add the following conjunct inside the square brackets of \eqref{eq:encoding1}:
\begin{eqnarray}
  \label{eq:encoding1_exact}
&&  (\neg\Fx \imp \exists \as. (\Pxa=
   g_{min} )),
\end{eqnarray}
which forces \Pxa to be 
exactly equal to the gap for at least one \as. 
Thus we conjoin the Shannon's expansion of \eqref{eq:encoding1_exact} to $\Phi(\ts)$ in \eqref{eq:encoding2-range}-\eqref{eq:encoding2-unrolled-end}:
\begin{eqnarray}
  \label{eq:encoding2-unrolled_out_exact}
... &\wedge&\bigwedge_{\set{\xs\in\set{-1,1}^n|F(\xs)=\bot}} \ \
\bigvee_{\as\in\set{-1,1}^h} (\Pxa= g_{min} ).
\end{eqnarray}


\subsection{Improving Efficiency and Scalability using Variable
  Elimination}
\label{sec:computing_offline_variableelimination}
In the SMT/OMT(\larat) formulation \eqref{eq:encoding2-unrolled}-\eqref{eq:encoding2-unrolled-end}, $\Phi(\ts)$ grows exponentially with the number of hidden variables $h$.
For practical purposes, this typically implies a limit on $h$ of about 10. 
Here, we describe an alternative formulation whose size dependence on
$h$ is $O(h2^{\textsf{\bf tw}})$, where  
{\sf{\bf tw}}  
is the treewidth of the subgraph of $G$ spanned by
the qubits corresponding to the ancillary variables, $G_a$. For the
Chimera graph, even when $h$ is as large as 32, 
{\sf{\bf tw}} 
is at most 8 and therefore still of tractable size.

The crux of the reformulation is based on the use of the variable elimination technique \cite{Dechter1998} to solve an Ising problem on $G_a$.
This method is a form of dynamic programming, storing tables in memory
describing all possible outcomes to the problem. When the treewidth
is 
{\sf{\bf tw}}, 
there is a variable elimination order guaranteeing that each table contains at most  $O(2^{\textsf{\bf tw}})$ entries. 
Rather than using numerical tables, our formulation replaces each of its entries with a continuous variable constrained by linear inequalities.
In principle, we need to parametrically solve an Ising problem for 
each $\xs \in \{-1,1\}^n$, generating $O(2^n h2^{\textsf{\bf tw}})$ continuous variables. However, by the local nature of the variable elimination process, many of these continuous variables are equal, leading to a reduced (as much as an order of magnitude smaller) and strengthened SMT formulation. \ignoreinlong{See \cite{bian2014discrete} for more details.}
\newcommand{\ip}[2]{\langle{#1},{#2}\rangle}
\newcommand{\vc}[1]{\ensuremath{\underline{\boldsymbol{#1}}}\xspace}
\newcommand{\bs}{\ensuremath{\underline{\mathbf{b}}}\xspace}

To describe the method, we first reformulate equations \eqref{eq:encoding2-unrolled_inout}-\eqref{eq:encoding2-unrolled-end} by introducing {\em witness} binary
variables $\vc{\beta}(\xs)\in\set{-1,1}^h$ to enforce the equality constraints \eqref{eq:encoding2-unrolled_in}, that is, 
$P_F(\xs,\vc{\beta}(\xs) | \theta)=0$.
Thus, we can rewrite $\Phi(\ts)$ as the SMT problem
$\Phi(\ts,\vc{\beta})$ defined by
\begin{eqnarray*}
\Phi(\ts,\vc{\beta}) &\defas& \eqref{eq:encoding2-range} \wedge \eqref{eq:encoding2-unrolled_inout} 
\wedge \eqref{eq:encoding2-unrolled_out}
%
\\
&\wedge&\bigwedge_{\set{\xs\in\set{-1,1}^n|F(\xs)=\top}} \ \ 
\bigvee_{\as\in\set{-1,1}^h} \big(\ (\vc{\beta}(\xs) \equiv \as) \wedge (\Pxa=0 ) \ \big). 
\footnotemark
\end{eqnarray*}

\footnotetext{For vectors $\as,\bs$, we use $\as\equiv \bs$ as a shorthand for $(a_1=b_1) \wedge (a_2 = b_2) \wedge (a_3=b_3) \wedge \ldots$.  }

Consider first the case when the graph $G_a$ has no edges. If, for $i=1,\ldots,h$, we define
\begin{equation*}
  f_i(a_i|\xs) = \theta_i a_i + a_i \sum_{j:ij \in E }\theta_{ij} \ 
  x_j, 
\end{equation*}
then we can write 
\begin{equation*}
\Pxa = c(\xs) + \sum_{i=1}^h f_i(a_i|\xs),
\end{equation*}
where $c(\xs)$ does not depend on the ancillary variables. Thus, 
\begin{equation}
\label{eq:f_i-min}
 \min_{\as} \Pxa = c(\xs) + \sum_{i=1}^h \min_{a_i\in\{-1,1\}} f_i(a_i|\xs).
\end{equation}
If $\ts$ is fixed, solving \eqref{eq:f_i-min} is straightforward.
However, since $\ts$ is a variable, the contribution $\min_{a_i\in\{-1,1\}} f_i(a_i|\xs)$ is a function of $\ts$, for each $i=1,\ldots,h$.
Each of these minimums will be associated with a continuous variable, denoted by $m_i(\emptyset|\xs)$, and referred to as a {\em message} variable (the naming will be clearer in the general case). 
To relate $m_i(\emptyset|\xs)$ with $\min_{a_i\in\{-1,1\}} f_i(a_i|\xs)$, we impose the constraints
\begin{equation*}
  m_i(\emptyset|\xs) \le f_i(-1|\xs) \quad \text{and} \quad  m_i(\emptyset|\xs) \le f_i(1|\xs)  .
\end{equation*}
Thus, if $F(\xs)=\bot$, since the message variables are lower bounds on the true minimums of \eqref{eq:f_i-min},
to enforce \eqref{eq:encoding2-unrolled-end} we need simply add the constraints
\begin{equation*}
  c(\xs) + \sum_{i=1}^h m_i(\emptyset|\xs) \ge g_{min}.
\end{equation*} 
When $F(\xs)=\top$, we need to ensure that the message variables take the minimums of   \eqref{eq:f_i-min}.
Note that variable $\beta_i(\xs)$  identifies the value of the ancillary variable $i$
that achieves the minimum in  \eqref{eq:f_i-min}. 
To relate the values of $\vc{\beta}(\xs)$ and the message variables $m(\emptyset|\xs)$
we add the SMT constraints
\begin{equation*}
  \beta_i(\xs) \Rightarrow \bigl(m_i(\emptyset|\xs) =  f_i(1|\xs) \bigr), 
\end{equation*}
\begin{equation*}
  \neg \beta_i(\xs) \Rightarrow \bigl(m_i(\emptyset|\xs) =  f_i(-1|\xs) \bigr). 
\end{equation*}
Finally, to impose \eqref{eq:encoding2-unrolled_in} and \eqref{eq:encoding2-unrolled_inout}, we need that
\begin{equation*}
  c(\xs) + \sum_{i=1}^h m_i(\emptyset|\xs) = 0.
\end{equation*} 
Since $G$ is usually sparse, it is likely that two binary states $\xs$ and $\xs'$ agree on the bits adjacent to a fixed ancillary variable $i$. In this case, it is clear that  $m_i(\emptyset|\xs) = m_i(\emptyset|\xs')$, and we can use a single message variable for both states. This observation
can be extended to the general case and will be valuable to reduce the size and strengthen the SMT problem formulation.


Next consider the general case when $|E(G_a)|>0$. In what follows, $c(\xs)$ and $f_i(a_i|\xs)$ are defined as above.
Assume first $\ts$ is fixed.
Given $\xs$, we want to solve the Ising model $\min_{\as}\Pxa$.
Variable elimination proceeds in order, eliminating one ancillary variable at a time. 
Suppose that ancillary variables are eliminated in the order $h,h~-~1,\ldots,1$.
Each ancillary variable $i$ is associated with a set $\mathcal{F}_i$ of \emph{factors}, which are functions that depend on ancillary variable $i$ and none or more ancillary variables with index less than $i$. The sets $\mathcal{F}_i$ are called \emph{buckets}, and are updated throughout the computation. Initially, each $\mathcal{F}_i$ consists of ancilla-ancilla edges\footnote{${G}_{a}$ is an undirected graph. An edge is defined by a pair of vertices, say $i$ and $k$; for convenience, in this section we associate this edge with the ordered pair $ik$ with $k<i$.
}
$f_{i,k}(a_i,a_k) = \theta_{ik}\ a_i a_k$ for $ik \in E({G}_{a})$, $k<i$.
Let $\mathcal{V}_i$ denote the set of ancillary variables involved in the factors of bucket $\mathcal{F}_i$ other than variable $i$ itself (thus, all variable indices in $\mathcal{V}_i$ are less than $i$, or $\mathcal{V}_i=\emptyset$). For a fixed $\as$ and a subset of ancillary variables $\mathcal{U}$, we use $\as_{\mathcal{U}}$ to denote $\{a_i:i\in \mathcal{U}\}$.
Variable $h$ is eliminated first. 
Note that once variables in $\mathcal{V}_h$ are instantiated to $\as_{\mathcal{V}_h}$, the optimal setting of variable $h$ is readily available by solving  
\begin{equation}
\label{eq:VE1}
   g_h(\as_{\mathcal{V}_h})=\min_{a_h} f_h(a_h|\xs) + \sum_{f\in \mathcal{F}_h} f(\as_{\mathcal{V}_h},a_h). 
\end{equation}  
Here $f=f_{i,h}\in \mathcal{F}_h$ represents an edge $ih$ between ancillary variables $i$ and $h$, $i<h$ (abusing notation  we write $f(a_i,a_h)$ as $f(\as_{\mathcal{V}_h},a_h)$), and $\mathcal{F}_h$ contains all edges adjacent to $h$.
The $2^{|\mathcal{V}_h|}$  possible settings of $\as_{\mathcal{V}_h}$ define $2^{|\mathcal{V}_h|}$ values \eqref{eq:VE1}.
These values define new factor $g_h$, a function of variables $\as_{\mathcal{V}_h}$,
that is added to the bucket $\mathcal{F}_i$ of variable $i$ with 
largest index in  $\mathcal{V}_h$. 
For each instantiation of $\as_{\mathcal{V}_h}$ we define the message 
$m_h(\as_{\mathcal{V}_h} | \xs)$ as $g_h(\as_{\mathcal{V}_h})$.
Iteratively, eliminating variable $i$ is accomplished by solving, for each  setting of $\as_{\mathcal{V}_i}$,
\begin{equation} 
\label{eq:VE2}
 g_i(\as_{\mathcal{V}_i})=\min_{a_i} f_i(a_i|\xs) + \sum_{f\in \mathcal{F}_i} f(\as_{\mathcal{V}_i},a_i) 
\end{equation}
generating a new factor $g_i$, a function of $\as_{\mathcal{V}_i}$.
For each one of the $2^{|\mathcal{V}_i|}$ possible values of $g_i$ we define message 
$m_i(\as_{\mathcal{V}_i} | \xs)$ to be $g_i(\as_{\mathcal{V}_i})$.
Factor $g_i$ is then added to bucket $\mathcal{F}_k$ where $k$ is the largest index in $\mathcal{V}_i$.
When $V_i=\emptyset$, \eqref{eq:VE2} takes the form
\begin{equation} 
 \min_{a_i} f_i(a_i|\xs) + \sum_{f\in \mathcal{F}_i} f({a_i}) 
\end{equation}
that determines the optimal value of $a_i$; the message corresponding to the value of
this minimum is $m_i(\emptyset | \xs)$.
All variables with $V_i=\emptyset$ can be eliminated at the same time, so that, at termination, the value of the Ising problem  $\min_{\as}\Pxa$ is equal to
\begin{equation*}
    c(\xs) + \sum_{i: \mathcal{V}_i=\emptyset} m_i(\emptyset | \xs).
\end{equation*}  
which will be equal to $\min_{\as}\Pxa$. 
Notice that the number of additional messages is $O(\sum_i  2^{|\mathcal{V}_i|})$, where each $\mathcal{V}_i$ corresponds to the time when variable $i$ is eliminated. When  ${G}_{a}$ has treewidth $t$, there is an elimination order for which each $|\mathcal{V}_i|\le t$, which typically, by our low treewidth assumption, will be much smaller than $2^h$.

When $\ts$ is not fixed, as in the case when there were no edges, the messages are variables. Since these message variable represent minimums, 
we upper bound the message variables adding the constraints
\begin{eqnarray*}
 m_i(\as_{\mathcal{V}_i}|\xs)&\le& f_i(-1|\xs) + \sum_{f\in \mathcal{F}_i} f(\as_{\mathcal{V}_i},-1)\\
 m_i(\as_{\mathcal{V}_i}|\xs)&\le& f_i(1|\xs) + \sum_{f\in \mathcal{F}_i} f(\as_{\mathcal{V}_i},1).
\end{eqnarray*}
As before, if $F(\xs)=\bot$, the constraint \eqref{eq:encoding2-unrolled_out} can be  replaced with
\begin{equation} \label{eq:LB}
  c(\xs) + \sum_{i:\mathcal{V}_i=\emptyset} m_i(\emptyset|\xs) \ge g_{min},
\end{equation} 
since the message variables provide a lower bound on  \eqref{eq:VE2}.
When $F(\xs)=\top$, we must ensure that all the message variables are tight. 
For a subset of ancillary variables $\mathcal{U}$, let $\vc{\beta}_\mathcal{U}(\xs) = \{\beta_i(\xs): i \in \mathcal{U} \}$. 
Thus, we must have that for all $\as_{\mathcal{V}_i}$
\begin{eqnarray*}
\bigl[\vc{\beta}_{\mathcal{V}_i}(\xs) \equiv \as_{\mathcal{V}_i} \wedge \beta_i(\xs)  \bigr] &\Rightarrow&
\bigl[ m_i(\as_{\mathcal{V}_i}|\xs) = 
f_i(1|\xs) + \sum_{f\in \mathcal{F}_i} f(\as_{\mathcal{V}_i},1) 
 \bigr] \\
  \bigl[\vc{\beta}_{\mathcal{V}_i}(\xs) \equiv \as_{\mathcal{V}_i} \wedge \neg \beta_i(\xs) \bigl] &\Rightarrow& \bigl[ m_i(\as_{\mathcal{V}_i}|\xs)= 
f_i(-1|\xs) + \sum_{f\in \mathcal{F}_i} f(\as_{\mathcal{V}_i}-1) 
\bigr] . 
\end{eqnarray*}
In this way, we can enforce that $\min_{\as}\Pxa=0$ (that is, constraints \eqref{eq:encoding2-unrolled_in} 
 and \eqref{eq:encoding2-unrolled_inout}), with the constraint 
\begin{equation}
  c(\xs) + \sum_{i:\mathcal{V}_i=\emptyset} m_i(\emptyset|\xs) = 0. \label{feasSoln}
\end{equation} 

As noted in the case when $G_a$ has no edges, some message variables will always have the same values. In fact, 
significant additional model reduction can be accomplished by identifying message variables that have to be the same across many states $\xs$. 
For instance, $m_i(\as_{\mathcal{V}_i}|\xs) = m_i(\as_{\mathcal{V}_i}|\xs')$ if their corresponding upper bounds are the same (propagating from $h$ down to $i$).
Because $G$ is sparse, the number of message variables can typically be reduced by an order of magnitude or more in this way.

In many cases, for counter-models $\xs$, $F(\xs)=\bot$, some constraints \eqref{eq:encoding2-unrolled_out} may be dropped or relaxed
without altering the optimal solution of the original SMT problem. For instance, we could include only  constraints \eqref{eq:encoding2-unrolled_out} for counter-models $\xs$ that are within Hamming distance at most $d$ from all models of $F$.
In our experiments, using $d\le 3$ sufficed in most cases. 

Alternatively, also  for counter-models, the variable elimination lower bounds \eqref{eq:LB} can be relaxed by weaker lower bounds such as a linear programming relaxation of the corresponding Ising problem, that requires $O(|V|+|E|)$ continuous variables and inequalities per $\xs$, $F(\xs)=\bot$. 
For instance, a linear programming lower bound on the QUBO formulation
\begin{equation*} 
 \min_{y_i\in\{0,1\}}\sum_{i\in V} c_i y_i + \sum_{e=\{i,j\}\in E} q_e \ y_i y_j \ ,
\end{equation*}
is the following:
\begin{align}
  &\text{Minimize\quad}
        \sum_{i\in V} c_i x_i + \sum_{e\in E} q_e \ z_e\\
\intertext{subject to}
     z_e - y_i -y_j  &\geq -1 & &\text{for each $e=ij \in E$, $i<j$} && (\lambda_e)\\
 -z_e + y_i  &\geq 0 & &\text{for each $e=ij\in E$,$i<j$}&& (\lambda_{e,i}^h)\\
-z_e + y_j  &\geq 0 & &\text{for each $e=ij\in E$,$i<j$}&& (\lambda_{e,j}^t)\\
-y_i  &\geq -1 & &\text{for each $i\in V$} &&(\alpha_{i})\\
y_i,z_{e} &\geq 0 & && 
\end{align}
Its linear programming dual is given by
\begin{align}
  \text{Maximize\quad}
        - \sum_{e\in E} \lambda_e - \sum_{i\in V} \alpha_i\\
\intertext{subject to}
     \lambda_e - \lambda_{e,i}^h - \lambda_{e,j}^t  &\leq q_{e} & \text{for each $e=ij\in E$,$i<j$} \label{eqn:D1}\\
-\sum_{e:i \in e} \lambda_e + \sum_{e=ik\in E,i<k} \lambda_{e,i}^h + \sum_{e=ki\in E,k<i} \lambda_{e,i}^t - \alpha_i   &\leq c_i & \text{for each $i\in V$} \label{eqn:D2}\\
\lambda_e, \lambda_{e,i}^h,\lambda_{e,i}^t, \alpha_i &\geq 0 & \label{eqn:D3}
\end{align}
Notice that if $c$ and $q$ are variables, the dual problem is still linear in the dual variables, $c$ and $q$. 
Thus, we can guarantee (in one direction only) that the value of the QUBO is at least $g$ with the set of linear inequalities
\begin{align}
- \sum_{e\in E} \lambda_e - \sum_{i\in V} \alpha_i &\geq g &\\
\eqref{eqn:D1}, \eqref{eqn:D2}, \eqref{eqn:D3}
\end{align}
Note that we can always take
\[
(-\sum_{e:i \in e} \lambda_e +  \sum_{e=ik\in E,i<k} \lambda_{e,i}^h + \sum_{e=ki\in E,k<i} \lambda_{e,i}^t- c_i )^{+} = \alpha_i .
\]
To make this work for an Ising problem, the $c$ and $q$ have to be written as linear functions of $\ts$, which is straightforward.

%
%



\subsection{Inequivalent Variable Placements and Exploiting Symmetries}
\label{sec:inequivalent_variable_placement}
Recall that a variable placement is a mapping from the input and ancilla variables $\xs\cup\as$ onto the vertices $V$; the formula $\Phi(\ts)$ in
\eqref{eq:encoding2-unrolled}-\eqref{eq:encoding2-unrolled_out_exact} 
can be built only after each $z_i\in\xs\cup\as$ has been placed. In general there will be many such placements, but by exploiting symmetry and the automorphism group of $G$, we can reduce the number of placements that need be considered.

Let $\qs \defas (\qz{1},...,\qz{{n+h}})$ denote a variable placement, so $\qz{i}$ is the vertex of $V$ onto which $z_i$ is placed. Two variable placements $\qs$ and $\qs' \defas (\qz{1}',...,\qz{{n+h}}')$ are \emph{equivalent} if there is a graph isomorphism $\phi$ of $G$ that point-wise maps the input variables ($x_i$) in $\qs$ to the input variables in $\qs'$; that is, $\qz{i} = \phi(\qz{i}')$ for all $i \leq n$. If $\qs$ and $\qs'$ are equivalent, then a penalty function for $\qs$ can be transformed into a penalty function for $\qs'$ by applying $\phi$. Therefore, in order to find a penalty function of maximal gap among all variable placements, it suffices to consider only inequivalent ones. 

\begin{example}
\label{ex:vp_chimera}
Suppose we want to encode a penalty function with $n+h = 8$ variables into an 8-qubit Chimera tile. There are $8!=40320$ candidate variable placements.
However, the tile structure is highly symmetric: any permutation of $\qs$ that either 
\begin{renumerate}
\item swaps horizontal qubits with vertical qubits, or 
\item maps horizontal qubits to horizontal qubits and vertical qubits to vertical qubits
\end{renumerate}
is an automorphism.
This fact can be exploited to reduce number of candidate placements to only
$\binom{7}{3}=35$ as follows. 
Let $1,...,4$ and $5,...,8$ be the indexes of the horizontal and
vertical qubits respectively. 
By (i), we assume w.l.o.g. that
$z_1$ is mapped into an horizontal qubit, and by (ii) we assume
w.l.o.g. that $\qz{1}=1$. 
Next, consider some size-3 subset $S$ of \set{\qz{2},...,\qz{8}}.
By (ii), all placements that map $S$ into the
remaining 3 horizontal qubits and map $\set{\qz{2},...,\qz{8}}\backslash S$ into the vertical qubits are equivalent. Since there are $\binom{7}{3}=35$ such subsets $S$, there are at most 35 inequivalent placements to consider. 
\end{example}

This notion of equivalence of variable placements can be coarsened slightly by taking advantage of NPN-equivalence. We define variables $x_1$ and $x_2$ in a Boolean function $F$ to be \emph{NPN-symmetric} if swapping the variables, and negating some subset of variables, produces an equivalent formula. For example, consider $F(x_1,x_2,x_3) \stackrel{\mathrm{def}}{=} x_3 \leftrightarrow (x_1 \wedge \neg x_2)$. Variables $x_1$ and $x_2$ are NPN-symmetric because $F(x_1,x_2,x_3) \leftrightarrow F(\neg x_2,\neg x_1,x_3)$. This symmetry defines an equivalence relation on the variables: for $x_i$ and $x_j$ the same equivalence class, there is a permutation and negation of the variables that does not change $F$ but maps $x_i$ to $x_j$ while not permuting variables outside the equivalence class.

We say that two variable placements $\qs$ and $\qs'$ are \emph{equivalent up to NPN-symmetry} if there is a graph isomorphism $\phi$ of $G$ that maps the input variables in $\qs$ to the input variables in $\qs'$ up to NPN-symmetry classes.  That is, for all $i \leq n$, there exists a $j \leq n$ such that $x_i$ and $x_j$ are NPN-symmetric and $\qz{i} = \phi(\qz{j}')$. Again, penalty functions for $\qs$ and can be transformed into penalty functions for $\qs'$ and vice versa.

\begin{example}
Consider placing the function $\mathrm{AND}(x_1,\ldots,x_4) = x_1 \wedge x_2 \wedge x_3 \wedge x_4$ with $h=4$ auxiliary variables on the $8$-qubit Chimera tile. From Example \ref{ex:vp_chimera}, it suffices to consider $35$ variable placements. However, the variables $x_1,\ldots,x_4$ in $\mathrm{AND}$ are all NPN-symmetric. Therefore any two variable placements $\qs$ and $\qs'$ that map the same number $x_i$'s to horizontal qubits are equivalent, since there is an automorphism that will map the horizontal $x_i$'s in $\qs$ to the horizontal $x_i$'s in $\qs'$. Moreover, a placement mapping $k \leq 4$ of the $x_i$'s to horizontal qubits is equivalent to one mapping $4-k$ of the $x_i$'s to horizontal qubits, by swapping horizontal and vertical qubits. As a result, there are only $3$ inequivalent variable placements to consider, in which $0$, $1$ or $2$ of the $x_i$'s are mapped to horizontal qubits. 
\end{example}

One way to to check for equivalent variable placements is to use vertex-coloured graph isomorphisms. Two vertex-coloured graphs $(G,c)$ and $(G',c')$ are \emph{vertex-coloured graph-isomorphic} if there is a permutation $\phi$ mapping $V(G)$ to $V(G')$ that preserves edges and maps every vertex of $G$ to a vertex of the same colour in $G'$ (for all $v \in V$, $c'(\phi(v)) = c(v)$). Using a variable placement $\qs$ and NPN-symmetry, define a vertex-coloring $c$ of $G$ as follows:
\[
c(g) = \begin{cases}
s & \text{ if $\qz{i} = g$ and $x_i$ is in the $s$-th equivalence class of NPN-symmetry},\\
0 & \text{ if $g$ is not in $\{\qz{1},\ldots,\qz{n}\}$.}
\end{cases}
\]
Similarly define a vertex coloring $c'$ for variable placement $\qs'$. From these definitions, $\qs$ and $\qs'$ are equivalent up to NPN-symmetry if and only if the vertex colored graphs $(G,c)$ and $(G,c')$ are vertex-colored graph-isomorphic. 

In practice, we can use the graph package \textsc{Nauty} \cite{mckay14nauty} to compute a canonical form for each vertex-colored graph and check if two are the same. \textsc{Nauty} works with vertex-coloured canonical forms natively as part of its graph isomorphism algorithm, and can compute canonical forms for graphs with thousands of vertices.


\subsection{Placing Variables \& Computing Penalty Functions via SMT/OMT(\laeuf).}
\label{sec:computing_offline_withmapping}
  \begin{figure}[t]
\begin{eqnarray}
\label{eq:encoding3-unrolled}
\Phi(\offset{},\bias,\coup,\qs) &\defas& 
{\sf Range}(\offset{},\bias,\coup,\qs)
\wedge
{\sf Distinct}(\qs)
\wedge
{\sf Graph}()
\\
\label{eq:encoding3-unrolled_inout}
&\wedge&{\bigwedge_{\set{\xs\in\set{-1,1}^n|F(\xs)=\top}} \ \ \bigwedge_{\as\in\set{-1,1}^h} (\Pxauf \ge 0 )\
}
\\
\label{eq:encoding3-unrolled_in}
&\wedge&\bigwedge_{\set{\xs\in\set{-1,1}^n|F(\xs)=\top}} \ \ \bigvee_{\as\in\set{-1,1}^h} (\Pxauf=0 )\  
\\
\label{eq:encoding3-unrolled_out}
&\wedge&\bigwedge_{\set{\xs\in\set{-1,1}^n|F(\xs)=\bot}} \ \  \bigwedge_{\as\in\set{-1,1}^h} (\Pxauf\ge
g_{min} )
\\
\label{eq:encoding3-unrolled_out_exact}
&\wedge&
\bigwedge_{\set{\xs\in\set{-1,1}^n|F(\xs)=\bot}} \ \  \bigvee_{\as\in\set{-1,1}^h} (\Pxauf=
g_{min} )\qquad
\end{eqnarray}
\noindent where:
\begin{eqnarray}
\label{eq:encoding3-range}
{\sf Range}(\offset{},\bias,\coup,\qs)
&\defas&
\bigwedge_{1\le j \le n+h}
\includedin{\qz{j}}{1}{n+h}
\\
&\wedge&
\bigwedge_{1\le j \le n+h} 
\includedin{\bias(j)}{\lowb}{\upb}
\\
\label{eq:euf-simmetry}
&\wedge&
\bigwedge_{\substack{1\le j \le n+h}} \hspace{-.4cm}
(\coup(j,j)=0)
\wedge 
\bigwedge_{\substack{1\le i<j \le n+h}}
\hspace{-.4cm}
         (\coup(i,j)=\coup(j,i))
\\
&\wedge&
\bigwedge_{\substack{1\le i<j \le n+h}} 
\includedin{\coup(i,j)}{\lowc}{\upc}
\\
 {\sf Distinct}(\qz{1},...,\qz{{n+h}})
 &\defas&
 \bigwedge_{1\le i < j \le n+h} \neg (\qz{i}=\qz{j})
 \\
{\sf Graph}()
&\defas&
\wedge 
\bigwedge_{\substack{1\le i < j \le n+h\\\tuple{i,j}\not\in E}}
(\coup(i,j)=0) 
\\
\label{eq:encoding3-penalty}
\Pxauf
&\defas&
\Pxaufexpanded.
\end{eqnarray}
\caption{\label{fig:eufla-encoding} \smt(\laeuf) encoding with automatic placement.}  
\end{figure}


As an alternative to identifying equivalent variable placements, 
for small formulae \Fx, we can combine the generation of the penalty function with an automatic
variable placement by means of SMT/OMT(\laeuf), 
\laeuf being the combined theories of linear arithmetic over
rationals and integers plus uninterpreted function symbols (\sref{sec:background_smt}).
This works as follows.

Suppose we want to produce the penalty function of some relatively small
function (e.g., so $n+h\le 8$, which fits into a single Chimera tile).
We index the $n+h$ 
vertices in the set $V$ into which we want to place the variables
as $V\defas\set{1,...,n+h}$, and we
introduce a set of $n+h$ {\em integer} 
variables $\qs \defas \set{\qz{1},...,\qz{{n+h}}}$ such that 
$\qz{j}\in V$ 
is (the
index of) the vertex into which $z_j$ is placed. 
(For example, ``$v_3=5$" means that variable 
$z_3$ is placed in vertex $\#5$.)
Then we add the standard 
SMT constraint
${\sf Distinct}(\qz{1},...,\qz{{n+h}})$ to the formula to guarantee
the injectivity of the map.~%
%
Then, instead of using variables 
$\theta_{i}$ and $\theta_{ij}$ for biases and couplings, we introduce
the {\em uninterpreted function symbols} 
$\bias: V \longmapsto \mathbb{Q}$ (``bias'')
and
$\coup: V\times V \longmapsto \mathbb{Q}$ (``coupling''),
so that we can rewrite 
 each bias $\theta_{j}$ as $\bias(\qz{{j}})$ and each
coupling $\theta_{ij}$ as $\coup(\qz{{i}},\qz{{j}})$ s.t $\qof{i},\qof{j}\in
[1,..,n+h]$ and ${\sf Distinct}(\qz{1},...,\qz{{n+h}})$.
 
This rewrites the \smtlarat problem
\eqref{eq:encoding2-unrolled}-\eqref{eq:encoding2-unrolled_out} 
into the $SMT(\laeuf)$ problem 
\eqref{eq:encoding3-unrolled}-\eqref{eq:encoding3-penalty} in Figure~\ref{fig:eufla-encoding}.
Equation \eqref{eq:encoding3-unrolled_out_exact} must be used if and only if we need an exact 
penalty function. (Notice that \eqref{eq:euf-simmetry} is necessary because we
could have $\coup(\qz{{i}},\qz{{j}})$ s.t. $\qz{{i}}>\qz{{j}}$.)
By solving \tuple{\Phi(\offset{},\bias,\coup,\qs),g_{min}} 
we not only find the best values of the
biases $\bias$ and couplings $\coup$, but also the best placement $\qs$ of the variables into (the indexes of) the qubits.




\begin{example}
    \label{es:example-lriauf}
  Consider $\xs\defas\set{x_1,x_2,x_3}$, $\as\defas
  \set{a_1}$ and $F(\xs)\defas (x_3 \iff (x_1 \wedge x_2))$,
  and 4-qubit fraction of a tile with 2 horizontal and 2 vertical
  qubits. 
 {Let $z_1$, $z_2$, $z_3$ and $z_4$,  
  denote $x_1$, $x_2$, $x_3$ and $a_1$ respectively, so that each
  $v_j$ denotes the vertex into which $z_j$ is placed.}
We consider the encoding 
\eqref{eq:encoding3-unrolled}-\eqref{eq:encoding3-penalty}, in
particular we have that:
\begin{eqnarray}
\nonumber
\Pxauf  
&\defas&
\offset{}+
\bias(\qof{1})x_1+ 
\bias(\qof{2})x_2+ 
\bias(\qof{3})x_3+ 
\bias(\qof{4})a_1+ \\
\nonumber
&&
\coup(\qof{1},\qof{2})x_1x_2+
\coup(\qof{1},\qof{3})x_1x_3+
\coup(\qof{1},\qof{4})x_1a_1+\\
\nonumber
&&
\coup(\qof{2},\qof{3})x_2x_3+
\coup(\qof{2},\qof{4})x_2a_1+
\coup(\qof{3},\qof{4})x_3a_1
%
\\
\nonumber
  {\sf Graph}() 
&\defas&
\coup(1,2)=0 \wedge \coup(2,1)=0 \wedge
\coup(3,4)=0 \wedge \coup(4,3)=0 
\end{eqnarray}
One possible solution is given in the following tables:
%
$$
\begin{array}{c}
\begin{array}{|l||l|l|l|l||}
\hline
g & \qof{1} & \qof{2} & \qof{3} & \qof{4} \\
2 & 1 & 3 & 2 & 4 \\
\hline
\end{array}
\\ \ \\

\begin{array}{|l||l|l|l|l|}
\hline
\offset{} & 
\bias(\qof{1}) & \bias(\qof{2}) & 
\bias(\qof{3}) & \bias(\qof{4}) 
\\
& 
\bias(1) & \bias(3) & 
\bias(2) & \bias(4) 
\\
5/2 &
- 1/2 &
- 1/2 &
1 & 
0 \\
\hline
\end{array}
\\ \ \\

\begin{array}{|l|l|l|l|l|l|l|l|l|l|l|l|l|l|l}
\hline
\coup(\qof{1},\qof{2}) &
\coup(\qof{1},\qof{3}) &
\coup(\qof{1},\qof{4}) &
\coup(\qof{2},\qof{3}) &
\coup(\qof{2},\qof{4}) &
\coup(\qof{3},\qof{4}) 
\\
\coup(1,3) &
\coup(1,2) &
\coup(1,4) &
\coup(3,2) &
\coup(3,4) &
\coup(2,4) 
\\ 
1/2 & 
0 & 
-1 & 
-1 & 
0 & 
-1 \\
\hline
\end{array}  
\end{array}
$$
which corresponds to the placing in Figure~\ref{fig:4tile} (center).
\end{example}
\begin{figure}[t]
  \centering
  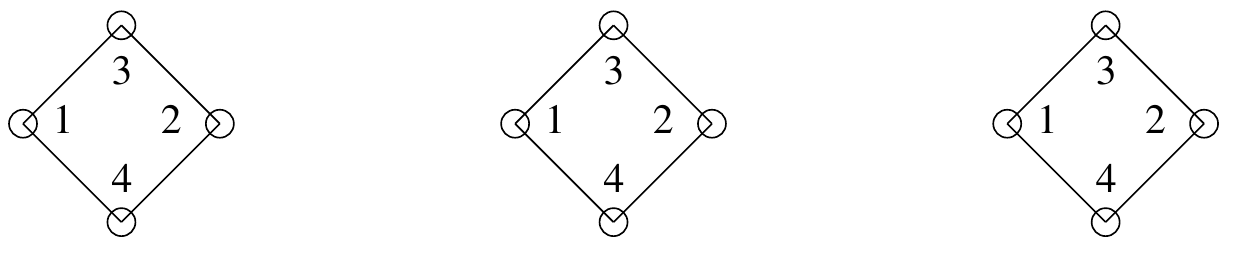
\caption{\label{fig:4tile}
3 possible placements of $\zs\defas\set{x_1,x_2,x_3}\cup\set{a_1}$ into a
4-qubit tile fraction with 2 horizontal and 2 vertical qubits. 
All $4!=24$ combinations are equivalent to one of them.  }
\end{figure}

\subsubsection{Exploiting symmetries.}
When using an SMT/OMT solver to search for penalty functions across all variable placements as in \eqref{eq:encoding3-unrolled}-\eqref{eq:encoding3-penalty}, we may restrict the search space by considering only one variable placement from each equivalence class under the automorphisms of $G$.

\begin{example}
In Example \ref{ex:vp_chimera}, when encoding a penalty function with $n+h = 8$ variables into a Chimera tile, automorphisms reduced the number of variable placements under consideration from $8!=40320$ to $\binom{7}{3}=35$. We can force the SMT/OMT solver to restrict the search to only
35 maps by adding the following constraint to
\eqref{eq:encoding3-unrolled}-\eqref{eq:encoding3-penalty}, consisting
into the disjunction of 35 cubes, each representing one placement. 

\begin{eqnarray}
  \label{eq:simmetrybraking}
\nonumber
&&
(
\overbrace{\qz{1}=1}^{Fixed} \wedge 
\overbrace{\qz{2}=2 \wedge \qz{3}=3 \wedge \qz{4}=4}^{\substack{size\mbox{-}3\
  subset\ of\ \set{\qz{2},...,\qz{8}} \\ mapped\ to\ horizontal\ qubits}} 
\wedge 
\overbrace{\qz{5}=5 \wedge \qz{6}=6 \wedge \qz{7}=7 \wedge\qz{8}=8
}^{\substack{complement\ of\ the\ previous\ subset \\ mapped\ to\ vertical\ qubits} } 
) \vee \\
\nonumber
&&
(
\qz{1}=1 \wedge 
\qz{2}=2 \wedge \qz{3}=3 \wedge \qz{5}=4 \wedge 
\qz{4}=5 \wedge \qz{6}=6 \wedge \qz{7}=7 \wedge\qz{8}=8  
) \vee \\
\nonumber
&&...\\
\nonumber
&&
(
\qz{1}=1 \wedge 
\qz{6}=2 \wedge \qz{7}=3 \wedge \qz{8}=4 \wedge 
\qz{2}=5 \wedge \qz{3}=6 \wedge \qz{4}=7 \wedge\qz{5}=8  
)  .
\end{eqnarray}
If we add this constraint, the first conjunction in \eqref{eq:encoding3-range} can be dropped.
\end{example}

\begin{example}
In Example~\ref{es:example-lriauf} we have $4!=24$ possible placements on to a tile of 2 horizontal and 2 vertical qubits. If we exploit symmetries as above, we have only $\binom{3}{1}=3$ 
inequivalent placements, which are described in
Figure~\ref{fig:4tile}. These can be obtained by adding the
constraint:
\begin{eqnarray}
\nonumber
&&
  \label{eq:symmetrybreaking_4tile1}
(\qof{1}=1 \wedge \qof{2}=2 \wedge \qof{3}=3 \wedge \qof{4}=4
) \vee
\\
\nonumber
&&
  \label{eq:symmetrybreaking_4tile2}
(\qof{1}=1 \wedge \qof{3}=2 \wedge \qof{2}=3 \wedge \qof{4}=4
) \vee
\\
\nonumber
&&
  \label{eq:symmetrybreaking_4tile3}
(\qof{1}=1 \wedge \qof{4}=2 \wedge \qof{2}=3 \wedge \qof{3}=4
).
\end{eqnarray}
\end{example}

 

\section{Encoding Larger Boolean Formulas}
\label{sec:larger-boolean} 
As pointed out in Section \ref{sec:problems_decomposition}, encoding
large Boolean functions using the SMT formulations of the previous
section is computationally intractable, as the number of constraints in the model increases roughly exponentially with the number of variables in the Boolean function. In this section, we describe the natural approach of pre-computing a library of encoded Boolean
functions and rewriting a larger Boolean function $\Fx$ as a set of
pre-encoded ones $\bigwedge_{k=1}^K F_k(\xsk)$. The penalty functions
$P_{F_k}(\xsk,\ask|\ts^k)$ for these pre-encoded functions may then be
combined using chains as described in Section
\ref{sec:problems_embedding}. This schema is shown in Figure
\ref{procgraph}. 
In terms of QA performance, this method has been shown experimentally  to
outperform other encoding methods for certain problem classes
\cite{bian2016mapping}. We will describe each of the
stages in turn (see also
  \cite{bian2014discrete,bian2016mapping,su2016sat}).  

\tikzset{
  font={\fontsize{14pt}{14}\selectfont}}
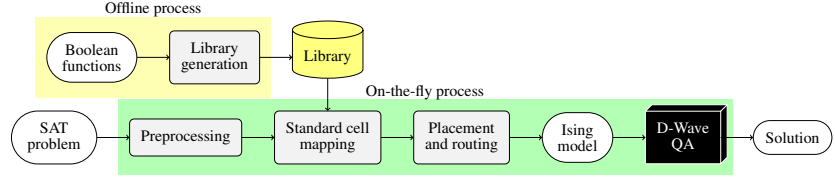
\begin{figure*}[tb]
  \centering{
\begin{adjustbox}{max width=0.9\textwidth}

  \begin{tikzpicture}[
    every node/.style={inner sep=1em, align=center},
    every edge/.append style={->, thick},
    datum/.style={fill=white, rounded rectangle,draw},
    oper/.style={rectangle, fill=gray!20, rounded corners, draw},
    db/.style={cylinder, cylinder uses custom fill,
      cylinder body fill=yellow!50,
      cylinder end fill=yellow!50,
      shape border rotate=90,
      aspect=0.25, draw},
    mach/.style={parallelepiped, draw=white, fill=black, text=white}]

\node [oper] (synth) {Standard cell \\ mapping};
\node [db] (library) [above=of synth] {Library};
\node [oper] (minim) [left= of synth] {Preprocessing};
\node [datum] (satp) [left=of minim]  {SAT \\ problem};
\node [oper] (encoding) [left=of library]{Library \\ generation};
\node [datum] (boolf) [left=of encoding, align=center] {Boolean \\ functions};
\node [oper] (placeroute) [right=of synth, align=center] {Placement \\ and routing};
\node [datum] (isingm) [right=of placeroute, align=center] {Ising \\ model};
\node [mach] (dwave) [right=of isingm, align=center] {D-Wave \\ QA};
\node [datum] (sol) [right= of dwave] {Solution};

\path (boolf) edge (encoding);
\path (encoding) edge (library);
\path (library) edge (synth);
\path (satp) edge (minim);
\path (minim) edge (synth);
\path (synth) edge (placeroute);
\path (placeroute) edge (isingm);
\path (isingm) edge (dwave);
\path (dwave) edge  (sol);

\begin{scope}[on background layer]
  \node [fill=yellow!30, fit=(boolf) (encoding), inner ysep= 1.2em] (box) {};
  \node [anchor=south] [above=-1em of box] {Offline process};
  \node [fill=green!30, fit=(minim) (dwave)] (box2) {};
  \node [anchor=north] [above=-1em of box2] {On-the-fly process};
\end{scope}

\end{tikzpicture}
\end{adjustbox}}
\caption{Graph of the encoding process. \label{procgraph}}
\end{figure*}

\subsection{Library generation}
In this stage, we find effective encodings of common small Boolean functions, using the SMT methods in Section \ref{sec:small_boolean} or by other means, and store them in a library for later use. Finding these encodings may be computationally expensive, but this task may be performed offline ahead of time, as it is independent of the problem input, and it need only be performed once for each NPN-inequivalent Boolean function.

Note that there exist many different penalty functions $\Pxa$ for any small Boolean function $\Fx$. Penalty functions with more qubits may have larger gaps, but using those functions may result in longer chains, so it is not always the case that larger gaps lead to better QA hardware performance. Choosing the most appropriate function may be a nontrivial problem. A reasonable heuristic is to choose penalty functions with gaps of similar size to the gap associated with a chain, namely $g_{min} =  2$.

\subsection{Preprocessing}
Preprocessing, or Boolean formula minimization, consists of
simplifying the input formula $\Fx$ to reduce its size or complexity. While not strictly necessary, it not only improves QA performance by reducing the size of $\Pxa$ but also reduces the computational expense of the encoding process. Moreover, the graphical representation commonly used in preprocessing, the \emph{AND-Inverter Graph} (AIG), is necessary for the subsequent phase of encoding. 

An AIG encodes $\Fx$ as a series of $2$-input $\mathrm{AND}$ gates and negations. More precisely, a directed acyclic graph $D$ on vertex set $\zs = \xs \cup \as = (x_1,\ldots,x_n,a_1,\ldots,a_m)$ is an \emph{AIG representing $\Fx$} if it has the following properties:
\begin{enumerate}
\item Each $x_i$ has no incoming arcs and each $a_k$ has $2$ incoming arcs (the \emph{inputs} to $a_k$), and there is a unique $a_o$ with no outgoing arcs (the \emph{primary output}). 
\item Each arc $z \rightarrow a$ is labelled with a sign $+$ or $-$ indicating whether or not $z$ should be negated as an input to $a$; define a literal $l_a(z) = z$ for an arc with sign $+$ and $l_a(z) = \neg z$ for an arc with sign $-$. 
\item For each node $a_k$ with arcs incoming from $z_1$ and $z_2$, there is an $\mathrm{AND}$ function $A_k(a_k,z_1,z_2)~=~a_k \leftrightarrow l_{a_k}(z_1) \wedge l_{a_k}(z_2)$, such that 
\begin{equation}
\Fx \leftrightarrow \bigwedge_{k=1}^m A_k(\zs) \wedge (a_o=\true).
\label{eqn:and_gate}
\end{equation}
\end{enumerate}
For example, the function $\Fx = x_1 \wedge x_2 \wedge \neg x_3$ is represented by both of the And-Inverter Graphs in Figure \ref{fig:aig}.

\begin{figure}
\includegraphics[scale=0.8]{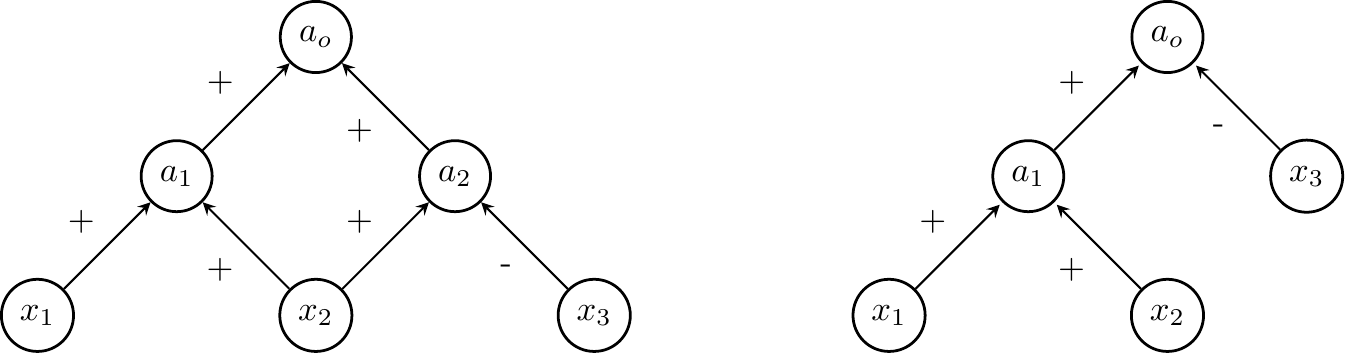}
\caption{Two And-Inverter Graphs representing the function $\Fx = x_1 \wedge x_2 \wedge \neg x_3$.}
\label{fig:aig}
\end{figure}

There are many And-Inverter Graphs representing a given $\Fx$. Is $\Fx$ is in CNF form, we can construct an AIG by rewriting each $\mathrm{OR}$ clause as an $\mathrm{AND}$ function via De Morgan's Law, and then rewriting each $\mathrm{AND}$ function with more than $2$ inputs as a sequence of $2$-input $\mathrm{AND}$ functions. 

Preprocessing is a well-studied problem with mature algorithms available \cite{Mishchenko06dag-awareaig,mishchenko2005fraigs}; here, we use \emph{DAG-aware minimization} as implemented by the logic optimizer ABC.\footnote{see https://github.com/berkeley-abc/abc and https://people.eecs.berkeley.edu/~alanmi/abc/.} 
DAG-aware minimization attempts to find an AIG with a minimal number of nodes by repeatedly identifying a small subgraph that can be replaced with another, smaller subgraph without changing the truth assignments of $\Fx$. 

More precisely, a \emph{cut} $C$ of node $z$ in $D$ is a subset of vertices such that every directed path from an input $x_i$ to $z$ must pass through $C$. The subgraph of $D$ induced by all paths from $C$ to $z$ is a candidate to be replaced by a smaller subgraph, since the Boolean value of $z$ is determined by $C$. We call this value of $z$ as a function of $C$ the \emph{Boolean function represented by $C$}. Cut $C$ is \emph{$k$-feasible} if $|C| \leq k$ and \emph{non-trivial} if $C \neq \{z\}$. For fixed $k$, there is an $O(n)$-time algorithm to identify all $k$-feasible cuts in an AIG: traverse the graph from the inputs $\xs$ to the primary output, identifying the $k$-feasible cuts of node $a_i$ by combining $k$-feasible cuts of $a_i$'s inputs. During traversal, DAG-aware minimization identifies a $4$-feasible cut $C$ and replaces the subgraph induced by $C$ with the smallest subgraph representing the same Boolean function. (There are $222$ NPN-inequivalent $4$-input Boolean functions, and smallest subgraph representing each one is pre-computed.) See \cite{een2007applying} for more details.


\subsection{Standard cell mapping}
\label{sec:onthefly-decomposition}
\newcommand{\zsk}{\ensuremath{\underline{\mathbf{z}^k}}\xspace}

In the standard cell mapping phase, $\Fx$ is decomposed into component functions $\bigwedge_{k=1}^K F_k(\xsk)$ that are available in the library of penalty functions. For SAT or constraint satisfaction problems, this mapping may be performed na{\"i}vely: given a set of constraints $\{F_k(\xsk)\}_{k=1}^K$ on the variables, each $F_k(\xsk)$ is found in the library (possibly combining small constraints into larger ones \cite{bian2014discrete}). However, more advanced techniques have been devised in the digital logic synthesis literature. \emph{Technology mapping} is the process of mapping a technology-independent circuit representation to the physical gates used in a digital circuit \cite{een2007applying,mishchenko2005technology}. Usually technology mapping is used to reduce circuit delay and load, and performs minimization as an additional step. Delay and load do not play a role in the context of QAs, but minimization is important to simplify the placement and routing phase that follows. 

In order to find an efficient decomposition, a technology mapping algorithm takes as input costs for small $F_k(\xsk)$ and attempts to minimize the sum of the costs of the components in $\bigwedge_{k=1}^K F_k(\xsk)$. We define the cost of $F_k$ to be the number of qubits used by the penalty model $P_{F_k}$, so that the cost of $\Fx = \bigwedge_{k=1}^K F_k(\xsk)$ is the total number of qubits used to represent $\Fx$, prior to adding chains. 

Here, we apply the technology mapping algorithm in \cite{een2007applying}: the idea is to decompose the AIG representing $\Fx$ into a collection of cuts such that each cut represents a small function $F_k(\zsk)$ that can be found in the penalty library. A \emph{mapping} $M$ of an AIG $D$ is a partial function that maps a node $a_i$ of $D$ to a non-trivial, $k$-feasible cut $M(a_i)$. We say $a_i$ is \emph{active} when $M(a_i)$ is defined and \emph{inactive} otherwise. Mapping $M$ is \emph{proper} if:
\begin{enumerate}
\item the primary output $a_o$ is active;
\item if $a_i$ is active; then every $a_j \in M(a_i)$ is active; and
\item if $a_j \neq a_o$ is active; then $a_j \in M(a_i)$ for some active $a_i$.
\end{enumerate}

For each active node $a_k$ in a proper mapping $M$, there is a Boolean function $F_k(\zsk)$ represented by the cut $M(a_k)$, and the original Boolean function $\Fx$ decomposes as 
\[
\Fx \leftrightarrow \bigwedge_{k=1}^K F_k(\zsk) \wedge (a_o=\true).
\]
Therefore, choosing $k$-feasible cuts with small $k$, proper mappings provide decompositions of $\Fx$ into small Boolean functions that can be found in the penalty library. One example of a proper mapping is the trivial mapping, in which each $a_i$ is mapped to the cut consisting of its two input nodes. Under the trivial mapping, $\Fx$ is decomposed into a collection $2$-input $\mathrm{AND}$'s.

The algorithm in \cite{een2007applying} iteratively refines mapping $M$ in order to improve the cost of the decomposition, in the following way. For each node $a_i$, maintain a list $L(a_i)$ of $k$-feasible cuts, ordered by their cost. (The cost of a cut is a function of the cost of the Boolean function it represents, taking into account the anticipated recursive effects of having a new set of active nodes: see \cite{een2007applying} for details.) Traverse the graph from inputs $\xs$ to primary output $a_o$. At each $a_i$, first update the costs of the cuts in $L(a_i)$ based on the changes to the costs of earlier nodes in the traversal. Next, if $a_i$ is active and the current cut $M(a_i)$ is not the cut in $L(a_i)$ of lowest cost, update $M(a_i)$. To do this, first inactivate $a_i$ (which recursively inactivates nodes in $M(a_i)$ if they are no longer necessary) and then reactivate $a_i$ (which reactivates nodes in $M(a_i)$, also recursively). This process of refining the mapping by traversing the graph is repeated several times.  

Given the connectivity of the Chimera hardware graph, a natural choice
is to decompose into Boolean functions that can be modelled with a
single 8-qubit tile. In particular all $3$-input, $1$-output Boolean
functions (all $3$-feasible cuts) can be modelled in one tile. 




\subsection{Placement and routing}
\label{sec:onthefly-placementrouting}
Once $\Fx$ is decomposed into smaller functions $\bigwedge_{k=1}^K F_k(\xsk)$ with penalty functions $P_{F_k}(\xsk,\ask|\ts^k)$, it remains to embed the entire formula onto the QA hardware as in equation \eqref{eqn:distinct_qubits}. This process has two parts: \emph{placement}, in which each $P_{F_k}(\xsk,\ask|\ts^k)$ is assigned to a disjoint subgraph of the QA hardware graph; and \emph{routing}, in which chains of qubits are built to ensure that distinct qubits $x_i$ and $x'_i$ representing the same variable take consistent values (using equivalence constraints with penalty functions of the form $1-x_ix'_i$). Both placement and routing are very well-studied in design of digital circuits \cite{betz1997vpr}. Nevertheless, this stage is a computational bottleneck for encoding large Boolean functions.


\subsubsection{Placement}
During placement, chain lengths can be minimized by placing penalty functions that share common variables close together. Current QA processors have a nearly 2-dimensional structure, which lets us measure distance between variables using planar coordinates. (For example, for the 2048-qubit Chimera graph in Fig.~\ref{fig:full16chimera}, define the planar coordinates of a unit cell to be its row and column index in the $16 \times 16$ grid.) One common objective function from digital circuit design is ``half-perimeter wire length" \cite{Kahng11}. Define the \emph{location} of a Boolean function $F_k(\xsk)$ to be the subgraph of $G$ onto which $F_k(\xsk)$ is placed, and define a placement function $p:\{1,...,K\} \rightarrow \mathbb{R}^2$ which maps each $k$ to the planar coordinates $p(k) = (a_k,b_k)$ of the location of $F_k(\xsk)$. The \emph{half-perimeter wire length (HPWL)}  of a variable $x_i$ is the total length and width of the smallest box that can be drawn around the locations of functions containing $x$. That is, for $S_i = \{k: x_i \in \xsk\}$,
\[
HPWL(x_i) \defas (\max_{k\in S_i}a_k - \min_{k\in S_i}a_k ) + (\max_{k\in S_i}b_k - \min_{k\in S_i}b_k ).
\]
A placement algorithm attempts to find a placement that minimizes $\sum_{i=1}^n HPWL(x_i)$.

Heuristic methods for placement include simulated annealing \cite{sun1995timberwolf}, continuous optimization \cite{chan00co_placement}, and recursive min-cut partitioning \cite{roy06capo}. These algorithms can be applied in the present context, but require some modification as current QA architectures do not distinguish between qubits used for penalty functions and qubits used for chains. For example, in some algorithms, a placement is optimized on the assumption that the resulting routing problem is feasible (possibly by expanding the planar area made available for routing). 
This assumption may not necessarily hold using a fixed QA hardware graph of limited size and connectivity. If unit cells are packed tightly with Boolean functions, then there will be few remaining qubits available for routing. On the other hand, reserving too many qubits for routing will have a negative impact on hardware performance in the form of longer chains.

In the experiments in \sref{sec:expaval} we made use of mPL\footnote{Available at \url{http://cadlab.cs.ucla.edu/cpmo/}}, a publicly available academic placement tool \cite{chan00co_placement}. mPL is multilevel method in which the placement problem is repeatedly coarsened (so that several $P_{F_k}$ are clustered and treated as one), placed, and uncoarsened with local improvements. At the coarsest level, placement is performed using a customized non-linear programming algorithm which maps penalty functions to real coordinates minimizing a quadratic distance function between shared variables. 


\subsubsection{Routing}
During routing, literals are chained together using as few qubits possible; this problem may be formalized as follows. Assume a single variable $x_i$ has been assigned to a set of vertices $T_i \subseteq V$, its \emph{terminals}, during the placement of small Boolean functions. To create a valid embedding, the chain of vertices representing $x_i$, call it $C_i$, must contain $T_i$ and induce a connected subgraph in $G$. Finding $C_i$ with a minimum number of vertices is an instance of the \emph{Steiner tree problem} \cite{Byrka2013} and $C_i$ is a Steiner tree. 
Given variables $(x_1,\ldots,x_n)$ assigned to terminals $(T_1,\ldots,T_n)$, the \emph{routing} problem demands a set of chains $(C_1,\ldots,C_n)$ such that each $C_i$ contains $T_i$, every chain is connected, and all chains are pairwise disjoint. Among routing solutions, we try to minimize the total number of vertices of $G$ used or the size of the largest chain.

Routing to minimize the total number of vertices used is NP-hard, but polynomial-time approximation algorithms exist \cite{gester13bonn}. In practice, heuristic routing algorithms scale to problem sizes much larger than current QA architectures \cite{xu09fast,roy08fgr,chen09ntugr,cho07box,chang08nthu}. 

Routing in the current context differs from routing used in digital circuit design in the sense that vertices (qubits) are the sparse resource that variables compete for, rather than edges. As a result, we make use vertex-weighted Steiner tree algorithms rather than edge-weighted ones. This makes the problem harder, as the edge-weighted Steiner tree problem is (1.39)-approximable in polynomial time \cite{Byrka10}, while vertex-weighted Steiner-tree is only $(\log k)$-approximable for $k$ terminals in polynomial time unless P=NP \cite{Klein95}. Nevertheless, in practice, simple 2-approximation algorithms for edge-weighted Steiner tree such as the MST algorithm \cite{Vazirani01} or Path Composition \cite{Gester13} also work very well for the vertex-weighted problem. In this section, we describe a modification of the routing algorithm BonnRoute \cite{Gester13} for vertex-weighted Steiner trees.

We first solve a continuous relaxation of the routing problem called \emph{min-max resource allocation}. Given a set of vertices $C \subseteq V$, the \emph{characteristic vector} of $C$ is the vector $\chi(C) \in \{0,1\}^{|V|}$ such that $\chi(C)_v = 1$ if $v \in C$ and $0$ otherwise. Let $H_i$ be the convex hull of all characteristic vectors of Steiner trees of $T_i$ in $G$. Then the \emph{min-max resource allocation} problem for terminals $T_1,\ldots,T_n$ is to minimize, over all $z_i \in H_i$, $i \in \{1,\ldots,n\}$,
\[
\lambda(z_1,\ldots,z_n) \defas \max_{v \in V} \sum_{i=1}^n (z_i)_v.
\]
The vertices $v$ are the \emph{resources}, which are allocated to \emph{customers} $(z_1,\ldots,z_n)$\footnote{The original BonnRoute algorithm uses min-max resource allocation with edges rather than vertices as resources.} To recover the routing problem, note that if each $z_i$ is a characteristic vector of a single Steiner tree, then $\sum_{i=1}^n (z_i)_v$ the number of times vertex $v$ is used in a Steiner tree. In that case, $\lambda(x) \leq 1$ if and only if the Steiner trees are a solution to the routing problem. 

To solve the min-max resource allocation, we iteratively use a weighted-Steiner tree approximation algorithm to generate a probability distribution over the Steiner trees for each $x_i$. After a Steiner tree is generated, the weights of the vertices in that Steiner tree are increased to discourage future Steiner trees from reusing them (see Algorithm \ref{fig:Bonn} for details). This algorithm produces good approximate solutions in reasonable time. More precisely, given an oracle that computes vertex-weighted Steiner tree approximations within a factor $\sigma$ of optimal, for any $\omega>0$ Algorithm \ref{fig:Bonn} computes a $\sigma(1+\omega)$-approximate solution to min-max resource allocation problem using $O(( \log |V|) (n + |V|) (\omega^{-2} + \log \log |V|))$ calls to the oracle \cite{Muller11}.

\begin{algorithm}
\begin{algorithmic}
\Require{Graph $G$, Steiner tree terminals $\{T_1,\ldots,T_n\}$, number of iterations $t$, weight penalty $\alpha > 1$}
\Ensure{For each $i$, a probability distribution $p_{i,S_i}$ over all Steiner trees $S_i$ for terminals $T_i$} 
\Statex
\Function{BonnRoute}{$G$,$\{T_1,\ldots,T_n\}$}
\For{each $v \in V(G)$}
	\State{$w_v \gets 1$}
	\EndFor
\For{each Steiner tree $S_i$ for terminals $T_i$, $i \in [n]$}
	\State{$z_{i,S_i} \gets 0$}
	\EndFor
\For{$j$ from $1$ to $t$}
	\For{each $i \in [n]$}
		\State{Find a Steiner tree $S_i$ for terminals $T_i$ with vertex-weights $w_v$}
	    \State{$z_{i,S_i} \gets z_{i,S_i} + 1$}
	    \State{$w_v \gets w_v * \alpha$ for all $v \in S_i$}
	    \EndFor
	\EndFor
\State{Return $p_{i,S_i} \gets z_{i,S_i}/t$}
\EndFunction			
\end{algorithmic}
\caption{BonnRoute Resource Sharing Algorithm \cite{Gester13}.}
\label{fig:Bonn}
\end{algorithm}

Once a solution to the min-max resource allocation has been found, we recover a solution to the original routing problem by formulating an integer linear program (IP), which may be solved via \omlarat.\footnote{The original BonnRoute algorithm uses randomized rounding to recover a routing solution from min-max resource allocation, but at current QA hardware scales this is not necessary.} For each Steiner tree $S_i$ with non-zero probability in the distribution returned from min-max resource allocation, define a binary variable as follows:
\[
x_{i,S_i} = \begin{cases}
1, & \text{if $S_i$ is the selected Steiner tree for variable $i$}; \\
0, & \text{otherwise}.
\end{cases}
\]
Then minimize the number of qubits selected, subject to selecting one Steiner tree for each $i$ and using each vertex at most once. That is,
\begin{align*}
\min \qquad & \sum_i \sum_{S_i} |S_i|x_{i,S_i}  \\
\text{s.t. } \qquad  & \sum_{S_i} x_{i,S_i} = 1 \text{ for all } i \\
& x_{i,S_i} + x_{j,S_j} \leq 1 \text{ for all } S_i, S_j \text{ s.t. } S_i \cap S_j \neq \emptyset.
\end{align*}

When applying routing to the Chimera graph, because of the symmetry within each unit tile, it is convenient to work with a \emph{reduced} graph in which
the horizontal qubits in each unit tile are identified as a single qubit, and similarly for the vertical qubits. As a result the scale of the routing problem is reduced by a factor of $4$. This necessitates the use of vertex capacities within the routing algorithm (each reduced vertex has a capacity of $4$), and variables are assigned to individual qubits within a tile during a secondary, detailed routing phase.


In the digital circuit literature, the placement and routing stages of embedding are typically performed separately. However, because of current limited number of qubits and the difficulty in allocating them to either placement or routing, a combined place-and-route algorithm can be more effective. This approach is discussed in detail in \cite{bian2016mapping}.


\section{Related work}
\label{sec:related}
There have been several previous efforts to map specific small Boolean functions (usually in the guise of constraint satisfaction problems) to Ising models. Most of those mappings have been ad hoc, but some were more systematic (beyond  \cite{bian2014discrete} and \cite{bian2016mapping} as previously discussed). Lucas \cite{lucas14ising} and Chancellor et al. \cite{chancellor2016ksat} developed Ising models for several specific NP-hard problems, while Su et al. \cite{su2016sat} and Pakin \cite{pakin2016qmasm, pakin2018} decomposed Boolean functions into common primitives.

There are have also been several attempts to map large Boolean functions or more generally large constrained Boolean optimization problems to D-Wave hardware. Most of these efforts (e.g. \cite{venturelli2015jobshop, rosenberg2015trading, dridi2016prime, perdomo2015quantum, rieffel2015planning, ogorman2016, zick15iso, bian13ramsey, jiang2018}) have used global embedding, in which an entire Ising model is minor-embedded heuristically \cite{cai2014practical} or a fixed embedding is used \cite{boothby2016fast, zaribafiyan2016sys}. However Su et al. \cite{su2016sat} used a general place-and-route approach, while Trummer et al.\cite{trummer2016mqo}, Chancellor et al. \cite{chancellor2016ksat}, Zaribafiyan et al. \cite{zaribafiyan2016sys}, and Andriyash et al.~\cite{ andriyash2017factoring} used a placement approach optimized for the specific constraints at hand.

Looking at SAT instances in particular, there have been at least two previous attempts at benchmarking D-Wave hardware performance: McGeoch et al. \cite{mcgeoch2013evaluation} and Santra et al. \cite{Santra2014} looked at (weighted) Max2SAT problems, and Douglass et al. \cite{douglass2015filters} and Pudenz et al. \cite{Pudenz2016} looked at SAT problems with the goal of sampling diverse solutions. Farhi et al. \cite{Farhi2012} and Hen and Young \cite{Hen2011} studied the performance of quantum annealing on SAT problems more generally. The applicability of QAs for various SAT formulations has also been discussed in \cite{choi2011different,king2015range}.

\section{Preliminary Experimental Evaluation}
\label{sec:expaval}




We
have implemented and made publicly available prototype encoders built
on top of the SMT/OMT tool \optimathsat{} \cite{st_cav15}. 
%
  In particular each \sattoqubo-specific step outlined in Figure \ref{procgraph} has been implemented as a Python library. For preprocessing we rely on the ABC tool suite \cite{brayton2010abc}. The same software is capable of performing technology mapping, though a Python version is available in the \textsc{techmapping} library\footnote{Available at \url{https://bitbucket.org/StefanoVt/tech_mapping}}. Finally the \textsc{placeandroute} library\footnote{Available at \url{https://bitbucket.org/StefanoVt/placeandroute}} performs the combined placement and routing step.
%
%
  Regarding the off-line part of the process, the \textsc{gatecollector} library\footnote{Available at \url{https://bitbucket.org/StefanoVt/gatecollector}} extracts the most common gates in a dataset of functions and generates a function library in the \textsc{ABC}-compatible \textsc{genlib} format. The \textsc{pfencoding} library\footnote{Available at \url{https://bitbucket.org/StefanoVt/pfencoding}} is then used to call \optimathsat{} to encode them for later use.
  Currently the most expensive step in the on-the-fly process is the placement and routing step. In the current setup we use $\approx$20 minutes  on a Intel i7-5600U CPU when we encode the problems used in the experimental evaluation. The software run-time is heavily tunable in order to trade off efficiency and effectiveness of the place-and-route process.

We offer preliminary empirical validation of
the proposed methods for solving SAT via \sattoqubo{} encoding by
evaluating the performance of D-Wave's 2000Q system in solving certain
hard SAT problems (\sref{sec:sat-exper}); we perform a similar
evaluation also on MaxSAT problems (\sref{sec:maxsat-exper}), despite 
the limitations highlighted in \sref{sec:maxsat}.

This task is subject to some limitations. 
First, we require instances that can be entirely encoded in a quantum annealer of 2000 qubits (although algorithms for solving much larger constraint satisfaction problems have been proposed; see \cite{bian2014discrete,bian2016mapping}). Furthermore, SAT
solvers are already quite effective on the average case, so we need concrete
worst-case problems.
Another important consideration in solving [Max]SAT instances is that the QA hardware cannot be made aware of the optimality of
solution; for example, the algorithm cannot terminate when all clauses in a SAT
problem are satisfied. In this way, QA hardware behaves more like an
SLS  solver than a CDCL-based one.  
To this extent, and
in order to evaluate the significance of the testbed, we solved the
same problems with the state-of-the-art \textsc{UBCSAT} SLS SAT solver using the best performing
algorithm, namely SAPS \cite{tompkins_ubcsat:_2004}. 
\textsc{UBCSAT} was run on a computer using a 8-core Intel\textsuperscript{\textregistered} Xeon\textsuperscript{\textregistered}
E5-2407 CPU, at 2.20GHz.

\begin{remark}
\label{remark:fairness}
The results reported in this section are not intended as a performance
comparison between D-Wave's 2000Q system and \textsc{UBCSAT}, or any other
classic computing tool. It is difficult
to make a reasonable comparison for many reasons, including issues of
specialized vs. off-the-shelf hardware, different timing mechanisms
and timing granularities, and costs of encoding. Instead we aim to
provide an empirical assessment  
of QA's potential for [Max]SAT solving, based on currently available
systems. 
\end{remark}

\paragraph{Reproducibility of results}
To make the results reproducible to those who have access to
a D-Wave system, we have set a website 
where experimental data,  problem files, translation files, 
demonstration 
code and supplementary material can be accessed.~\footnote{\label{footnote:website}
  \url{https://bitbucket.org/aqcsat/aqcsat}.} 
%
Notice that public access to a D-Wave 2000Q machine is possible through
D-Wave's Leap
cloud service~\footnote{\url{https://cloud.dwavesys.com/leap/}.}.


\subsection{SAT}
\label{sec:sat-exper}

\begin{figure}
  \includegraphics[width=\textwidth]{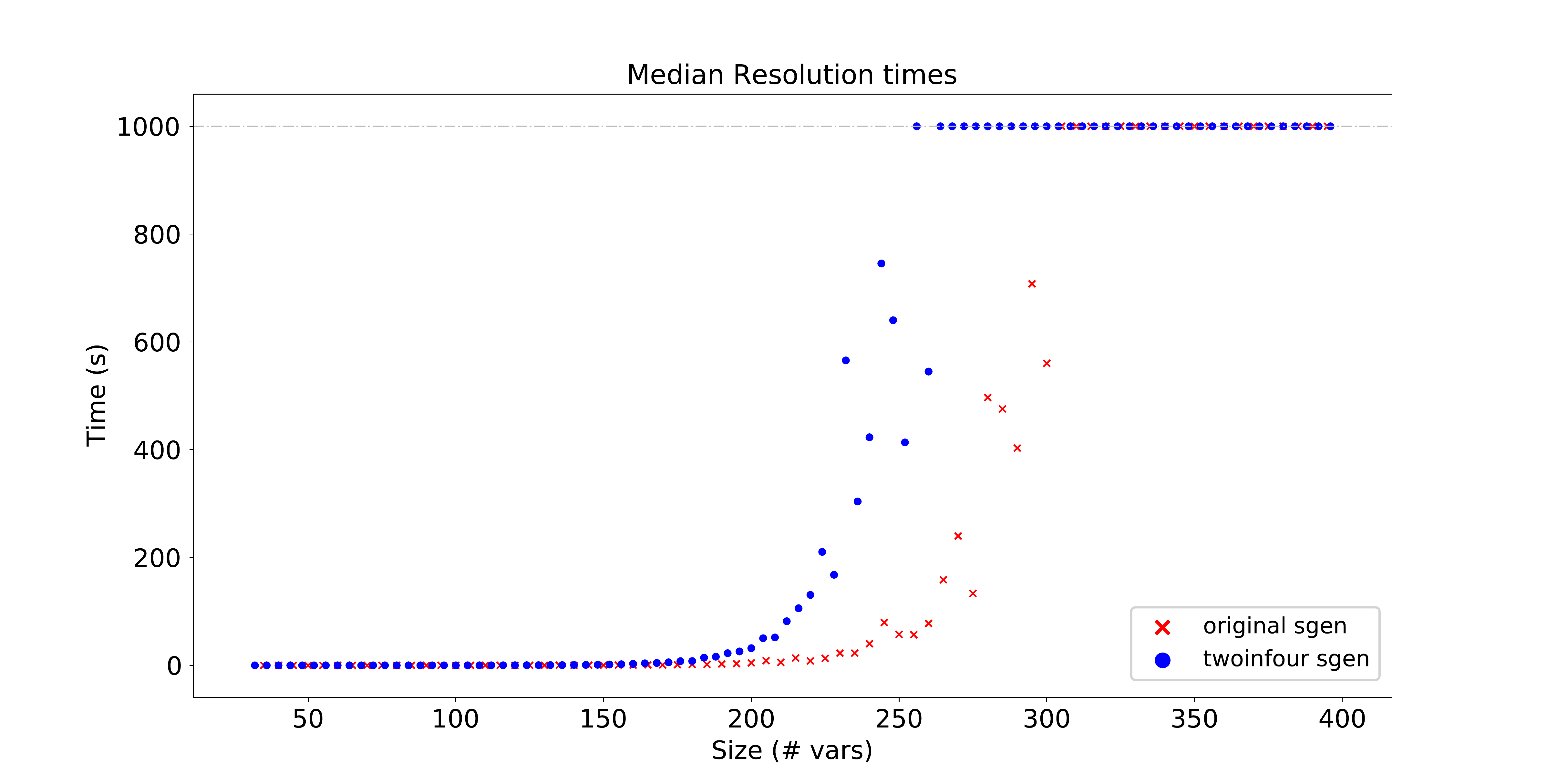}
\caption{Median times for the best-performing SLS algorithm on two
    different variants of the \sgen{} problem on \textsc{UBCSAT} (SAPS). Timeout
    is marked with a gray line.
The figure report times on a computer with a 8-core
Intel\textsuperscript{\textregistered}
Xeon\textsuperscript{\textregistered} E5-2407 CPU, at 2.20GHz. 
\label{fig:asymp_sat}}
 \end{figure}

\paragraph{Choosing the benchmark problems}
\label{sec:choosing-benchmarks}
In order to provide a significant empirical evaluation, 
and due to the limitations in size and connectivity of current QA systems,
we require SAT problems which have a low number of variables but are
nevertheless hard for standard SAT solvers. 

\begin{IGNORE} 
Another useful characteristic is having
a regular structure, thus easing embedding. If the problem is composed
by a low number of similar complex constraints, we can exploit the
SMT-based techniques of \sref{sec:small_boolean} to have an efficient
encoding of the constraint.  We can use SAT solver competitions
\cite{satcomp} as a reference for finding hard SAT problem
instances. In these yearly competitions, different SAT solvers are
tested on a set of problem instances. Some of these problems are
inspired from industry use-cases, others are crafted to be
particularly hard. 
 
We are interested in small hard problems, so we focused on the smallest
problems between the crafted category. 
In the previous edition of the SATCOMP,
the smallest crafted problems are generated using a tool called \sgen{}, described
in \cite{spence_sgen1:_2010}.  
\end{IGNORE} 
%
To this end we chose and modified the tool \sgen{}
\cite{spence_sgen1:_2010}, which has been used to generate the
smallest unsolvable problems in recent SAT competitions.
The problems share a structure that is suited for
the problem embedding, as it contains multiple clones of slightly complex
constraints, and even problems with few hundreds variables are considerably
hard.
The \sgen{} family of random generators received many improvements over
the years, but 
the method to generate satisfiable instances has remained the same
\cite{gelder_zeroone_2010,spence_weakening_2015}. \sgen{} works by setting
cardinality constraints over different partitions of the variable set. The
generator operates as follows:

\begin{enumerate} 
\item The user decides the number of Boolean variables in the
problem. 

\item The tool partitions the variable set into sets of 5 elements.

\item For satisfiable problem instances, the desired solution contains exactly
one true variable for each subset. For each subset we guarantee that at most
one variable is true (10 2-CNF clauses).

\item The partition is shuffled. The tool ensures that each new subset contain
exactly one true variable, and minimizes the similarity with the previous
partition. 

\item For each new subset we ensure that at least one variable is true (a
single CNF clause).

\item The previous two steps are repeated one more time, further restricting
the solution space.

\end{enumerate}

In Figure~\ref{fig:asymp_sat} (red plot) 
we can see how \textsc{UBCSAT} SAPS performs on
these random \sgen{} problems. Notice that with $>300$ variables the solver
reaches the timeout of 1000s. 
In our experiments, we modify the tool by
using exactly-2-in-4 constraints 
on partitions with sets of size 4 with exactly two true variables per subset.
This kind of constraint has a more efficient  embedding and the modified
problems are harder (see Figure \ref{fig:asymp_sat}, blue plot, where
\textsc{UBCSAT} reaches the timeout with $>270$ variables).

\begin{table}[t]
\small
\begin{center}
\subtable[\label{tab:dw_sat_solved}]{
\begin{tabular}{||l||r|r|r||r||}
\hline
\multicolumn{5}{|c|}{\textbf{D-Wave 2000Q}}\\\hline
\textbf{Problem size}&\textbf{\begin{tabular}{@{}c@{}}\# solved\\5 samples\end{tabular}}& \textbf{\begin{tabular}{@{}c@{}}\# solved\\10 samples\end{tabular}} & \textbf{\begin{tabular}{@{}c@{}}\# solved\\20 samples\end{tabular}} &\textbf{\begin{tabular}{@{}c@{}}\% optimal\\samples\end{tabular}}\\\hline
\textbf{32 vars}&100&100&100&97.4\\\hline
\textbf{36 vars}&100&100&100&96.4\\\hline
\textbf{40 vars}&100&100&100&94.8\\\hline
\textbf{44 vars}&100&100&100&93.8\\\hline
\textbf{48 vars}&100&100&100&91.4\\\hline
\textbf{52 vars}&100&100&100&93.4\\\hline
\textbf{56 vars}&100&100&100&91.4\\\hline
\textbf{60 vars}&100&100&100&88.2\\\hline
\textbf{64 vars}&100&100&100&84.6\\\hline
\textbf{68 vars}&100&100&100&84.4\\\hline
\textbf{72 vars}&98 &100&100&84.6\\\hline
\textbf{76 vars}&99 &99&100&86.6\\\hline
\textbf{80 vars}&100&100&100&86.0\\\hline
\end{tabular}}
\qquad
\subtable[\label{tab:ubcsat-saps-avgtimes}]{
\begin{tabular}{|l|r|}
\hline
\multicolumn{2}{|c|}{\textbf{\textsc{UBCSAT} (SAPS)}}\\\hline
\textbf{Problem size}&\textbf{Avg time (ms)}\\\hline
\textbf{32 vars}&0.1502\\\hline
\textbf{36 vars}&0.2157\\\hline
\textbf{40 vars}&0.3555\\\hline
\textbf{44 vars}&0.5399\\\hline
\textbf{48 vars}&0.8183\\\hline
\textbf{52 vars}&1.1916\\\hline
\textbf{56 vars}&1.4788\\\hline
\textbf{60 vars}&2.2542\\\hline
\textbf{64 vars}&3.1066\\\hline
\textbf{68 vars}&4.8058\\\hline
\textbf{72 vars}&6.2484\\\hline
\textbf{76 vars}&8.2986\\\hline
\textbf{80 vars}&12.4141\\\hline
\end{tabular}}
\caption{\label{tab:satresults}%
(a) Number of \sattoqubo{} problem instances (out of 100) solved by the QA hardware
using 5 samples \resp{10 and 20} and average fraction of samples from the QA hardware
that are optimal solutions. Annealing was executed at a rate of
\SI{10}{\micro\second} per sample, for a total of
\SI{50}{\micro\second}, \resp{\SI{100}{\micro\second} and
\SI{200}{\micro\second}}    of anneal time per instance respectively. Total time used by
the D-Wave processor includes programming and readout; this amounts to
about \SI{150}{\micro\second} per sample, plus a constant
\SI{10}{\milli\second} of overhead. \newline
(b) Run-times in
\SI{}{\milli\second} for SAT instances solved by \textsc{UBCSAT} using SAPS,
averaged over 100 instances of each problem size. Computations were
performed using an 8-core Intel\textsuperscript{\textregistered}
Xeon\textsuperscript{\textregistered} E5-2407 CPU, at 2.20GHz. 
}
\end{center}
\end{table}

\paragraph{Experiments and Results}
To solve these SAT instances, we encode and embed them as in
\sref{sec:small_boolean}-\sref{sec:larger-boolean}
 and then draw a fixed number of samples/instance (5, 10, 20)
at an annealing rate of \SI{10}{\micro\second} per sample. Table
\ref{tab:dw_sat_solved} shows the results from the D-Wave 2000Q QA hardware.

The QA hardware solves almost all problems with 5 samples (i.e. within
\SI{50}{\micro\second} of total anneal time), and all of them with 20
samples (i.e. within \SI{200}{\micro\second} of total anneal time),
and the rates of sampling optimal solutions remain relatively stable
at this scale of problem.

In order to evaluate the significance of the testbed, we also report the
results of solving the
same problems with the \textsc{UBCSAT} SLS SAT solver using SAPS
\cite{tompkins_ubcsat:_2004}. Remark~\ref{remark:fairness} applies here.
Table~\ref{tab:ubcsat-saps-avgtimes}
shows that the problems are nontrivial despite the small number of
variables, and the run-times increase significantly with the size of
the problem. (See also Figure~\ref{fig:asymp_sat}.)


\begin{IGNORE}
\subsection{Propositional Satisfiability(SAT)}
\label{sec:sat-exper-long}

We generated problem instances with various problem sizes and constraint
configurations. Problem sizes ranged from 32 variables to 80, the biggest size
on which embedding has been successfully performed. Problems have been
generated in the original version (with at-least-one-in-five and
at-most-one-in-five) and a two-in-four version (using exactly-two-in-four
constraints). For each problem configuration, 100 instances were generated
using different random seeds.

We want to solve these problems using the D-Wave machine. In order to do that
we performed the encoding as explained in \sref{sec:computing}. We took
\TODO{how many} samples, using $10 \mu s$ for each anneal.

We compare the results using the \textsc{UBCSAT} SLS SAT solver, using the SAPS
algorithm \cite{tompkins_ubcsat:_2004}. The SAPS algorithm proved to be the
best performer between all SLS algorithms in \textsc{UBCSAT}. The tests were performed
on a computer using a 8-core Intel\textsuperscript{\textregistered} Xeon\textsuperscript{\textregistered}
E5-2407 CPU, at 2.20GHz.

SAT solvers and quantum annealers works in fundamentally different ways: while
the software solvers keep working for as much time the user wants, annealing
times for the D-Wave machine are very short. Thus, performing a comparison
between them is not obvious. 

For quantum annealers, we can estimate the expected time required to receive a
sample satisfying the formula using the success rate and anneal time. We can
also ignore initialization times, as we do not want to consider conversion and
hardware I/O issues.

Despite the use of \sgen{} problems, the size of the hardware is still not
sufficient to contain encoded problems significantly hard for SAT solvers. So
far, the encoding times of the examined problem instances dwarf the resolution
times. Due the exponential growth of the resolution time, improvements of the
quantum annealer capacity that allow problems above a certain threshold will be
hard to match with software improvements.

\begin{table}[t]
\centering

\caption{SAT problem instances solved by \textsc{UBCSAT} (SAPS) and D-Wave 2000Q in less than
a millisecond. Missing entries are problem sizes that cannot be partitioned
evenly in groups of the required size. \label{tab:ubcsat-saps}}
\small
\begin{tabular}{|l|c c||c c|}
\hline
size & \multicolumn{4}{|c|}{problems solved in $<1 ms$}  \\ \hline
  &  \multicolumn{2}{|c||}{\textsc{UBCSAT}(SAPS)} & \multicolumn{2}{c|}{D-Wave  2000Q(wall-clock)} \\ 
  &  one-in-five & two-in-four & one-in-five & two-in-four  \\ \hline
\hline
32 &   & 100  &   &100      \\ \hline
35 & 100 &    &   &     \\ \hline
36 &   & 100  &   &100     \\ \hline
40 & 100 & 97 &   &100     \\ \hline
44 &   & 90   &   &100     \\ \hline
45 & 99 &     &   &     \\ \hline
48 &   & 76   &   &100     \\ \hline
50 & 97 &     &   &     \\ \hline
52 &   & 56   &   &100     \\ \hline
55 & 83 &     &   &     \\ \hline
56 &   & 42   &   &100     \\ \hline
60 & 65 & 6   &   &100      \\ \hline
64 &   & 3    &   &100      \\ \hline
65 & 36 &     &   &      \\ \hline
68 &   & 0    &   &100      \\ \hline
70 & 11 &     &   &      \\ \hline
72 &   & 0    &   &100      \\ \hline
75 & 0 &      &   &      \\ \hline
76 &   & 0    &   &100      \\ \hline
80 & 0 & 0    &   & 100         \\ \hline
\end{tabular}
\end{table}
\begin{table}[]
\centering
\caption{Average resolution times of SAT problems solved by \textsc{UBCSAT} (SAPS) and D-Wave 2000Q. Missing entries are problem sizes that cannot be partitioned evenly in groups of the required size. \label{tab:ubcsat-saps-avgtimes-old}}
\begin{tabular}{|l|c|c||c|c|}
\hline
size & \multicolumn{4}{|c|}{average solution time (ms)}  \\ \hline
  &  \multicolumn{2}{|c||}{\textsc{UBCSAT}(SAPS)} & \multicolumn{2}{c|}{D-Wave  2000Q(wall-clock)} \\ 
  &  one-in-five & two-in-four & one-in-five & two-in-four  \\ \hline
\hline
32 &   & 0.1502  & &        0.1696   \\ \hline
35 & 0.1105 &  & &                      \\ \hline
36 &   & 0.2157  & &        0.1727   \\ \hline
40 & 0.1798 & 0.3555  & &0.1805     \\ \hline
44 &   & 0.5399  & &        0.1805   \\ \hline
45 & 0.2924 &  & &                      \\ \hline
48 &   & 0.8183 &  &        0.1929   \\ \hline
50 & 0.4632 &  & &                      \\ \hline
52 &   & 1.1916 & &         0.1789   \\ \hline
55 & 0.7515 &  & &                      \\ \hline
56 &   & 1.4788 & &         0.1913   \\ \hline
60 & 1.0243 & 2.2542 & & 0.2099   \\ \hline
64 &   & 3.1066 & &         0.2316   \\ \hline
65 & 1.3661 &  & &                      \\ \hline
68 &   & 4.8058 & &         0.2177   \\ \hline
70 & 2.0313 &  & &                      \\ \hline
72 &   & 6.2484 & &         0.2193   \\ \hline
75 & 3.0195 &  & &                      \\ \hline
76 &   & 8.2986 & &         0.2177   \\ \hline
80 & 3.9424 & 12.4141 & & 0.2099   \\ \hline
\end{tabular}
\end{table}

In Tables \ref{tab:ubcsat-saps} and \ref{tab:ubcsat-saps-avgtimes-old}, we see
that, at least at the millisecond scale, software SAT solve take progressively
more time, while for the D-Wave2x quantum annealer times remain stable.
\end{IGNORE}


\subsection{Weighted MaxSAT solving and sampling}
\label{sec:maxsat-exper}

\paragraph{Choosing the benchmarks}
To demonstrate the performance of the QA hardware in this regime, 
we generated MaxSAT instances that have many distinct optimal solutions. These problems were generated from the
2-in-4-SAT instances described above by removing a fraction of the constraints and then adding constraints on single variables with smaller weight\ignoreinlong{\hphantom{ }(details in \cite{bian17_frocos17extended})}. %
More precisely: 
\begin{enumerate}
\item Beginning with the 2-in-4-SAT instances of the previous section, we remove one of the partitions of the variable set, and change one 2-in-4 constraint to 1-in-4. (This makes the SAT problem unsatisfiable: for an $n$ variable problem, the first partition demands exactly $n/2$ true variables, while the second demands exactly $n/2-1$.)
\item We change the SAT problem into a weighted MaxSAT problem by assigning existing constraints a soft weight of 3 and randomly assigning each variable or its negation a soft constraint of weight 1.
\item We repeatedly generate MaxSAT instances of this form, until we find an instance in which the optimal solution has exactly one violated clause of weight 3 and at least $n/3$ violated clauses of weight 1, and at least $200$ distinct optimal solutions exist.
\end{enumerate}

\begin{table}[p]
\begin{center}
\subtable[\label{table:maxsat_firstsol}]{
\begin{tabular}{|l|r|r|}
\hline
\multicolumn{3}{|c|}{\textbf{D-Wave 2000Q}}\\\hline
\textbf{Problem size}&\textbf{\# solved}&\textbf{\begin{tabular}{@{}c@{}}\% optimal\\samples\end{tabular}}\\\hline

\textbf{32 vars}&100&78.7\\\hline
\textbf{36 vars}&100&69.0\\\hline
\textbf{40 vars}&100&60.2\\\hline
\textbf{44 vars}&100&49.9\\\hline
\textbf{48 vars}&100&40.4\\\hline
\textbf{52 vars}&100&35.2\\\hline
\textbf{56 vars}&100&24.3\\\hline
\textbf{60 vars}&100&22.3\\\hline
\textbf{64 vars}&99&17.6\\\hline
\textbf{68 vars}&99&13.0\\\hline
\textbf{72 vars}&98&9.6\\\hline
\textbf{76 vars}&94&6.6\\\hline
\textbf{80 vars}&93&4.3\\\hline
\end{tabular}}
\qquad
\subtable[\label{table:maxsat_firstsol_timing}]{
\begin{tabular}{|l|r|r|r|r|}
\hline
\multicolumn{5}{|c|}{\textbf{MaxSAT solvers: avg time (ms)}}\\\hline
\textbf{Problem
  size}&\textbf{g2wsat}&\textbf{rots}&\textbf{maxwalksat}&\textbf{novelty}\\\hline
\textbf{32 vars}&0.020&0.018&0.034&0.039\\\hline
\textbf{36 vars}&0.025&0.022&0.043&0.060\\\hline
\textbf{40 vars}&0.039&0.029&0.056&0.119\\\hline
\textbf{44 vars}&0.049&0.043&0.070&0.187\\\hline
\textbf{48 vars}&0.069&0.054&0.093&0.311\\\hline
\textbf{52 vars}&0.122&0.075&0.115&0.687\\\hline
\textbf{56 vars}&0.181&0.112&0.156&1.319\\\hline
\textbf{60 vars}&0.261&0.130&0.167&1.884\\\hline
\textbf{64 vars}&0.527&0.159&0.207&4.272\\\hline
\textbf{68 vars}&0.652&0.210&0.270&8.739\\\hline
\textbf{72 vars}&0.838&0.287&0.312&14.118\\\hline
\textbf{76 vars}&1.223&0.382&0.396&18.916\\\hline
\textbf{80 vars}&1.426&0.485&0.430&95.057\\\hline
\end{tabular}} 
\caption{\label{table:maxsat}%
(a) Number of \maxsattoqubo{} problem instances (out of 100) solved by
the QA hardware using 100 samples, and average fraction of samples
from the QA hardware that are optimal solutions. Annealing was
executed at a rate of \SI{10}{\micro\second} per sample, for a total
of \SI{1}{\milli\second} of anneal time per instance. \newline (b) Time in \SI{}{\milli\second} taken to find an optimal solution by various inexact weighted MaxSAT solvers, averaged over 100 MaxSAT instances of each problem size. Classical computations were performed on an Intel i7 2.90GHz $\times$ 4 processor. The solvers gw2sat\ignoreinshort{ \cite{li2005g2wsat}}, rots\ignoreinshort{ \cite{smyth2003rots}}, and novelty\ignoreinshort{ \cite{mcallester1997novelty}} are as implemented in \textsc{UBCSAT} \cite{tompkins_ubcsat:_2004}. All classical algorithms are performed with the optimal target weight specified; in the absence of a target weight they are much slower. }
\end{center}
\end{table}
%
\begin{table}
\begin{center}
\subtable[\label{table:maxsat_cutoff_dw}]{
\begin{tabular}{|l|r|r|}
\hline
\multicolumn{3}{|c|}{\textbf{D-Wave 2000Q}}\\\hline
\textbf{Size}&\textbf{anneal only}&\textbf{wall-clock}\\\hline
\textbf{32 vars}&448.5&443.9\\\hline
\textbf{36 vars}&607.0&579.9\\\hline
\textbf{40 vars}&1007.9&922.0\\\hline
\textbf{44 vars}&1322.6&1066.6\\\hline
\textbf{48 vars}&1555.4&1111.8\\\hline
\textbf{52 vars}&3229.0&1512.5\\\hline
\textbf{56 vars}&2418.9&1147.4\\\hline
\textbf{60 vars}&4015.3&1359.3\\\hline
\textbf{64 vars}&6692.6&1339.1\\\hline
\textbf{68 vars}&6504.2&1097.1\\\hline
\textbf{72 vars}&3707.6&731.7\\\hline
\textbf{76 vars}&2490.3&474.2\\\hline
\textbf{80 vars}&1439.4&332.7\\\hline
\end{tabular}}
\qquad
\subtable[\label{table:maxsat_cutoff_classical}]{
\begin{tabular}{|l|r|r|r|r|}
\hline
\multicolumn{5}{|c|}{\textbf{MaxSAT solvers}}\\\hline
\textbf{Size} &\textbf{g2wsat}&\textbf{rots}&\textbf{maxwalksat}&\textbf{novelty}\\\hline
\textbf{32 vars}&448.5&448.5&448.5&448.5\\\hline
\textbf{36 vars}&607.0&606.9&606.9&606.8\\\hline
\textbf{40 vars}&1007.7&1006.3&1005.3&1005.0\\\hline
\textbf{44 vars}&1313.8&1307.1&1311.7&1255.5\\\hline
\textbf{48 vars}&1515.4&1510.7&1504.9&1320.5\\\hline
\textbf{52 vars}&2707.5&2813.0&2854.6&1616.2\\\hline
\textbf{56 vars}&2021.9&2106.2&2186.6&969.8\\\hline
\textbf{60 vars}&2845.6&3061.7&3289.0&904.4\\\hline
\textbf{64 vars}&3100.0&4171.0&4770.0&570.6\\\hline
\textbf{68 vars}&2742.2&3823.3&4592.4&354.8\\\hline
\textbf{72 vars}&1841.1&2400.2&2943.4&212.6\\\hline
\textbf{76 vars}&1262.5&1716.0&2059.2&116.4\\\hline
\textbf{80 vars}&772.2&1111.1&1363.9&66.7\\\hline
\end{tabular}}
\caption{\label{table:maxsat_numsols}%
Number of distinct optimal solutions found in 1 second by various MaxSAT solvers, averaged across 100 instances of each problem size. \newline(a) ``anneal only" accounts for only the \SI{10}{\micro\second} per sample anneal time used by the D-Wave processor. ``wall-clock" accounts for all time used by the D-Wave processor, including programming and readout. \newline(b) Classical computations were performed as in Table \ref{table:maxsat_firstsol_timing}.}
\end{center}
\end{table}

As discussed in \sref{sec:maxsat}, determining an appropriate gap for chains in MaxSAT problems is more complicated than for SAT problems, and finding the smallest viable chain gap may be difficult analytically. However, a gap may be found experimentally by sweeping over a range of values and choosing one that results in optimal performance. Chain gaps that are too small result in a large number of broken chains, while chain gaps that are too large result in gaps for problem constraints that are smaller than the noise levels of the hardware, yielding solutions that are far from optimal. For the MaxSAT experiments in this section, the chosen chain gap was always in the range $g_{chain} \in [2,6]$ (relative to penalty functions $\Pxa$ with $\theta_i \in [-2,2], \theta_{ij} \in [-1,1]$.)

\paragraph{Experiments and Results}
Table \ref{table:maxsat} summarizes the performance of the D-Wave
processor in generating a single optimal MaxSAT solution, as well as
the run-times for various high-performing SLS MaxSAT solvers. The QA
hardware solves almost all problems with 100 samples/instance (i.e. within \SI{1}{m\second} of anneal time). Remark~\ref{remark:fairness} also applies here.
One of the strengths of D-Wave's processor is its ability to rapidly sample the near-optimal solutions: current systems typically anneal at a rate of \SI{10}{\micro\second} or \SI{20}{\micro\second} per sample and are designed to take thousands of samples during each programming cycle. As a result, the first practical benefits of QAs will likely come from applications which require many solutions rather than a single optimum. 

To this extent, 
Table \ref{table:maxsat_numsols} considers generating distinct optimal solutions. For each solver and problem size, the table indicates the number of distinct solutions found in 1 second, averaged across 100 problem instances of that size. For the smallest problems, 1 second is sufficient for all solvers to generate all solutions, while the diversity of solutions found varies widely as problem size increases. Although the D-Wave processor returns a smaller fraction of optimal solutions for MaxSAT instances than for the SAT instances, it is still effective in enumerating distinct optimal solutions because its rapid sampling rate.




\begin{table}[p]
\begin{center}
\subtable[\label{table:maxsat_hack_firstsol}]{
\begin{tabular}{|l|r|r|}
\hline
\multicolumn{3}{|c|}{\textbf{D-Wave 2000Q}}\\\hline
\textbf{Problem size}&\textbf{\# solved}&\textbf{\begin{tabular}{@{}c@{}}\% optimal\\samples\end{tabular}}\\\hline
\textbf{32 vars}&100&97.5\\\hline
\textbf{36 vars}&100&95.7\\\hline
\textbf{40 vars}&100&92.9\\\hline
\textbf{44 vars}&100&91.1\\\hline
\textbf{48 vars}&100&88\\\hline
\textbf{52 vars}&100&86.1\\\hline
\textbf{56 vars}&100&83.5\\\hline
\textbf{60 vars}&100&83.1\\\hline
\textbf{64 vars}&100&80.8\\\hline
\textbf{68 vars}&100&81\\\hline
\textbf{72 vars}&100&79.5\\\hline
\textbf{76 vars}&100&79\\\hline
\textbf{80 vars}&100&75.1\\\hline
\end{tabular}}
\qquad
\subtable[\label{table:maxsat_hack_firstsol_timing}]{
\begin{tabular}{|l|r|r|r|r|}
\hline
\multicolumn{5}{|c|}{\textbf{MaxSAT solvers: avg time (ms)}}\\\hline
\textbf{Problem size}&\textbf{g2wsat}&\textbf{rots}&\textbf{maxwalksat}&\textbf{novelty}\\\hline
\textbf{32 vars}&0.018&0.013&0.025&0.012\\\hline
\textbf{36 vars}&0.024&0.019&0.036&0.018\\\hline
\textbf{40 vars}&0.037&0.030&0.052&0.024\\\hline
\textbf{44 vars}&0.049&0.041&0.076&0.038\\\hline
\textbf{48 vars}&0.070&0.064&0.115&0.056\\\hline
\textbf{52 vars}&0.102&0.099&0.176&0.080\\\hline
\textbf{56 vars}&0.153&0.161&0.262&0.117\\\hline
\textbf{60 vars}&0.217&0.252&0.403&0.171\\\hline
\textbf{64 vars}&0.303&0.383&0.598&0.241\\\hline
\textbf{68 vars}&0.434&0.604&0.938&0.362\\\hline
\textbf{72 vars}&0.620&0.964&1.448&0.551\\\hline
\textbf{76 vars}&0.914&1.536&2.262&0.829\\\hline
\textbf{80 vars}&1.364&2.567&3.618&1.312\\\hline
\end{tabular}} 
\caption{\label{table:maxsat_hack}%
(a) Number of \maxsattoqubo{} problem instances (out of 100) solved by the QA hardware using 100 samples, and average fraction of samples from the QA hardware that are optimal solutions, for the ``unbiased" MaxSAT instances. Annealing was executed at a rate of \SI{10}{\micro\second} per sample, for a total of \SI{1}{\milli\second} of anneal time per instance. \newline(b) Time in \SI{}{\milli\second} taken to find an optimal solution by various inexact weighted MaxSAT solvers, averaged over 100 MaxSAT instances of each problem size. Classical computations were performed on an Intel i7 2.90GHz $\times$ 4 processor. gw2sat\ignoreinshort{ \cite{li2005g2wsat}}, rots\ignoreinshort{ \cite{smyth2003rots}}, and novelty\ignoreinshort{ \cite{mcallester1997novelty}} are as implemented in \textsc{UBCSAT} \cite{tompkins_ubcsat:_2004}. All classical algorithms are performed with the optimal target weight specified; in the absence of a target weight they are much slower. }
\end{center}
\end{table}

\paragraph{Alternative penalty functions}

Different penalty functions can result in different QA performance, even when those penalty functions have the same gap between ground and excited states. As an example of this, we describe another set of MaxSAT instances which result in better performance on the D-Wave 2000Q processor relative to classical solvers, even though the penalty functions they use are less theoretically justified. 

We call these instances ``unbiased" to distinguish them from the MaxSAT instances of the previous section. They are generated as follows. Beginning with the \sgen{} 2-in-4-SAT instances, we first change one 2-in-4 constraint to 1-in-4, making the SAT problem unsatisfiable. We then remove $5$ constraints from one partition of the variable set. This increases the total number of optimal solutions. Finally, we treat the resulting constraints as a MaxSAT problem in which each 1-in-4 or 2-in-4 constraint has the same weight.  Despite having many solutions, these problems become difficult for MaxSAT solvers with a relatively small number of variables. 

When solving these instances, we represent each 2-in-4-MaxSAT constraint by the following penalty function: 
$\Pxa = 4 + x_1x_2 + x_1x_4 + x_2x_3 + x_3x_4 - x_1a_1 - x_2a_2 + x_3a_1 + x_4a_2$. This model satisfies:
\[
\min_{\as} \Pxa = \begin{cases}
0, & \sum_i x_i = 0; \\
2, & |\sum_i x_i | = 1; \\
8, & |\sum_i x_i | = 2.
\end{cases}
\]
Because the unsatisfiable states $|\sum_i x_i | = 1$ and $|\sum_i x_i | = 2$ have different minimal energy configurations, this is not an \emph{exact} penalty function as required for MaxSAT as in \eqref{eq:encoding2-unrolled_out_exact}. Nevertheless, this model performs well in practice, because for the unbiased MaxSAT instances only configurations with $|\sum_i x_i | \leq 1$ are of interest. 
 
Table \ref{table:maxsat_hack} summarizes the performance of the D-Wave hardware and classical solvers in finding an optimal solution for the unbiased MaxSAT instances. It is instructive to compare these results to the ``biased" MaxSAT instances in Table \ref{table:maxsat}. The unbiased instances require more time for the best classical solvers to solve, yet result in better D-Wave hardware performance, despite the fact that the penalty function used is not exact.

\section{Ongoing and Future Work}
\label{sec:future}
\begin{figure}[t]
\includegraphics[width=\textwidth]{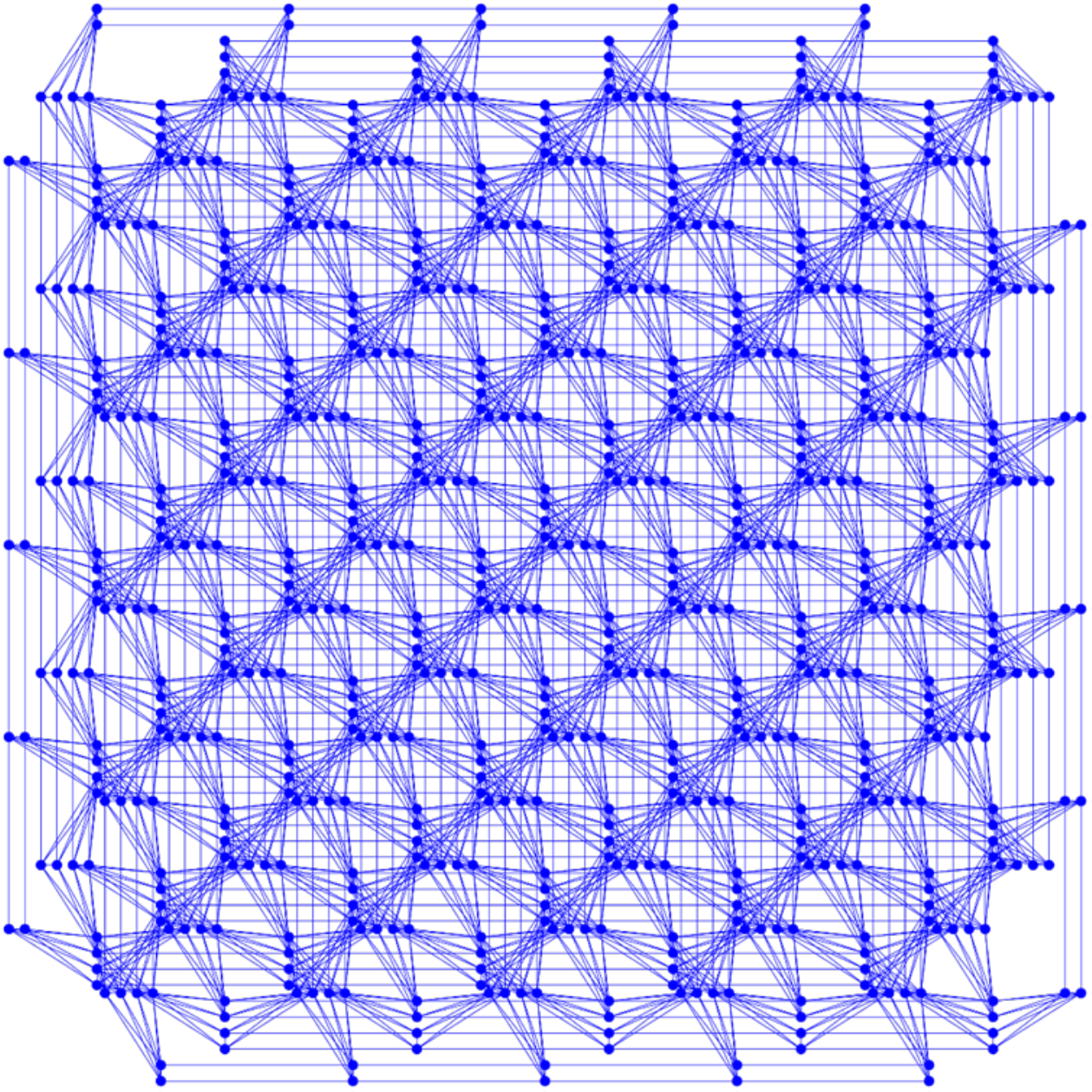}
\caption{\label{fig:[egasus} ``Pegasus", the hardware graph of an
  experimental QA system under development at D-Wave (720-qubit version). Qubits have maximum degree $15$ rather than $6$, and qubits do not fall into well-defined unit tiles as in Chimera.}
  \label{fig:pegasus}
\end{figure}

Future QA architectures will be larger and more connected, enabling more efficient encodings of larger and more difficult SAT problems. Faster and more scalable SMT-based encoding methods for small Boolean functions is currently an important direction of research. The ability to increase the number of ancillary variables can lead to larger gaps, which in turn can make quantum annealing more reliable. 
Among the encoding challenges presented in this paper, a few are of particular interest and relevance to SMT research:
\begin{itemize}
\item \emph{Variable placement.} Methods for simultaneously placing variables and computing penalty functions are currently less scalable, and have been less studied, than those for fixed variable placements.
\item \emph{Augmenting penalty functions.} For large Boolean functions, generating penalty functions directly from SMT becomes difficult because the number of constraints grows much more quickly than the number of available parameters. Function decomposition and chains provide one way around this, but chains limit the resulting gaps. There may be other methods of recombining a decomposed function that are not so restrictive. Alternatively, it may be possible to augment an existing penalty function with additional qubits for the purposes of increasing its gap. SMT formulations of these problems have not yet been explored.

\item \emph{Solving \eqref{eq:encoding1} directly.} In the field of
  automated theorem proving and SMT, novel techniques for solving
  {\em quantified} SMT formulas are emerging. It is thus possible to
  investigate these techniques for solving directly the quantified
  formulas \eqref{eq:encoding1}, avoiding thus the expensive Shannon
  expansion of \eqref{eq:encoding2-unrolled}-\eqref{eq:encoding2-unrolled_out}. 

\item \emph{Better function decompositions.} While Boolean function
  decomposition and minimization are mature classical subjects, those
  algorithms can probably be improved by taking into consideration the
  specifics of the embedding (placement and routing onto a QA hardware
  graph) that follow them.
\item \emph{More connected topologies}. Future QA hardware graphs will be larger, have higher per-qubit connectivity, and have less separation between clusters (tiles) of qubits. An example of a next-generation hardware graph under development at D-Wave is shown in Figure \ref{fig:pegasus}. While these changes will result in the ability to solve larger and more difficult Ising problems, they will also require new encoding strategies. In particular, new methods for problem decomposition, placing small Boolean functions, and penalty modelling that take advantage of additional connectivity will significantly improve the encoding process.

\end{itemize}
Furthermore, we believe the problems presented here are not only
practical, but also complex enough to be used to challenge new SMT
solvers. To encourage the use of these problems as SMT benchmarks, we
have provided example .smt files on the website of supplementary
materialì\footnote{See Footnote \ref{footnote:website}.}.

\section*{References}
\bibliography{rs_refs,rs_ownrefs,rs_specific,sathandbook,citations,main-blx,related_refs,sv_refs}

\end{document}